\documentclass[preprintnumbers,superscriptaddress,amsmath,amssymb,nofootinbib,twocolumn,showpacs]{revtex4-1}

\usepackage{dcolumn}
\usepackage{bm}
\usepackage{graphicx}
\usepackage{subcaption}

\usepackage{color}
\usepackage{natbib}
\usepackage{twoopt}
\newcommand{\beq}{\begin{equation}}
\newcommand{\eeq}{\end{equation}}
\newcommand{\ben}{\begin{eqnarray}}
\newcommand{\een}{\end{eqnarray}}
\newcommand{\bi}{\begin{itemize}}
\newcommand{\ei}{\end{itemize}}
\newcommand{\nn}{\nonumber}

\newcommand{\eg}{\textit{e.g.}}

\newcommand{\etal}{\textit{et al.}}

\newcommand{\citeeq}[1]{Eq.~(\ref{#1})}

\newcommand{\citefig}[1]{Fig.~\ref{#1}}

\newcommand{\fsol}{{\ifmmode f_{\odot} \else $f_{\odot}$\fi}}

\newcommand{\sqgev}{{\ifmmode {\rm GeV}^2 \else ${\rm GeV}^2$\fi}}

\newcommand{\sr}{{\ifmmode {\rm sr} \else ${\rm sr}$\fi}}
\newcommand{\invsr}{{\ifmmode {\rm sr}^{-1} \else ${\rm sr}^{-1}$\fi}}

\newcommand{\scnd}{{\ifmmode {\rm s} \else ${\rm s}$\fi}}
\newcommand{\invscnd}{{\ifmmode {\rm s}^{-1} \else ${\rm s}^{-1}$\fi}}

\newcommand{\kpc}{{\ifmmode {\rm kpc} \else ${\rm kpc}$\fi}}
\newcommand{\invkpc}{{\ifmmode {\rm kpc}^{-1} \else ${\rm kpc}^{-1}$\fi}}

\newcommand{\sqkpc}{{\ifmmode {\rm kpc}^{2} \else ${\rm kpc}^{2}$\fi}}
\newcommand{\invsqkpc}{{\ifmmode {\rm kpc}^{-2} \else ${\rm kpc}^{-2}$\fi}}

\newcommand{\cm}{{\ifmmode {\rm cm} \else ${\rm cm}$\fi}}
\newcommand{\invcm}{{\ifmmode {\rm cm}^{-1} \else ${\rm cm}^{-1}$\fi}}

\newcommand{\sqcm}{{\ifmmode {\rm cm}^2 \else ${\rm cm}^2$\fi}}
\newcommand{\invsqcm}{{\ifmmode {\rm cm}^{-2} \else ${\rm cm}^{-2}$\fi}}

\newcommand{\meter}{{\ifmmode {\rm m} \else ${\rm m}$\fi}}
\newcommand{\invmeter}{{\ifmmode {\rm m}^{-1} \else ${\rm m}^{-1}$\fi}}

\newcommand{\sqmeter}{{\ifmmode {\rm m}^2 \else ${\rm m}^2$\fi}}
\newcommand{\invsqmeter}{{\ifmmode {\rm m}^{-2} \else ${\rm m}^{-2}$\fi}}

\newcommand{\lcdm}{{\ifmmode \Lambda{\rm CDM} \else $\Lambda{\rm CDM}$\fi}}

\newcommand{\Rvirh}{{\ifmmode R_{\rm vir}^{\rm h} \else 
    $R_{\rm vir}^{\rm h}$\fi}}

\newcommand{\Ncl}{{\ifmmode N_{\rm cl} \else $N_{\rm cl}$\fi}}
\newcommand{\ncl}{{\ifmmode n_{\rm cl} \else $n_{\rm cl}$\fi}}
\newcommand{\phicl}{{\ifmmode \phi_{\rm cl} \else $\phi_{\rm cl}$\fi}}
\newcommand{\phicltot}{{\ifmmode \phi_{\rm cl}^{\rm tot} \else 
    $\phi_{\rm cl}^{\rm tot}$\fi}}
\newcommand{\Beff}{{\ifmmode B_{\rm eff} \else $B_{\rm eff}$\fi}}

\newcommand{\mchi}{\mbox{$m_{\chi}$}}

\newcommand{\sigv}{\mbox{$\langle \sigma v\rangle$}}

\newcommand{\aj}{Astron.~J.}
\newcommand{\aap}{Astron.~Astroph.}
\newcommand{\apjl}{Astrophys.~J.~Lett.}

\newcommand{\apss}{Astrophys.~Space~Sci.}

\newcommand{\jgr}{J.~Geophys.~Res.}
\newcommand{\jcap}{JCAP}
\newcommand{\mnras}{MNRAS}

\newcommand{\physrep}{Phys.~Rept.}
\newcommand{\ssr}{Space~Science~Rev.}

\usepackage[pdftex,
            breaklinks=true,%
            colorlinks=true,%
            linkcolor=red,
            urlcolor=blue,
            citecolor=blue,
            pdfauthor={Stref, Lavalle},%
            pdftitle={Template for article in XXX}%
           ]{hyperref}

\graphicspath{{./}{Figs/}}

\begin{document}

\title{Robust cosmic-ray constraints on $p$-wave annihilating MeV dark matter}
\author{Mathieu Boudaud}
\email{boudaud@lpthe.jussieu.fr}
\affiliation{Laboratoire de Physique Th\'eorique et Hautes \'Energies (LPTHE),
  UMR 7589 CNRS \& UPMC, 4 Place Jussieu, F-75252 Paris -- France}

\author{Thomas Lacroix}
\email{thomas.lacroix@lupm.in2p3.fr}
\affiliation{Instituto de F\'isica Te\'orica UAM/CSIC,
  Universidad Aut\'onoma de Madrid, E-28049 Madrid, Spain;
  Departamento de F\'isica Te\'orica, M-15, Universidad Aut\'onoma de Madrid,
  E-28049 Madrid, Spain}
\affiliation{Laboratoire Univers \& Particules de Montpellier (LUPM),
  CNRS \& Universit\'e de Montpellier (UMR-5299),
  Place Eug\`ene Bataillon,
  F-34095 Montpellier Cedex 05 --- France}

\author{Martin Stref}
\email{martin.stref@umontpellier.fr}
\affiliation{Laboratoire Univers \& Particules de Montpellier (LUPM),
  CNRS \& Universit\'e de Montpellier (UMR-5299),
  Place Eug\`ene Bataillon,
  F-34095 Montpellier Cedex 05 --- France}

\author{Julien Lavalle}
\email{lavalle@in2p3.fr}
\affiliation{Laboratoire Univers \& Particules de Montpellier (LUPM),
  CNRS \& Universit\'e de Montpellier (UMR-5299),
  Place Eug\`ene Bataillon,
  F-34095 Montpellier Cedex 05 --- France}

\begin{abstract}
  We recently proposed a method to constrain $s$-wave annihilating MeV dark matter from
  a combination of the Voyager 1 and the AMS-02 data on cosmic-ray electrons and positrons.
  Voyager 1 actually provides an unprecedented probe of dark matter annihilation to
  cosmic rays down to $\sim 10$ MeV in an energy range where the signal is mostly immune to
  uncertainties in cosmic-ray propagation. In this article, we derive for the first time new
  constraints on $p$-wave annihilation down to the MeV mass range using cosmic-ray data.
  To proceed, we derive a self-consistent velocity distribution for the dark matter across the
  Milky Way by means of the Eddington inversion technique and its extension to anisotropic systems.
  As inputs, we consider state-of-the-art Galactic mass models including baryons and constrained
  on recent kinematic data. They allow for both a cored or a cuspy halo. We then calculate the flux
  of cosmic-ray electrons and positrons induced by $p$-wave annihilating dark matter and
  obtain very stringent limits in the MeV mass range, robustly excluding cross sections greater
  than $\sim 10^{-22}{\rm cm^3/s}$ (including theoretical uncertainties), about 5 orders of
  magnitude better than current CMB constraints. This limit assumes that dark matter annihilation
  is the sole source of cosmic rays and could therefore be made even more stringent when reliable
  models of astrophysical backgrounds are included.
\end{abstract}

\pacs{12.60.-i,95.35.+d,96.50.S-,98.35.Gi,98.70.Sa}
\maketitle
\preprint{LUPM:18-046}
Thermal dark matter (DM) is one of the most appealing DM scenarios owing to its simplicity and
to the fact that it can be experimentally or observationally tested. It predicts that DM is
made of exotic particles with couplings to known elementary particles, such that they can be
produced in the early universe and driven to thermal equilibrium before their comoving abundance
is frozen as expansion takes over annihilation \cite{LeeEtAl1977a,BondEtAl1982}. If this decoupling
occurs when DM is nonrelativistic, we are left with cold DM (CDM), leading to a compelling
cosmological structure formation scenario \cite{Peebles1982}. A prototypical candidate is the WIMP
(weakly-interacting massive particle), which is currently actively searched for by a series of
experiments. If the dark sector is not overly complex, the typical relevant mass range for DM
particles exhibiting a thermal spectrum is $\sim10$ keV-100 TeV, which is bound from below by
structure formation \cite{ColombiEtAl1996,VielEtAl2005,VielEtAl2013,BoseEtAl2016,BaurEtAl2017},
and from above by unitarity limits \cite{GriestEtAl1990}. The lower mass bound can be raised up
to the MeV scale for WIMPs arising in minimal dark sectors \cite{WilkinsonEtAl2014}, still leaving
a wide range of possibilities \cite{LeaneEtAl2018}. The GeV-TeV scale is already under assault
thanks to the direct and indirect detection techniques (for reviews, see
\eg~\cite{LewinEtAl1996,FreeseEtAl2013,LavalleEtAl2012,BringmannEtAl2012c}),
and also thanks to particle colliders (\eg~\cite{FairbairnEtAl2007,Kahlhoefer2017}). However, the
sub-GeV and multi-TeV mass ranges are much less constrained and represent very interesting windows
yet to be explored, with the former potentially leading to interesting cosmological signatures
\cite{WilkinsonEtAl2014,Lopez-HonorezEtAl2016}. In this letter, we will mostly focus on the
sub-GeV scale.

The annihilation properties of WIMPs usually help define the most relevant search strategy. The
annihilation rate, proportional to the average velocity-weighted annihilation cross section
$\sigv$, is constrained at the time of chemical decoupling by the cosmological DM abundance
\cite{LeeEtAl1977a,BinetruyEtAl1984a,SrednickiEtAl1988,GondoloEtAl1991}. In the CDM scenario,
WIMPs decouple when nonrelativistic in the early universe at a temperature
$T_{\rm f}= \mchi/x_{\rm f}$, where $\mchi$ is the WIMP mass and $x_{\rm f}\sim 20$. In most cases,
the annihilation cross section can be expanded in powers of
$x^{-1}\equiv(\mchi/T)^{-1}\propto v^{2} \ll 1$ (see some exceptions in~\cite{GriestEtAl1991a}).
Making the units explicit, we may write this expansion as
\ben
\label{eq:sigv}
\sigv &=& \sigv_{s\text{-wave}} + \sigv_{p\text{-wave}} + \text{higher orders}\\
&=&\sigma_0 \, c + \sigma_1\,c\, \langle \frac{v_{\rm r}^2}{c^2} \rangle +
      {\cal O}\left( \frac{v_{\rm r}^4}{c^4}\right)\,,\nn
\een
where $\sigma_0$ and $\sigma_1$ are model-dependent cross-section terms that encode the WIMP
interaction properties, $v_{\rm r}\ll c$ is the relative WIMP speed (in a 2-particle system),
$c$ is the speed of light, and $\langle \rangle$ denotes an average over the velocity
distribution.\footnote{In the context of the relic density calculation where a
  Maxwell-Boltzmann distribution is assumed for WIMPs, the expansion is often made in terms of
  inverse powers of $x\equiv m_\chi/T$, $T$ being the WIMP temperature. The correspondence with
  \citeeq{eq:sigv} is $\sigma_1 \langle v_{\rm r}^2 \rangle \leftrightarrow (3/2)
  \sigma_1 /x$, here in natural units. For a computation in galactic halos,
  the speed $v_{\rm r}$ to average over is the {\em relative} speed between annihilating particles.}
This form is particularly well suited to consistently compare the constraints coming from very
different probes. The speed-independent term $\propto\sigma_0$ is called $s$-wave
annihilation in analogy with the partial-wave expansion technique. WIMPs annihilating through
$s$-wave terms can easily be probed by indirect searches because they efficiently annihilate
in regions and/or epochs where DM is locally dense enough. This is for example the case at the
time of recombination when the cosmic microwave background (CMB) was emitted, or in the centers of
galactic halos in the present universe. The next annihilation term $\propto \sigma_1$ is called
$p$-wave annihilation. In the following, we will concentrate on the latter and assume that
$\sigma_0=0$.

The WIMP relic abundance sets constraints on the annihilation cross section at a speed ---or
inverse temperature--- in the early universe ($v^2\sim 3\,T/m_\chi \sim 0.15$), typically much
larger than that in galactic halos ($v^2\sim 10^{-6}$), and even much larger than at the
recombination epoch ($v^2\sim 10^{-9}$). This has no impact on the $s$-wave annihilation rate which
only depends on the squared DM density, but makes $p$-wave annihilation much more difficult to
probe with indirect searches. Instead, $p$-wave annihilating WIMPs can be more efficiently probed
by direct searches, because rotated annihilation Feynmann diagrams correspond to elastic scattering
which is usually not velocity-suppressed when annihilation is. A classical example is that of
fermionic WIMPs annihilating into standard model fermions through a neutral scalar mediator in the
$s$-channel \cite{AbdallahEtAl2015}. However, direct DM searches are currently mostly efficient in
the GeV-TeV mass range \cite{MarrodanUndagoitiaEtAl2015}, leaving the sub-GeV mass range unexplored
(but see \cite{EssigEtAl2012,KouvarisEtAl2017}).

We recently derived \cite{BoudaudEtAl2017} strong constraints on $s$-wave annihilating MeV
DM from measurements of MeV cosmic-ray (CR) electrons and positrons by the famous Voyager 1
(V1) spacecraft, launched in 1977 \cite{KrimigisEtAl1977}. V1 crossed the heliopause in 2012,
which has allowed it to collect interstellar sub-GeV CRs prevented by the solar magnetic field from
reaching the Earth \cite{StoneEtAl2013,CummingsEtAl2016}. Our bounds were extended to $\sim$TeV
energies thanks to the AMS-02 data on positron CRs \cite{AguilarEtAl2014}. These limits are nicely
complementary to those extracted from the CMB data
\cite{AdamsEtAl1998,ChenEtAl2004,Slatyer2016,LiuEtAl2016,PlanckCollaboration2018a}, a
completely different probe. In the present article, we go beyond these results and compute the
V1 and AMS-02 constraints on $p$-wave annihilation in detail. As we will see, in contrast to the
$s$-wave case, these constraints will be much more stringent than those inferred from the CMB
\cite{LiuEtAl2016} or the diffuse extragalactic gamma-ray background (EGB)
\cite{EssigEtAl2013a,MassariEtAl2015}.

In the $p$-wave annihilation rate, the cross section no longer factorizes out of the volume
integral of the squared WIMP mass density $\rho^2$. Indeed, the cross section has an explicit
relative-speed dependence which is itself expected to vary across the Galactic halo. Therefore,
the source term for the injection of CR electrons and positrons becomes:
\ben
\label{eq:source}
    {\cal Q}^{\text{$p$-wave}}_{e^+/e^-}(\vec{x},E) =&
    \delta_\chi \frac{\sigma_1\,c}{2}\left\{\frac{\rho(\vec{x})}{m_\chi}\right\}^2
    \frac{dN_{e^+/e^-}}{dE}\\
     \times \int d^3\vec{v}_1\int d^3\vec{v}_2 &
    \frac{|\vec{v}_2-\vec{v}_1|^2}{c^2} 
    f_{\vec{v}}(\vec{v}_1,\vec{x})\,f_{\vec{v}}(\vec{v}_2,\vec{x})\nn \\
    =&\delta_\chi \frac{\sigma_1\,c}{2}\frac{dN_{e^+/e^-}}{dE}
    \left\{\frac{\rho_{\rm eff}(\vec{x})}{m_\chi}\right\}^2
    \,,\nn
\een
with
\ben
\rho^2_{\rm eff}(\vec{x})&\equiv& \rho^2(\vec{x}) \int d^3\vec{v}_1\int d^3\vec{v}_2
\frac{|\vec{v}_2-\vec{v}_1|^2}{c^2} f_{\vec{v}}(\vec{v}_1,\vec{x})\,f_{\vec{v}}(\vec{v}_2,\vec{x})\nn\\
&=& \rho^2(\vec{x}) \langle\frac{v_{\rm r}^2}{c^2}\rangle_{\vec{v}_1,\vec{v}_2}(\vec{x})\,,
\label{eq:rho2eff}
\een
where $\delta_\chi=1$ (1/2) for Majorana (Dirac) DM fermions, $dN_{e^+/e^-}/dE$ is the injected
electron-positron spectrum, and $f_{\vec{v}}(\vec{v},\vec{x})$ is the normalized WIMP velocity
distribution that depends on the position in the Milky Way (MW). For each annihilation final
state, the injected CR spectrum will be determined from the Micromegas numerical package
\cite{BelangerEtAl2018}, based on the Pythia Monte Carlo generator \cite{SjoestrandEtAl2015}. All
allowed final-state radiation processes are included.

Equation~(\ref{eq:rho2eff}) shows that an important input in the $p$-wave signal is the
velocity distribution function (DF) of WIMPs in the system of interest. In many $p$-wave studies,
the latter is often assumed to be a Maxwell-Boltzmann (MB) distribution, either with a constant
velocity dispersion, or using the circular velocity as a proxy for the velocity dispersion. While
the MB approximation is perfectly sound in the early universe up to CMB times, it is much more
dubious in galaxies, which do not behave as isothermal spheres, especially in the densest central
regions \cite{MoEtAl2010,BinneyEtAl2008}. In this work, we adopt a more
theoretically motivated approach based on the Eddington inversion method \cite{Eddington1916a},
which relates the phase-space DF of WIMPs to their mass density profile and the total potential of
the MW (including baryons) -- see Ref.~\cite{LacroixEtAl2018} for an extensive critical review, to
which we refer the reader for all technical details, \eg~the calculation of
\citeeq{eq:rho2eff} (see also
\cite{FerrerEtAl2013,Hunter2014,BoddyEtAl2018}). This approach allows us to describe isotropic as
well as anisotropic systems \cite{Osipkov1979,Merritt1985a,Cuddeford1991}, assuming spherical
symmetry for the dark halo and the total gravitational potential. This actually provides a much
better {\em predictive} description of hydrodynamical cosmological simulations than the MB
approximation, even in the maximally symmetric approximation (spherical symmetry and isotropy),
especially in the central regions of galactic halos \cite{LacroixEtAl2018a}.

To compute the phase-space DF of WIMPs from the Eddington inversion method, we use the Galactic
mass model of Ref.~\cite{McMillan2017}, which is constrained against a series of recent kinematic
data. It includes a spherical DM halo (scaling in radius $r$ as $\propto r^{-\gamma}$ in the center,
with $\gamma\in[0,1]$, and as $\propto r^{-3}$ at large radii), and baryonic components comprising
a bulge and three disks (the thin and thick stellar disks, and a gaseous disk). All baryonic
components are ``sphericized'' to compute the DM DF \cite{LacroixEtAl2018}. To account for
uncertainties in the DM anisotropy, we considered both isotropic and anisotropic DFs. In the latter
case, we explore a wide range of possibilities by using both the radially anisotropic
Osipkov-Merritt (OM) model with an anisotropy radius $r_a$ set to the scale radius $r_s$ of the DM
halo profile, and a tangentially anisotropic model with a constant anisotropy parameter
$\beta=-0.3$. To further account for uncertainties in the dark halo shape, we consider two
inner-profile indices, $\gamma=1/4$ (coredlike profile) and $\gamma=1$ (cuspy profile \`a la
Navarro-Frenk-White (NFW) \cite{NavarroEtAl1996a,Zhao1996}). Taking a fully cored profile
would break necessary conditions for dynamical stability of the DF \cite{LacroixEtAl2018}.
Disregarding stability of the DF and forcing $\gamma=0$ would anyway provide results very similar
to $\gamma=0.25$, which is therefore a very conservative case.

Equation~(\ref{eq:source}) is the source term of a steady-state CR transport equation
\cite{GinzburgEtAl1964,BerezinskiiEtAl1990}, whose parameters (related to spatial diffusion,
energy losses, reacceleration, and convection) are standardly calibrated on secondary-to-primary
ratios \cite{StrongEtAl1998,JonesEtAl2001,MaurinEtAl2001,StrongEtAl2007,LavalleEtAl2012,Kissmann2014,EvoliEtAl2017,Maurin2018}. In the context of electron and positron CRs, we can solve this equation
by means of the {\em pinching} semianalytical method introduced in Ref.~\cite{BoudaudEtAl2017a},
compatible with the USINE framework \cite{Maurin2018}. This method capitalizes over previous
analytical developments optimized for energies beyond the GeV
\cite{BulanovEtAl1974,BaltzEtAl1998,LavalleEtAl2007}, but improves on the low-energy part
where radiative energy losses in the disk, diffusive reacceleration, and convection can dominate
over spatial diffusion. It basically allows us to recast an equation where part of the energy losses
occurs all over the magnetic halo by another analytically solvable equation where all losses are
pinched into an infinitely thin disk (see also \cite{DelahayeEtAl2009}) -- this limit is justified
as the Galactic disk half-height $h\sim 100$ pc is much smaller than that of the magnetic halo,
$L\gtrsim 5$ kpc \cite{LavalleEtAl2014,BoudaudEtAl2017a,ReinertEtAl2018}. Both equations have
solutions strictly equivalent {\em in the disk}, and the latter can hence be used in the context
of local DM searches. In this work (and \cite{BoudaudEtAl2018}), we slightly modify it to get more
accurate results when the propagation length gets smaller than $h$, in which case the pinching
approximation breaks down. In this regime, however, the spatial boundaries of the magnetic halo
become irrelevant such that the infinite three-dimensional solution
\cite{LavalleEtAl2007,DelahayeEtAl2009,DelahayeEtAl2010} safely applies. This typically occurs at
energies close to the injected energy. This correction is therefore important to accurately
compute the local flux induced by a quasimonochromatic injection, like in the process
$\chi\bar \chi\to e^+ e^-(\gamma)$.  A nonsingular transition is further easily implemented
between the two regimes. The same approach is used to predict the secondary positron background
induced by the scattering of CR nuclei off the interstellar medium (ISM), and provides better
precision than previous similar calculations
\cite{DelahayeEtAl2009,DelahayeEtAl2010,Lavalle2011b,BoudaudEtAl2017a}. While we expect additional
primary contributions in the sub-GeV range from electron-positron sources like pulsars
\cite{Shen1970,AharonianEtAl1995,HooperEtAl2009,DelahayeEtAl2010,Profumo2011,BoudaudEtAl2015},
likely responsible for the rise in the positron fraction beyond a few GeV
\cite{DuVernoisEtAl2001,AdrianiEtAl2009,AguilarEtAl2013}, we will not include them here
because associated predictions are still plagued with large theoretical uncertainties. Therefore,
our limits on DM annihilation can be considered as very conservative.

\begin{figure*}[!t]
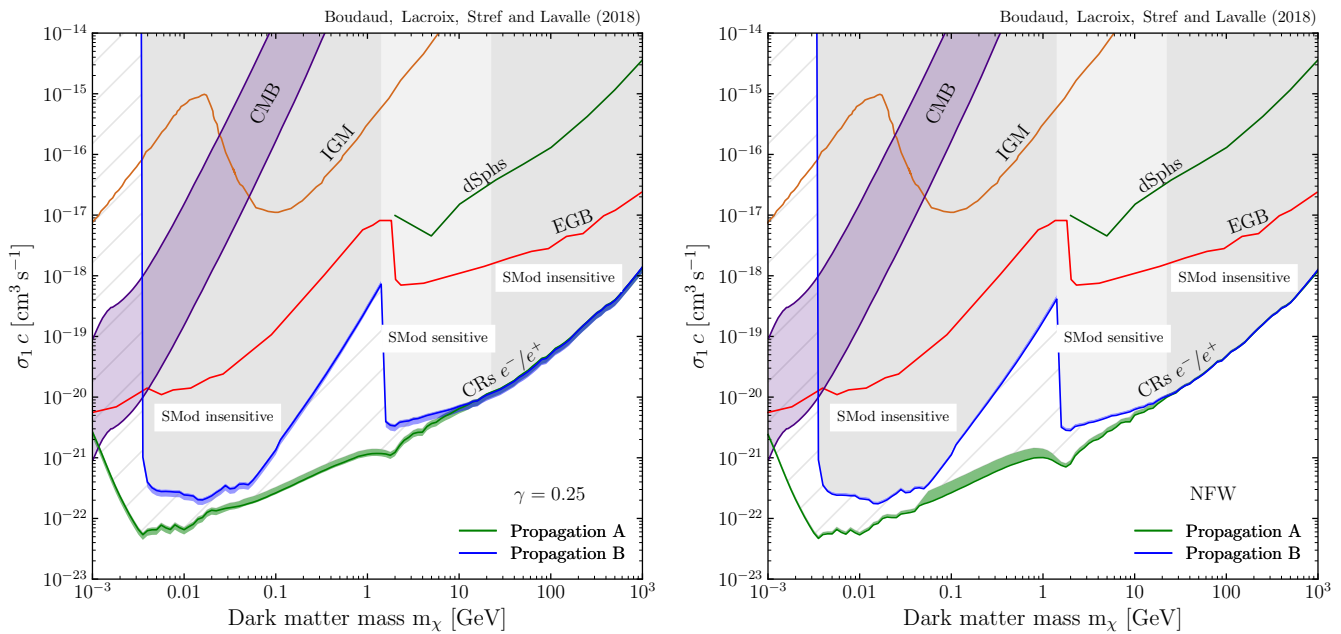

\centering
\includegraphics[width = 0.495\textwidth]{{{sigmav_bounds_gamma_pt25}}}
\includegraphics[width = 0.495\textwidth]{{{sigmav_bounds_NFW}}}
\caption{\small Limits on the $p$-wave cross section as a function of the
  WIMP mass \mchi, in the 100\% $\chi\bar \chi\to e^+ e^- (\gamma)$ channel, for a
  high-reacceleration propagation model (model $A$---green curve) or a reacceleration-less,
  very conservative, propagation model (model $B$---blue curve). Uncertainty bands account
  for uncertainties in the anisotropy of the WIMP velocity DF. Also shown are the limits obtained
  with the CMB \cite{Slatyer2016,LiuEtAl2016,PlanckCollaboration2018a},
  IGM \cite{LiuEtAl2016}, EGB \cite{EssigEtAl2013a,MassariEtAl2015}, and gamma-ray observations
  of dwarf galaxies \cite{ZhaoEtAl2016}. {\bf Left panel}: coredlike DM
  density profile with $\gamma=0.25$ \cite{McMillan2017}. {\bf Right panel}: cuspy DM density
  profile with $\gamma=1$ \cite{McMillan2017}.}
\label{fig:excl_pwave}
\end{figure*}

For CR propagation, we use two extreme cases identified in Ref.~\cite{BoudaudEtAl2017}: one with
strong reacceleration (model $A$), allowing CR electrons and positrons to get energies higher than
the WIMP mass in the MeV range; another with negligible reacceleration (model $B$). In both cases,
energy losses in the MeV range have a timescale much smaller than the then subdominant spatial
diffusion and convection processes (no longer true in the GeV range). Model $A$ is the MAX model
proposed in \cite{MaurinEtAl2001,DonatoEtAl2004}, whose main feature beside a pseudo-Alfven velocity
$V_{\rm a}\sim 100$ km/s is a large magnetic halo with $L=15$ kpc, making it a very optimistic
setup for DM signal predictions. Though calibrated on old secondary-to-primary CR data, this setup
is still valid for its general characteristics \cite{BoudaudEtAl2018a}. Model $B$ is the model of
Ref.~\cite{ReinertEtAl2018} best fitting the recent AMS-02 B/C data \cite{AguilarEtAl2016a}, and
accounting for a spectral break in the diffusion coefficient \cite{AhnEtAl2010,AdrianiEtAl2011a,AguilarEtAl2015,AguilarEtAl2015a,BlasiEtal2012a,Blasi2017,GenoliniEtAl2017,EvoliEtAl2018}. It is very
conservative because it assumes the smallest possible magnetic halo with $L=4.1$ kpc
\cite{LavalleEtAl2014,BoudaudEtAl2015a,ReinertEtAl2018} (hence minimizing the yield from DM
annihilation), and has negligible reacceleration, reducing the flux predictions below a few GeV.

Since losses in the MeV range are caused by radiative interactions with the ISM, whose
average properties over $\sim 100$-pc scales are well controlled
\cite{Ferriere2001,JohannessonEtAl2018}, uncertainties in the diffusion parameters have
no impact on predictions in this energy range. Model $B$ thus provides a robust and
conservative limit as far as interstellar CR propagation is concerned \cite{BoudaudEtAl2017}.
Moreover, for predictions associated with the energy range covered by V1, solar modulation of CRs
\cite{Potgieter2013} is irrelevant and does not contribute additional uncertainties. In the AMS-02
range ($\sim$GeV and above), we use the so-called force-field approximation
\cite{GleesonEtAl1968a,Fisk1971} to deal with solar modulation, with a conservative
Fisk potential estimate of $\phi = 830$ MV \cite{GhelfiEtAl2016}. In the GeV range, spatial
diffusion takes over, and propagation uncertainties can in principle be larger. However, above a
few tens of GeV, inverse Compton and synchrotron losses become the main transport processes and
propagation uncertainties reduce to those in the magnetic and interstellar radiation
fields (B-field \cite{SunEtAl2010,JanssonEtAl2012} and ISRF \cite{PorterEtAl2017})---in this
high-energy limit, both propagation models converge, and solar modulation becomes irrelevant again.
Uncertainties related to the DM density profile are estimated by considering both the coredlike
and NFW profiles introduced above, whose fits to kinematic data provide a very similar local DM
density $\rho_\odot\simeq 0.01 \,M_\odot/{\rm pc}^3$ \cite{McMillan2017}. We also evaluate
the impact of uncertainties in the velocity DF by considering both radially and tangentially
anisotropic DFs, beside a reference isotropic DF.

We get limits on the $p$-wave cross section of \citeeq{eq:sigv} as a function of the
WIMP mass assuming a full annihilation in $e^+e^-(\gamma)$. Our results are shown in
\citefig{fig:excl_pwave}, where the left (right) panel corresponds to the $\gamma=0.25$ (1)
halo profile. We also show complementary bounds obtained with the CMB
\cite{Slatyer2016,LiuEtAl2016} (purple) rescaled to the latest Planck results
\cite{PlanckCollaboration2018a}, the high-redshift intergalactic medium (IGM) temperature
\cite{LiuEtAl2016} (orange), the diffuse EGB \cite{EssigEtAl2013a,MassariEtAl2015} (red), and
gamma-ray observations of MW-satellite dwarf galaxies \cite{ZhaoEtAl2016} (dark green curve). The
CMB bound extrapolates the one obtained for $s$-wave annihilation by assuming a MB DF with a
temperature at redshift $\sim 600$ that depends on the WIMP kinetic decoupling temperature
$T_{\rm kd}$. We adopt two extreme values for the ratio $x_{\rm kd}= \mchi/T_{\rm kd}$, $10^2$ and
$10^4$, to cover most of the relevant parameter (the CMB limit is $\propto x_{\rm kd}^{-1}$).

Our limits are shown for both propagation models $A$ (green) and $B$ (blue curve), and the region
for which solar modulation of CRs is irrelevant (relevant) is indicated as ``SMod insensitive''
(``SMod sensitive''). The associated shaded areas account for uncertainties induced by the
unknown anisotropy of the WIMP DF. As stressed above, the limit obtained for model $B$ is
conservative. Moreover, associated propagation uncertainties in the V1 sub-GeV region reduce to
those in the low-energy energy losses, which are very small. This conservative result is strikingly
more constraining than complementary searches, by more than two orders of magnitude in the
5-100 MeV mass range, making CRs remarkable probes of $p$-wave annihilation. This contrasts
with constraints on $s$-wave annihilation, for which CMB bounds are stronger. Indeed, in the
$p$-wave case, the CMB probe is penalized by a DM ``temperature'' much lower at the recombination
epoch than in virialized halos today. Moreover, the CR probe has the advantage over gamma-ray
observations that predictions saturate the data with very small annihilation cross sections
without including any background. The secondary background is actually completely negligible in
the V1 energy range \cite{BoudaudEtAl2017}, while it gets close to the low-energy AMS-02 positron
data though in a regime where solar modulation matters.

The limits obtained with model $B$ relax around 0.1-1 GeV because there is no data available
between the V1 range and the AMS-02 one. This gap can still be probed by AMS-02 if propagation
is characterized by a significant re-acceleration, like in model $A$.  In that case, the low-energy
limits extend below the V1 energy threshold.

We note that our strongly improved limit lies now only at two orders of magnitude from the $p$-wave
cross section required for the correct WIMP abundance, $\sim 10^{-24}{\rm cm^3/s}$. Moreover, we
stress that it already excludes some interesting WIMP models with enriched dark sectors,
\eg~that of Ref.~\cite{ChoquetteEtAl2016}. Our work thus provides stringent constraints on particle
model building along this line. We give additional details about how we infer the prediction
uncertainties from the both those in the phase-space distribution and in the halo profile, as well
as predictions for other annihilation channels, in the appendices.
%
%

To summarize, we have used the electron and positron data from V1 and the positron data from
AMS-02 to constrain $p$-wave annihilating DM. We have obtained limits that are much more stringent
than those derived from complementary astrophysical messengers in the MeV--TeV energy range. Those
derived for our very conservative model $B$ are very robust in the V1 range (below the GeV),
because the flux predictions then only depend on the average ISM properties, on the halo model,
and on the anisotropy level in the DM DF. We have shown in \citefig{fig:excl_pwave} that using
a kinematically constrained cored vs. cuspy halo does not alter our result, nor does spanning
different anisotropy configurations. Moreover, above few tens of GeV, solar modulation gets
irrelevant again, and CR propagation is then set by inverse-Compton and synchrotron losses,
for which uncertainties reduce to those in the local ISRF and B-field. We emphasize that these
limits could be made even more severe if additional astrophysical primary contributions were
considered \cite{BoschiniEtAl2018}. This will likely be done in the future when more detailed
low-energy CR studies succeed in more reliably modeling the yield from these astrophysical sources.

\acknowledgments{
  We wish to thank Pierre Salati for early participation in this project. We also thank Alan
  C. Cummings and Martin Winkler for valuable exchanges. MB acknowledges
  support from the European Research Council (ERC) under the EU Seventh Framework
  Program (FP7/2007-2013)/ERC Starting Grant (agreement n. 278234 | NewDark project led
  by M. Cirelli). TL, JL, and MS are partly supported by the ANR project ANR-18-CE31-0006,
  the OCEVU Labex (ANR-11-LABX-0060), the CNRS IN2P3-Theory/INSU-PNHE-PNCG project
  ``Galactic Dark Matter'', and European Union's Horizon 2020 research and innovation program
  under the Marie Sk\l{}odowska-Curie grant agreements N$^\circ$ 690575 and No 674896 -- in
  addition to recurrent funding by CNRS and the University of Montpellier.
}

\appendix
\section{Effective density profile}
\label{app1}
In this section, we show the detailed results obtained for the effective squared density profile
defined in \citeeq{eq:rho2eff}. They are displayed on \citefig{fig:rho2eff}. The original
squared profile are shown in blue ($\gamma=0.25$) and red ($\gamma=1$) crosses (rescaled by
a factor of $\sim 10^{-6}\sim (\sigma_v/c)^2$, where $\sigma_v$ is the velocity dispersion in the
Milky Way), while the velocity-corrected profiles are shown in solid (dashed, dot-dashed, and
dotted) curves for the isotropic (constant tangential anisotropy, spatial-dependent radial
anisotropy, and Maxwellian) velocity DF. The bottom part of the plot shows the residuals with
respect to the isotropic, reference case. For the Maxwellian calculation, the velocity dispersion is
taken proportional to the circular velocity, consistently with the isothermal sphere approximation
\cite{BinneyEtAl2008}. We see that the velocity-weighted squared profiles lead to suppressed
annihilation luminosity with respect to the standard case. We also see that the Maxwell-Boltzmann
approximation strongly undershoots the luminosity arising from the other predictions, which
turn out to be much better supported both by theory and simulations
\cite{LacroixEtAl2018,LacroixEtAl2018a}. We have discarded the unrealistic Maxwellian case from
our limits.

Note that when going to CR flux predictions, diffusion plays the role of averaging the luminosity
over the CR horizon which is set by the dominant transport process at a given energy. Therefore,
to figure out the local CR yield induced by DM $p$-wave annihilation, one needs to average
the effective luminosity over the relevant volume around the observer, who sits at
$r\sim 8$ kpc from the Galactic center.

\begin{figure}[!t]
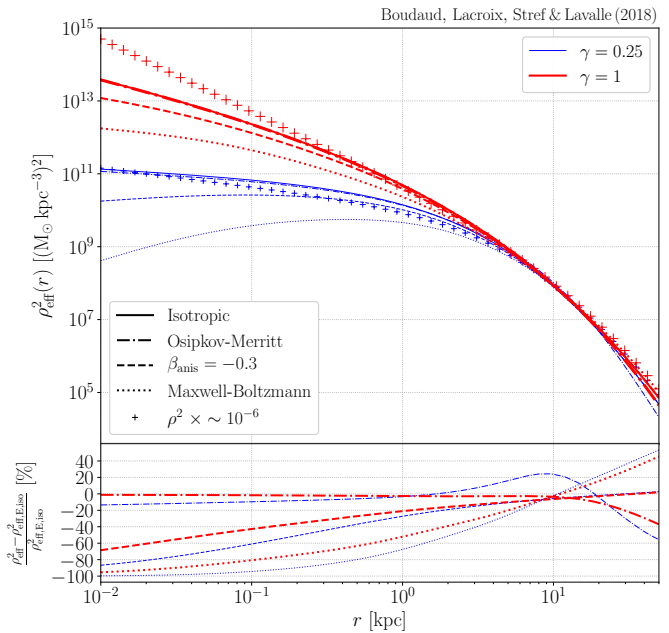

\centering
\includegraphics[width = 0.495\textwidth]{{{rho_sq_eff_all_profiles}}}
\caption{\small Effective squared density profiles, which translate the impact of the
  spatial-dependent average squared relative velocity. The effective profiles are shown for
  different assumptions for the WIMP velocity DF, made explicit in the legend.}
\label{fig:rho2eff}
\end{figure}

\section{Bounds on various annihilation channels}
\label{app2}
In \citefig{fig:channels}, we show the conservative limits we get for different annihilation
final states (propagation model $B$, and cored-like halo profile with $\gamma=0.25$). We see
that for final-state particle masses above a few GeV, limits are dominated by the AMS-02 data.
In contrast, the Voyager data are very powerful in constraining leptonic channels.

\begin{figure}[!t]
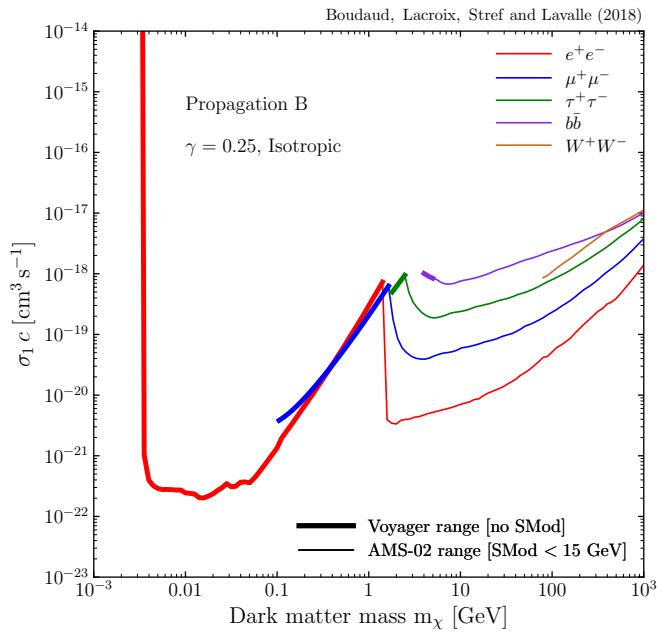

\centering
\includegraphics[width = 0.495\textwidth]{{{sigmav_bounds_gamma_pt25_channels}}}
\caption{\small Limits on the $p$-wave annihilation cross section obtained for different
  annihilation channels, using the most conservative setup for the CR flux predictions:
  propagation model $B$ (no reacceleration, minimal magnetic halo size $\sim 4$ kpc), and a
  cored-like halo profile ($\gamma=0.25$).}
\label{fig:channels}
\end{figure}

\section{Detailed view on theoretical uncertainties}
\label{app3}

In this section, we provide additional details about the origin of the uncertainties that
were featured as shaded areas around the limits shown in \citefig{fig:excl_pwave}. They
originate from the different assumptions made for the anisotropy in the velocity DF, which
is not firmly predicted by the non-linear theory and could therefore vary from a galaxy
to another. In order to span the most likely configurations, we adopted two contrasting cases,
one with constant tangential anisoptropy, another with spatial-dependent radial (OM)
anisotropy, on top on the isotropic Eddington case and the simplistic Maxwellian
approximation. The limits obtained for these different configurations are shown in
\citefig{fig:anis}. The top (bottom) panels correspond to the cored-like (NFW) halo with
$\gamma=0.25$ ($\gamma=1$). The left (right) panels are associated with CR predictions made
with optimistic (very conservative) propagation model $A$ ($B$) that is characterized by a
strong reacceleration and $L=15$ kpc (no reacceleration and $L=4.1$ kpc). All panels display
the limits obtained for all the velocity DFs mentioned above. These results can be easily
interpreted from the hierarchy in the luminosity yield shown in \citefig{fig:rho2eff}.

\begin{figure*}[!t]
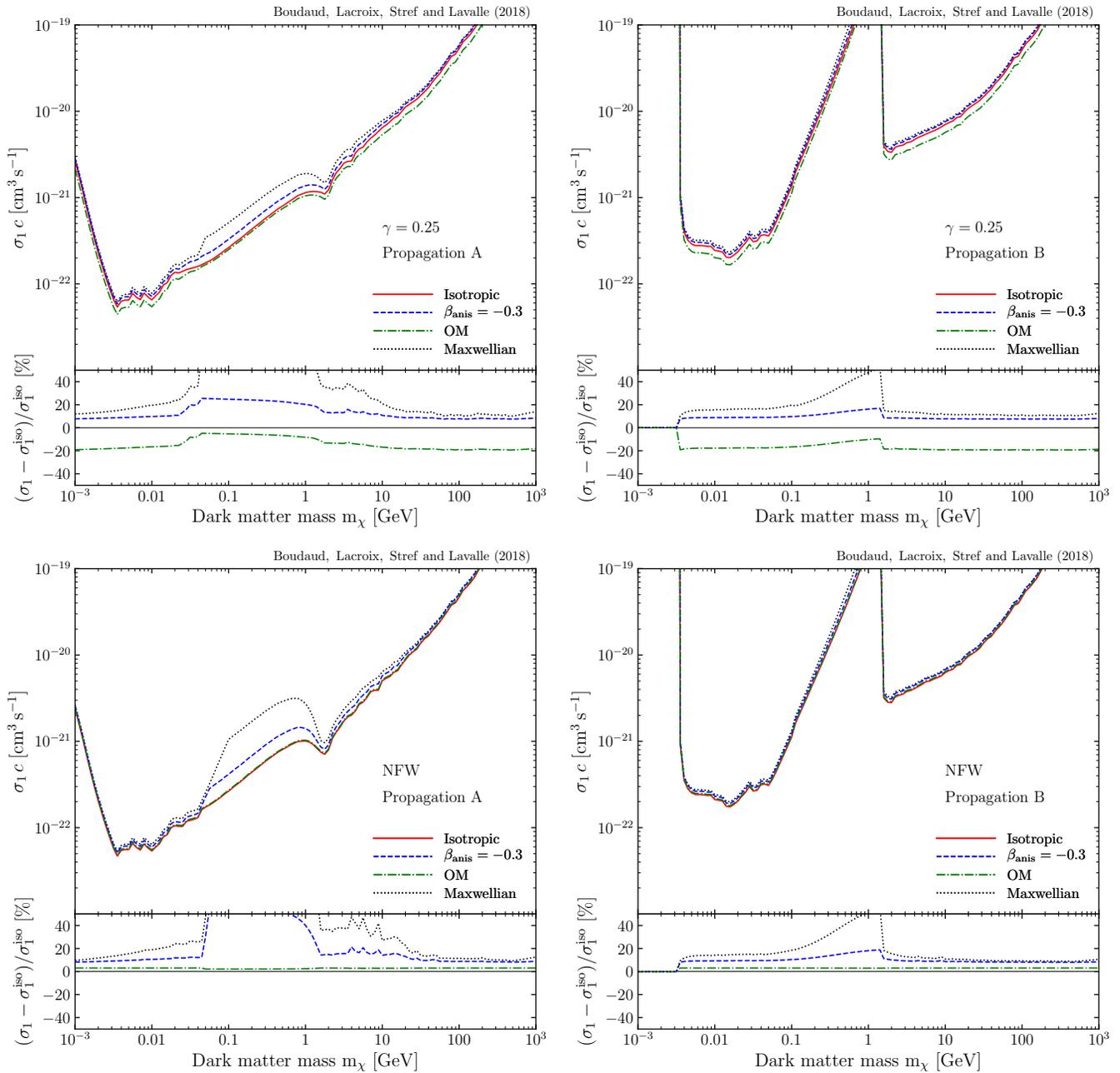

\centering
\includegraphics[width = 0.495\textwidth]{{{sigmav_bounds_ratios_e_AE_gamma=pt25_anis}}}
\includegraphics[width = 0.495\textwidth]{{{sigmav_bounds_ratios_e_BE_gamma=pt25_anis}}}
\includegraphics[width = 0.495\textwidth]{{{sigmav_bounds_ratios_e_AE_NFW_anis}}}
\includegraphics[width = 0.495\textwidth]{{{sigmav_bounds_ratios_e_BE_NFW_anis}}}
\caption{\small Impact of different assumptions for the anisotropy in the velocity
  DF, and relative comparison to the Eddington-inverted isotropic DF (solid red curves).
  The black dotted curves show the results obtained for the Maxwellian approximation, the
  blue dashed curves correspond to a constant anisotropy parameter $\beta=-0.3$ (tangential
  anisotropy), and the green dash-dotted curves correspond to the Osipkov-Merrit model
  with an anisotropy radius set to the scale radius of the DM halo. {\bf Left (right) panels}:
  propagation model $A$ ($B$). {\bf Top (bottom) panels}: cored (NFW) DM halo.
}
\label{fig:anis}
\end{figure*}

\bibliography{biblio_jabref}

\begin{thebibliography}{105}%
\makeatletter
\providecommand \@ifxundefined [1]{%
 \@ifx{#1\undefined}
}%
\providecommand \@ifnum [1]{%
 \ifnum #1\expandafter \@firstoftwo
 \else \expandafter \@secondoftwo
 \fi
}%
\providecommand \@ifx [1]{%
 \ifx #1\expandafter \@firstoftwo
 \else \expandafter \@secondoftwo
 \fi
}%
\providecommand \natexlab [1]{#1}%
\providecommand \enquote  [1]{``#1''}%
\providecommand \bibnamefont  [1]{#1}%
\providecommand \bibfnamefont [1]{#1}%
\providecommand \citenamefont [1]{#1}%
\providecommand \href@noop [0]{\@secondoftwo}%
\providecommand \href [0]{\begingroup \@sanitize@url \@href}%
\providecommand \@href[1]{\@@startlink{#1}\@@href}%
\providecommand \@@href[1]{\endgroup#1\@@endlink}%
\providecommand \@sanitize@url [0]{\catcode `\\12\catcode `\$12\catcode
  `\&12\catcode `\#12\catcode `\^12\catcode `\_12\catcode `\%12\relax}%
\providecommand \@@startlink[1]{}%
\providecommand \@@endlink[0]{}%
\providecommand \url  [0]{\begingroup\@sanitize@url \@url }%
\providecommand \@url [1]{\endgroup\@href {#1}{\urlprefix }}%
\providecommand \urlprefix  [0]{URL }%
\providecommand \Eprint [0]{\href }%
\providecommand \doibase [0]{http://dx.doi.org/}%
\providecommand \selectlanguage [0]{\@gobble}%
\providecommand \bibinfo  [0]{\@secondoftwo}%
\providecommand \bibfield  [0]{\@secondoftwo}%
\providecommand \translation [1]{[#1]}%
\providecommand \BibitemOpen [0]{}%
\providecommand \bibitemStop [0]{}%
\providecommand \bibitemNoStop [0]{.\EOS\space}%
\providecommand \EOS [0]{\spacefactor3000\relax}%
\providecommand \BibitemShut  [1]{\csname bibitem#1\endcsname}%
\let\auto@bib@innerbib\@empty
\bibitem [{\citenamefont {{Lee}}\ and\ \citenamefont
  {{Weinberg}}(1977)}]{LeeEtAl1977a}%
  \BibitemOpen
  \bibfield  {author} {\bibinfo {author} {\bibfnamefont {B.~W.}\ \bibnamefont
  {{Lee}}}\ and\ \bibinfo {author} {\bibfnamefont {S.}~\bibnamefont
  {{Weinberg}}},\ }\href {\doibase 10.1103/PhysRevLett.39.165} {\bibfield
  {journal} {\bibinfo  {journal} {\prl}\ }\textbf {\bibinfo
  {volume} {39}},\ \bibinfo {pages} {165} (\bibinfo {year} {1977})}\BibitemShut
  {NoStop}%
\bibitem [{\citenamefont {{Bond}}\ \emph {et~al.}(1982)\citenamefont {{Bond}},
  \citenamefont {{Szalay}},\ and\ \citenamefont {{Turner}}}]{BondEtAl1982}%
  \BibitemOpen
  \bibfield  {author} {\bibinfo {author} {\bibfnamefont {J.~R.}\ \bibnamefont
  {{Bond}}}, \bibinfo {author} {\bibfnamefont {A.~S.}\ \bibnamefont
  {{Szalay}}}, \ and\ \bibinfo {author} {\bibfnamefont {M.~S.}\ \bibnamefont
  {{Turner}}},\ }\href {\doibase 10.1103/PhysRevLett.48.1636} {\bibfield
  {journal} {\bibinfo  {journal} {\prl}\ }\textbf {\bibinfo
  {volume} {48}},\ \bibinfo {pages} {1636} (\bibinfo {year}
  {1982})}\BibitemShut {NoStop}%
\bibitem [{\citenamefont {{Peebles}}(1982)}]{Peebles1982}%
  \BibitemOpen
  \bibfield  {author} {\bibinfo {author} {\bibfnamefont {P.~J.~E.}\
  \bibnamefont {{Peebles}}},\ }\href {\doibase 10.1086/183911} {\bibfield
  {journal} {\bibinfo  {journal} {\apjl}\ }\textbf {\bibinfo {volume} {263}},\
  \bibinfo {pages} {L1} (\bibinfo {year} {1982})}\BibitemShut {NoStop}%
\bibitem [{\citenamefont {{Colombi}}\ \emph {et~al.}(1996)\citenamefont
  {{Colombi}}, \citenamefont {{Dodelson}},\ and\ \citenamefont
  {{Widrow}}}]{ColombiEtAl1996}%
  \BibitemOpen
  \bibfield  {author} {\bibinfo {author} {\bibfnamefont {S.}~\bibnamefont
  {{Colombi}}}, \bibinfo {author} {\bibfnamefont {S.}~\bibnamefont
  {{Dodelson}}}, \ and\ \bibinfo {author} {\bibfnamefont {L.~M.}\ \bibnamefont
  {{Widrow}}},\ }\href {\doibase 10.1086/176788} {\bibfield  {journal}
  {\bibinfo  {journal} {\apj}\ }\textbf {\bibinfo {volume} {458}},\ \bibinfo
  {pages} {1} (\bibinfo {year} {1996})},\ \Eprint
  {http://arxiv.org/abs/astro-ph/9505029} {astro-ph/9505029} \BibitemShut
  {NoStop}%
\bibitem [{\citenamefont {{Viel}}\ \emph {et~al.}(2005)\citenamefont {{Viel}},
  \citenamefont {{Lesgourgues}}, \citenamefont {{Haehnelt}}, \citenamefont
  {{Matarrese}},\ and\ \citenamefont {{Riotto}}}]{VielEtAl2005}%
  \BibitemOpen
  \bibfield  {author} {\bibinfo {author} {\bibfnamefont {M.}~\bibnamefont
  {{Viel}}}, \bibinfo {author} {\bibfnamefont {J.}~\bibnamefont
  {{Lesgourgues}}}, \bibinfo {author} {\bibfnamefont {M.~G.}\ \bibnamefont
  {{Haehnelt}}}, \bibinfo {author} {\bibfnamefont {S.}~\bibnamefont
  {{Matarrese}}}, \ and\ \bibinfo {author} {\bibfnamefont {A.}~\bibnamefont
  {{Riotto}}},\ }\href {\doibase 10.1103/PhysRevD.71.063534} {\bibfield
  {journal} {\bibinfo  {journal} {\prd}\ }\textbf {\bibinfo {volume} {71}},\
  \bibinfo {eid} {063534} (\bibinfo {year} {2005})},\ \Eprint
  {http://arxiv.org/abs/astro-ph/0501562} {astro-ph/0501562} \BibitemShut
  {NoStop}%
\bibitem [{\citenamefont {{Viel}}\ \emph {et~al.}(2013)\citenamefont {{Viel}},
  \citenamefont {{Becker}}, \citenamefont {{Bolton}},\ and\ \citenamefont
  {{Haehnelt}}}]{VielEtAl2013}%
  \BibitemOpen
  \bibfield  {author} {\bibinfo {author} {\bibfnamefont {M.}~\bibnamefont
  {{Viel}}}, \bibinfo {author} {\bibfnamefont {G.~D.}\ \bibnamefont
  {{Becker}}}, \bibinfo {author} {\bibfnamefont {J.~S.}\ \bibnamefont
  {{Bolton}}}, \ and\ \bibinfo {author} {\bibfnamefont {M.~G.}\ \bibnamefont
  {{Haehnelt}}},\ }\href {\doibase 10.1103/PhysRevD.88.043502} {\bibfield
  {journal} {\bibinfo  {journal} {\prd}\ }\textbf {\bibinfo {volume} {88}},\
  \bibinfo {eid} {043502} (\bibinfo {year} {2013})},\ \Eprint
  {http://arxiv.org/abs/1306.2314} {arXiv:1306.2314 [astro-ph.CO]} \BibitemShut
  {NoStop}%
\bibitem [{\citenamefont {{Bose}}\ \emph {et~al.}(2016)\citenamefont {{Bose}},
  \citenamefont {{Hellwing}}, \citenamefont {{Frenk}}, \citenamefont
  {{Jenkins}}, \citenamefont {{Lovell}}, \citenamefont {{Helly}},\ and\
  \citenamefont {{Li}}}]{BoseEtAl2016}%
  \BibitemOpen
  \bibfield  {author} {\bibinfo {author} {\bibfnamefont {S.}~\bibnamefont
  {{Bose}}}, \bibinfo {author} {\bibfnamefont {W.~A.}\ \bibnamefont
  {{Hellwing}}}, \bibinfo {author} {\bibfnamefont {C.~S.}\ \bibnamefont
  {{Frenk}}}, \bibinfo {author} {\bibfnamefont {A.}~\bibnamefont {{Jenkins}}},
  \bibinfo {author} {\bibfnamefont {M.~R.}\ \bibnamefont {{Lovell}}}, \bibinfo
  {author} {\bibfnamefont {J.~C.}\ \bibnamefont {{Helly}}}, \ and\ \bibinfo
  {author} {\bibfnamefont {B.}~\bibnamefont {{Li}}},\ }\href {\doibase
  10.1093/mnras/stv2294} {\bibfield  {journal} {\bibinfo  {journal} {\mnras}\
  }\textbf {\bibinfo {volume} {455}},\ \bibinfo {pages} {318} (\bibinfo {year}
  {2016})},\ \Eprint {http://arxiv.org/abs/1507.01998} {arXiv:1507.01998}
  \BibitemShut {NoStop}%
\bibitem [{\citenamefont {{Baur}}\ \emph {et~al.}(2017)\citenamefont {{Baur}},
  \citenamefont {{Palanque-Delabrouille}}, \citenamefont {{Y{\`e}che}},
  \citenamefont {{Boyarsky}}, \citenamefont {{Ruchayskiy}}, \citenamefont
  {{Armengaud}},\ and\ \citenamefont {{Lesgourgues}}}]{BaurEtAl2017}%
  \BibitemOpen
  \bibfield  {author} {\bibinfo {author} {\bibfnamefont {J.}~\bibnamefont
  {{Baur}}}, \bibinfo {author} {\bibfnamefont {N.}~\bibnamefont
  {{Palanque-Delabrouille}}}, \bibinfo {author} {\bibfnamefont
  {C.}~\bibnamefont {{Y{\`e}che}}}, \bibinfo {author} {\bibfnamefont
  {A.}~\bibnamefont {{Boyarsky}}}, \bibinfo {author} {\bibfnamefont
  {O.}~\bibnamefont {{Ruchayskiy}}}, \bibinfo {author} {\bibfnamefont
  {{\'E}.}~\bibnamefont {{Armengaud}}}, \ and\ \bibinfo {author} {\bibfnamefont
  {J.}~\bibnamefont {{Lesgourgues}}},\ }\href {\doibase
  10.1088/1475-7516/2017/12/013} {\bibfield  {journal} {\bibinfo  {journal}
  {\jcap}\ }\textbf {\bibinfo {volume} {12}},\ \bibinfo {eid} {013} (\bibinfo
  {year} {2017})},\ \Eprint {http://arxiv.org/abs/1706.03118}
  {arXiv:1706.03118} \BibitemShut {NoStop}%
\bibitem [{\citenamefont {{Griest}}\ and\ \citenamefont
  {{Kamionkowski}}(1990)}]{GriestEtAl1990}%
  \BibitemOpen
  \bibfield  {author} {\bibinfo {author} {\bibfnamefont {K.}~\bibnamefont
  {{Griest}}}\ and\ \bibinfo {author} {\bibfnamefont {M.}~\bibnamefont
  {{Kamionkowski}}},\ }\href {\doibase 10.1103/PhysRevLett.64.615} {\bibfield
  {journal} {\bibinfo  {journal} {\prl}\ }\textbf {\bibinfo {volume} {64}},\
  \bibinfo {pages} {615} (\bibinfo {year} {1990})}\BibitemShut {NoStop}%
\bibitem [{\citenamefont {Wilkinson}\ \emph {et~al.}(2014)\citenamefont
  {Wilkinson}, \citenamefont {Lesgourgues},\ and\ \citenamefont
  {B{\oe}hm}}]{WilkinsonEtAl2014}%
  \BibitemOpen
  \bibfield  {author} {\bibinfo {author} {\bibfnamefont {R.~J.}\ \bibnamefont
  {Wilkinson}}, \bibinfo {author} {\bibfnamefont {J.}~\bibnamefont
  {Lesgourgues}}, \ and\ \bibinfo {author} {\bibfnamefont {C.}~\bibnamefont
  {B{\oe}hm}},\ }\href {\doibase 10.1088/1475-7516/2014/04/026} {\bibfield
  {journal} {\bibinfo  {journal} {\jcap}\ }\textbf {\bibinfo {volume} {4}},\
  \bibinfo {eid} {026} (\bibinfo {year} {2014})},\ \Eprint
  {http://arxiv.org/abs/1309.7588} {arXiv:1309.7588} \BibitemShut {NoStop}%
\bibitem [{\citenamefont {{Leane}}\ \emph {et~al.}(2018)\citenamefont
  {{Leane}}, \citenamefont {{Slatyer}}, \citenamefont {{Beacom}},\ and\
  \citenamefont {{Ng}}}]{LeaneEtAl2018}%
  \BibitemOpen
  \bibfield  {author} {\bibinfo {author} {\bibfnamefont {R.~K.}\ \bibnamefont
  {{Leane}}}, \bibinfo {author} {\bibfnamefont {T.~R.}\ \bibnamefont
  {{Slatyer}}}, \bibinfo {author} {\bibfnamefont {J.~F.}\ \bibnamefont
  {{Beacom}}}, \ and\ \bibinfo {author} {\bibfnamefont {K.~C.~Y.}\ \bibnamefont
  {{Ng}}},\ }\href {\doibase 10.1103/PhysRevD.98.023016} {\bibfield  {journal}
  {\bibinfo  {journal} {\prd}\ }\textbf {\bibinfo {volume} {98}},\ \bibinfo
  {eid} {023016} (\bibinfo {year} {2018})},\ \Eprint
  {http://arxiv.org/abs/1805.10305} {arXiv:1805.10305 [hep-ph]} \BibitemShut
  {NoStop}%
\bibitem [{\citenamefont {{Lewin}}\ and\ \citenamefont
  {{Smith}}(1996)}]{LewinEtAl1996}%
  \BibitemOpen
  \bibfield  {author} {\bibinfo {author} {\bibfnamefont {J.~D.}\ \bibnamefont
  {{Lewin}}}\ and\ \bibinfo {author} {\bibfnamefont {P.~F.}\ \bibnamefont
  {{Smith}}},\ }\href {\doibase 10.1016/S0927-6505(96)00047-3} {\bibfield
  {journal} {\bibinfo  {journal} {Astroparticle Physics}\ }\textbf {\bibinfo
  {volume} {6}},\ \bibinfo {pages} {87} (\bibinfo {year} {1996})}\BibitemShut
  {NoStop}%
\bibitem [{\citenamefont {{Freese}}\ \emph {et~al.}(2013)\citenamefont
  {{Freese}}, \citenamefont {{Lisanti}},\ and\ \citenamefont
  {{Savage}}}]{FreeseEtAl2013}%
  \BibitemOpen
  \bibfield  {author} {\bibinfo {author} {\bibfnamefont {K.}~\bibnamefont
  {{Freese}}}, \bibinfo {author} {\bibfnamefont {M.}~\bibnamefont {{Lisanti}}},
  \ and\ \bibinfo {author} {\bibfnamefont {C.}~\bibnamefont {{Savage}}},\
  }\href {\doibase 10.1103/RevModPhys.85.1561} {\bibfield  {journal} {\bibinfo
  {journal} {Reviews of Modern Physics}\ }\textbf {\bibinfo {volume} {85}},\
  \bibinfo {pages} {1561} (\bibinfo {year} {2013})},\ \Eprint
  {http://arxiv.org/abs/1209.3339} {arXiv:1209.3339} \BibitemShut {NoStop}%
\bibitem [{\citenamefont {{Lavalle}}\ and\ \citenamefont
  {{Salati}}(2012)}]{LavalleEtAl2012}%
  \BibitemOpen
  \bibfield  {author} {\bibinfo {author} {\bibfnamefont {J.}~\bibnamefont
  {{Lavalle}}}\ and\ \bibinfo {author} {\bibfnamefont {P.}~\bibnamefont
  {{Salati}}},\ }\href {\doibase 10.1016/j.crhy.2012.05.001} {\bibfield
  {journal} {\bibinfo  {journal} {Comptes Rendus Physique}\ }\textbf {\bibinfo
  {volume} {13}},\ \bibinfo {pages} {740} (\bibinfo {year} {2012})},\ \Eprint
  {http://arxiv.org/abs/1205.1004} {arXiv:1205.1004 [astro-ph.HE]} \BibitemShut
  {NoStop}%
\bibitem [{\citenamefont {{Bringmann}}\ and\ \citenamefont
  {{Weniger}}(2012)}]{BringmannEtAl2012c}%
  \BibitemOpen
  \bibfield  {author} {\bibinfo {author} {\bibfnamefont {T.}~\bibnamefont
  {{Bringmann}}}\ and\ \bibinfo {author} {\bibfnamefont {C.}~\bibnamefont
  {{Weniger}}},\ }\href {\doibase 10.1016/j.dark.2012.10.005} {\bibfield
  {journal} {\bibinfo  {journal} {Physics of the Dark Universe}\ }\textbf
  {\bibinfo {volume} {1}},\ \bibinfo {pages} {194} (\bibinfo {year} {2012})},\
  \Eprint {http://arxiv.org/abs/1208.5481} {arXiv:1208.5481 [hep-ph]}
  \BibitemShut {NoStop}%
\bibitem [{\citenamefont {{Fairbairn}}\ \emph {et~al.}(2007)\citenamefont
  {{Fairbairn}}, \citenamefont {{Kraan}}, \citenamefont {{Milstead}},
  \citenamefont {{Sj{\"o}strand}}, \citenamefont {{Skands}},\ and\
  \citenamefont {{Sloan}}}]{FairbairnEtAl2007}%
  \BibitemOpen
  \bibfield  {author} {\bibinfo {author} {\bibfnamefont {M.}~\bibnamefont
  {{Fairbairn}}}, \bibinfo {author} {\bibfnamefont {A.~C.}\ \bibnamefont
  {{Kraan}}}, \bibinfo {author} {\bibfnamefont {D.~A.}\ \bibnamefont
  {{Milstead}}}, \bibinfo {author} {\bibfnamefont {T.}~\bibnamefont
  {{Sj{\"o}strand}}}, \bibinfo {author} {\bibfnamefont {P.}~\bibnamefont
  {{Skands}}}, \ and\ \bibinfo {author} {\bibfnamefont {T.}~\bibnamefont
  {{Sloan}}},\ }\href {\doibase 10.1016/j.physrep.2006.10.002} {\bibfield
  {journal} {\bibinfo  {journal} {\physrep}\ }\textbf {\bibinfo {volume}
  {438}},\ \bibinfo {pages} {1} (\bibinfo {year} {2007})},\ \Eprint
  {http://arxiv.org/abs/hep-ph/0611040} {hep-ph/0611040} \BibitemShut {NoStop}%
\bibitem [{\citenamefont {Kahlhoefer}(2017)}]{Kahlhoefer2017}%
  \BibitemOpen
  \bibfield  {author} {\bibinfo {author} {\bibfnamefont {F.}~\bibnamefont
  {Kahlhoefer}},\ }\href {\doibase 10.1142/S0217751X1730006X} {\bibfield
  {journal} {\bibinfo  {journal} {International Journal of Modern Physics A}\
  }\textbf {\bibinfo {volume} {32}},\ \bibinfo {eid} {1730006} (\bibinfo {year}
  {2017})},\ \Eprint {http://arxiv.org/abs/1702.02430} {arXiv:1702.02430
  [hep-ph]} \BibitemShut {NoStop}%
\bibitem [{\citenamefont {Lopez-Honorez}\ \emph {et~al.}(2016)\citenamefont
  {Lopez-Honorez}, \citenamefont {Mena}, \citenamefont {Molin{\'e}},
  \citenamefont {Palomares-Ruiz},\ and\ \citenamefont
  {Vincent}}]{Lopez-HonorezEtAl2016}%
  \BibitemOpen
  \bibfield  {author} {\bibinfo {author} {\bibfnamefont {L.}~\bibnamefont
  {Lopez-Honorez}}, \bibinfo {author} {\bibfnamefont {O.}~\bibnamefont {Mena}},
  \bibinfo {author} {\bibfnamefont {{\'A}.}~\bibnamefont {Molin{\'e}}},
  \bibinfo {author} {\bibfnamefont {S.}~\bibnamefont {Palomares-Ruiz}}, \ and\
  \bibinfo {author} {\bibfnamefont {A.~C.}\ \bibnamefont {Vincent}},\ }\href
  {\doibase 10.1088/1475-7516/2016/08/004} {\bibfield  {journal} {\bibinfo
  {journal} {\jcap}\ }\textbf {\bibinfo {volume} {8}},\ \bibinfo {eid} {004}
  (\bibinfo {year} {2016})},\ \Eprint {http://arxiv.org/abs/1603.06795}
  {arXiv:1603.06795} \BibitemShut {NoStop}%
\bibitem [{\citenamefont {{Bin{\'e}truy}}\ \emph {et~al.}(1984)\citenamefont
  {{Bin{\'e}truy}}, \citenamefont {{Girardi}},\ and\ \citenamefont
  {{Salati}}}]{BinetruyEtAl1984a}%
  \BibitemOpen
  \bibfield  {author} {\bibinfo {author} {\bibfnamefont {P.}~\bibnamefont
  {{Bin{\'e}truy}}}, \bibinfo {author} {\bibfnamefont {G.}~\bibnamefont
  {{Girardi}}}, \ and\ \bibinfo {author} {\bibfnamefont {P.}~\bibnamefont
  {{Salati}}},\ }\href {\doibase 10.1016/0550-3213(84)90161-5} {\bibfield
  {journal} {\bibinfo  {journal} {Nuclear Physics B}\ }\textbf {\bibinfo
  {volume} {237}},\ \bibinfo {pages} {285} (\bibinfo {year}
  {1984})}\BibitemShut {NoStop}%
\bibitem [{\citenamefont {{Srednicki}}\ \emph {et~al.}(1988)\citenamefont
  {{Srednicki}}, \citenamefont {{Watkins}},\ and\ \citenamefont
  {{Olive}}}]{SrednickiEtAl1988}%
  \BibitemOpen
  \bibfield  {author} {\bibinfo {author} {\bibfnamefont {M.}~\bibnamefont
  {{Srednicki}}}, \bibinfo {author} {\bibfnamefont {R.}~\bibnamefont
  {{Watkins}}}, \ and\ \bibinfo {author} {\bibfnamefont {K.~A.}\ \bibnamefont
  {{Olive}}},\ }\href {\doibase 10.1016/0550-3213(88)90099-5} {\bibfield
  {journal} {\bibinfo  {journal} {Nuclear Physics B}\ }\textbf {\bibinfo
  {volume} {310}},\ \bibinfo {pages} {693} (\bibinfo {year}
  {1988})}\BibitemShut {NoStop}%
\bibitem [{\citenamefont {{Gondolo}}\ and\ \citenamefont
  {{Gelmini}}(1991)}]{GondoloEtAl1991}%
  \BibitemOpen
  \bibfield  {author} {\bibinfo {author} {\bibfnamefont {P.}~\bibnamefont
  {{Gondolo}}}\ and\ \bibinfo {author} {\bibfnamefont {G.}~\bibnamefont
  {{Gelmini}}},\ }\href {\doibase 10.1016/0550-3213(91)90438-4} {\bibfield
  {journal} {\bibinfo  {journal} {Nuclear Physics B}\ }\textbf {\bibinfo
  {volume} {360}},\ \bibinfo {pages} {145} (\bibinfo {year}
  {1991})}\BibitemShut {NoStop}%
\bibitem [{\citenamefont {{Griest}}\ and\ \citenamefont
  {{Seckel}}(1991)}]{GriestEtAl1991a}%
  \BibitemOpen
  \bibfield  {author} {\bibinfo {author} {\bibfnamefont {K.}~\bibnamefont
  {{Griest}}}\ and\ \bibinfo {author} {\bibfnamefont {D.}~\bibnamefont
  {{Seckel}}},\ }\href {\doibase 10.1103/PhysRevD.43.3191} {\bibfield
  {journal} {\bibinfo  {journal} {\prd}\ }\textbf {\bibinfo {volume} {43}},\
  \bibinfo {pages} {3191} (\bibinfo {year} {1991})}\BibitemShut {NoStop}%
\bibitem [{\citenamefont {{Abdallah}}\ \emph {et~al.}(2015)\citenamefont
  {{Abdallah}}, \citenamefont {{Araujo}}, \citenamefont {{Arbey}},
  \citenamefont {{Ashkenazi}}, \citenamefont {{Belyaev}}, \citenamefont
  {{Berger}}, \citenamefont {{Boehm}}, \citenamefont {{Boveia}}, \citenamefont
  {{Brennan}}, \citenamefont {{Brooke}}, \citenamefont {{Buchmueller}},
  \citenamefont {{Buckley}}, \citenamefont {{Busoni}}, \citenamefont
  {{Calibbi}}, \citenamefont {{Chauhan}}, \citenamefont {{Daci}}, \citenamefont
  {{Davies}}, \citenamefont {{De Bruyn}}, \citenamefont {{De Jong}},
  \citenamefont {{De Roeck}}, \citenamefont {{de Vries}}, \citenamefont {{Del
  Re}}, \citenamefont {{De Simone}}, \citenamefont {{Di Simone}}, \citenamefont
  {{Doglioni}}, \citenamefont {{Dolan}}, \citenamefont {{Dreiner}},
  \citenamefont {{Ellis}}, \citenamefont {{Eno}}, \citenamefont {{Etzion}},
  \citenamefont {{Fairbairn}}, \citenamefont {{Feldstein}}, \citenamefont
  {{Flaecher}}, \citenamefont {{Feng}}, \citenamefont {{Fox}}, \citenamefont
  {{Genest}}, \citenamefont {{Gouskos}}, \citenamefont {{Gramling}},
  \citenamefont {{Haisch}}, \citenamefont {{Harnik}}, \citenamefont {{Hibbs}},
  \citenamefont {{Hoh}}, \citenamefont {{Hopkins}}, \citenamefont {{Ippolito}},
  \citenamefont {{Jacques}}, \citenamefont {{Kahlhoefer}}, \citenamefont
  {{Khoze}}, \citenamefont {{Kirk}}, \citenamefont {{Korn}}, \citenamefont
  {{Kotov}}, \citenamefont {{Kunori}}, \citenamefont {{Landsberg}},
  \citenamefont {{Liem}}, \citenamefont {{Lin}}, \citenamefont {{Lowette}},
  \citenamefont {{Lucas}}, \citenamefont {{Malgeri}}, \citenamefont {{Malik}},
  \citenamefont {{McCabe}}, \citenamefont {{Mete}}, \citenamefont {{Morgante}},
  \citenamefont {{Mrenna}}, \citenamefont {{Nakahama}}, \citenamefont
  {{Newbold}}, \citenamefont {{Nordstrom}}, \citenamefont {{Pani}},
  \citenamefont {{Papucci}}, \citenamefont {{Pataraia}}, \citenamefont
  {{Penning}}, \citenamefont {{Pinna}}, \citenamefont {{Polesello}},
  \citenamefont {{Racco}}, \citenamefont {{Re}}, \citenamefont {{Riotto}},
  \citenamefont {{Rizzo}}, \citenamefont {{Salek}}, \citenamefont {{Sarkar}},
  \citenamefont {{Schramm}}, \citenamefont {{Skubic}}, \citenamefont {{Slone}},
  \citenamefont {{Smirnov}}, \citenamefont {{Soreq}}, \citenamefont {{Sumner}},
  \citenamefont {{Tait}}, \citenamefont {{Thomas}}, \citenamefont {{Tomalin}},
  \citenamefont {{Tunnell}}, \citenamefont {{Vichi}}, \citenamefont
  {{Volansky}}, \citenamefont {{Weiner}}, \citenamefont {{West}}, \citenamefont
  {{Wielers}}, \citenamefont {{Worm}}, \citenamefont {{Yavin}}, \citenamefont
  {{Zaldivar}}, \citenamefont {{Zhou}},\ and\ \citenamefont
  {{Zurek}}}]{AbdallahEtAl2015}%
  \BibitemOpen
  \bibfield  {author} {\bibinfo {author} {\bibfnamefont {J.}~\bibnamefont
  {{Abdallah}}}, \bibinfo {author} {\etal},\ }\href {\doibase 10.1016/j.dark.2015.08.001}
  {\bibfield  {journal} {\bibinfo  {journal} {Physics of the Dark Universe}\
  }\textbf {\bibinfo {volume} {9}},\ \bibinfo {pages} {8} (\bibinfo {year}
  {2015})},\ \Eprint {http://arxiv.org/abs/1506.03116} {arXiv:1506.03116
  [hep-ph]} \BibitemShut {NoStop}%
\bibitem [{\citenamefont {Marrod{\'a}n~Undagoitia}\ and\ \citenamefont
  {Rauch}(2016)}]{MarrodanUndagoitiaEtAl2015}%
  \BibitemOpen
  \bibfield  {author} {\bibinfo {author} {\bibfnamefont {T.}~\bibnamefont
  {Marrod{\'a}n~Undagoitia}}\ and\ \bibinfo {author} {\bibfnamefont
  {L.}~\bibnamefont {Rauch}},\ }\href {\doibase 10.1088/0954-3899/43/1/013001}
  {\bibfield  {journal} {\bibinfo  {journal} {Journal of Physics G Nuclear
  Physics}\ }\textbf {\bibinfo {volume} {43}},\ \bibinfo {eid} {013001}
  (\bibinfo {year} {2016})},\ \Eprint {http://arxiv.org/abs/1509.08767}
  {arXiv:1509.08767 [physics.ins-det]} \BibitemShut {NoStop}%
\bibitem [{\citenamefont {{Essig}}\ \emph {et~al.}(2012)\citenamefont
  {{Essig}}, \citenamefont {{Mardon}},\ and\ \citenamefont
  {{Volansky}}}]{EssigEtAl2012}%
  \BibitemOpen
  \bibfield  {author} {\bibinfo {author} {\bibfnamefont {R.}~\bibnamefont
  {{Essig}}}, \bibinfo {author} {\bibfnamefont {J.}~\bibnamefont {{Mardon}}}, \
  and\ \bibinfo {author} {\bibfnamefont {T.}~\bibnamefont {{Volansky}}},\
  }\href {\doibase 10.1103/PhysRevD.85.076007} {\bibfield  {journal} {\bibinfo
  {journal} {\prd}\ }\textbf {\bibinfo {volume} {85}},\ \bibinfo {eid} {076007}
  (\bibinfo {year} {2012})},\ \Eprint {http://arxiv.org/abs/1108.5383}
  {arXiv:1108.5383 [hep-ph]} \BibitemShut {NoStop}%
\bibitem [{\citenamefont {{Kouvaris}}\ and\ \citenamefont
  {{Pradler}}(2017)}]{KouvarisEtAl2017}%
  \BibitemOpen
  \bibfield  {author} {\bibinfo {author} {\bibfnamefont {C.}~\bibnamefont
  {{Kouvaris}}}\ and\ \bibinfo {author} {\bibfnamefont {J.}~\bibnamefont
  {{Pradler}}},\ }\href {\doibase 10.1103/PhysRevLett.118.031803} {\bibfield
  {journal} {\bibinfo  {journal} {\prl}\ }\textbf {\bibinfo {volume} {118}},\
  \bibinfo {eid} {031803} (\bibinfo {year} {2017})},\ \Eprint
  {http://arxiv.org/abs/1607.01789} {arXiv:1607.01789 [hep-ph]} \BibitemShut
  {NoStop}%
\bibitem [{\citenamefont {{Boudaud}}\ \emph
  {et~al.}(2017{\natexlab{a}})\citenamefont {{Boudaud}}, \citenamefont
  {{Lavalle}},\ and\ \citenamefont {{Salati}}}]{BoudaudEtAl2017}%
  \BibitemOpen
  \bibfield  {author} {\bibinfo {author} {\bibfnamefont {M.}~\bibnamefont
  {{Boudaud}}}, \bibinfo {author} {\bibfnamefont {J.}~\bibnamefont
  {{Lavalle}}}, \ and\ \bibinfo {author} {\bibfnamefont {P.}~\bibnamefont
  {{Salati}}},\ }\href {\doibase 10.1103/PhysRevLett.119.021103} {\bibfield
  {journal} {\bibinfo  {journal} {\prl}\ }\textbf {\bibinfo {volume} {119}},\
  \bibinfo {pages} {021103} (\bibinfo {year} {2017}{\natexlab{a}})},\ \Eprint
  {http://arxiv.org/abs/1612.07698} {arXiv:1612.07698 [astro-ph.HE]}
  \BibitemShut {NoStop}%
\bibitem [{\citenamefont {{Krimigis}}\ \emph {et~al.}(1977)\citenamefont
  {{Krimigis}}, \citenamefont {{Bostrom}}, \citenamefont {{Armstrong}},
  \citenamefont {{Axford}}, \citenamefont {{Fan}}, \citenamefont
  {{Gloeckler}},\ and\ \citenamefont {{Lanzerotti}}}]{KrimigisEtAl1977}%
  \BibitemOpen
  \bibfield  {author} {\bibinfo {author} {\bibfnamefont {S.~M.}\ \bibnamefont
  {{Krimigis}}}, \bibinfo {author} {\bibfnamefont {C.~O.}\ \bibnamefont
  {{Bostrom}}}, \bibinfo {author} {\bibfnamefont {T.~P.}\ \bibnamefont
  {{Armstrong}}}, \bibinfo {author} {\bibfnamefont {W.~I.}\ \bibnamefont
  {{Axford}}}, \bibinfo {author} {\bibfnamefont {C.~Y.}\ \bibnamefont {{Fan}}},
  \bibinfo {author} {\bibfnamefont {G.}~\bibnamefont {{Gloeckler}}}, \ and\
  \bibinfo {author} {\bibfnamefont {L.~J.}\ \bibnamefont {{Lanzerotti}}},\
  }\href {\doibase 10.1007/BF00211545} {\bibfield  {journal} {\bibinfo
  {journal} {\ssr}\ }\textbf {\bibinfo {volume} {21}},\ \bibinfo {pages} {329}
  (\bibinfo {year} {1977})}\BibitemShut {NoStop}%
\bibitem [{\citenamefont {{Stone}}\ \emph {et~al.}(2013)\citenamefont
  {{Stone}}, \citenamefont {{Cummings}}, \citenamefont {{McDonald}},
  \citenamefont {{Heikkila}}, \citenamefont {{Lal}},\ and\ \citenamefont
  {{Webber}}}]{StoneEtAl2013}%
  \BibitemOpen
  \bibfield  {author} {\bibinfo {author} {\bibfnamefont {E.~C.}\ \bibnamefont
  {{Stone}}}, \bibinfo {author} {\bibfnamefont {A.~C.}\ \bibnamefont
  {{Cummings}}}, \bibinfo {author} {\bibfnamefont {F.~B.}\ \bibnamefont
  {{McDonald}}}, \bibinfo {author} {\bibfnamefont {B.~C.}\ \bibnamefont
  {{Heikkila}}}, \bibinfo {author} {\bibfnamefont {N.}~\bibnamefont {{Lal}}}, \
  and\ \bibinfo {author} {\bibfnamefont {W.~R.}\ \bibnamefont {{Webber}}},\
  }\href {\doibase 10.1126/science.1236408} {\bibfield  {journal} {\bibinfo
  {journal} {Science}\ }\textbf {\bibinfo {volume} {341}},\ \bibinfo {pages}
  {150} (\bibinfo {year} {2013})}\BibitemShut {NoStop}%
\bibitem [{\citenamefont {Cummings}\ \emph {et~al.}(2016)\citenamefont
  {Cummings}, \citenamefont {Stone}, \citenamefont {Heikkila}, \citenamefont
  {Lal}, \citenamefont {Webber}, \citenamefont {J{\'o}hannesson}, \citenamefont
  {Moskalenko}, \citenamefont {Orlando},\ and\ \citenamefont
  {Porter}}]{CummingsEtAl2016}%
  \BibitemOpen
  \bibfield  {author} {\bibinfo {author} {\bibfnamefont {A.~C.}\ \bibnamefont
  {Cummings}}, \bibinfo {author} {\bibfnamefont {E.~C.}\ \bibnamefont {Stone}},
  \bibinfo {author} {\bibfnamefont {B.~C.}\ \bibnamefont {Heikkila}}, \bibinfo
  {author} {\bibfnamefont {N.}~\bibnamefont {Lal}}, \bibinfo {author}
  {\bibfnamefont {W.~R.}\ \bibnamefont {Webber}}, \bibinfo {author}
  {\bibfnamefont {G.}~\bibnamefont {J{\'o}hannesson}}, \bibinfo {author}
  {\bibfnamefont {I.~V.}\ \bibnamefont {Moskalenko}}, \bibinfo {author}
  {\bibfnamefont {E.}~\bibnamefont {Orlando}}, \ and\ \bibinfo {author}
  {\bibfnamefont {T.~A.}\ \bibnamefont {Porter}},\ }\href {\doibase
  10.3847/0004-637X/831/1/18} {\bibfield  {journal} {\bibinfo  {journal}
  {\apj}\ }\textbf {\bibinfo {volume} {831}},\ \bibinfo {eid} {18} (\bibinfo
  {year} {2016})}\BibitemShut {NoStop}%
\bibitem [{\citenamefont {Aguilar}\ \emph {et~al.}(2014)\citenamefont
  {Aguilar}, \citenamefont {Aisa}, \citenamefont {Alvino}, \citenamefont
  {Ambrosi}, \citenamefont {Andeen}, \citenamefont {Arruda}, \citenamefont
  {Attig}, \citenamefont {Azzarello}, \citenamefont {Bachlechner},
  \citenamefont {Barao}, \citenamefont {Barrau}, \citenamefont {Barrin},
  \citenamefont {Bartoloni}, \citenamefont {Basara}, \citenamefont {Battarbee},
  \citenamefont {Battiston}, \citenamefont {Bazo}, \citenamefont {Becker},
  \citenamefont {Behlmann}, \citenamefont {Beischer}, \citenamefont {Berdugo},
  \citenamefont {Bertucci}, \citenamefont {Bigongiari}, \citenamefont {Bindi},
  \citenamefont {Bizzaglia}, \citenamefont {Bizzarri}, \citenamefont {Boella},
  \citenamefont {de~Boer}, \citenamefont {Bollweg}, \citenamefont {Bonnivard},
  \citenamefont {Borgia}, \citenamefont {Borsini}, \citenamefont {Boschini},
  \citenamefont {Bourquin}, \citenamefont {Burger}, \citenamefont {Cadoux},
  \citenamefont {Cai}, \citenamefont {Capell}, \citenamefont {Caroff},
  \citenamefont {Casaus}, \citenamefont {Cascioli}, \citenamefont {Castellini},
  \citenamefont {Cernuda}, \citenamefont {Cervelli}, \citenamefont {Chae},
  \citenamefont {Chang}, \citenamefont {Chen}, \citenamefont {Chen},
  \citenamefont {Cheng}, \citenamefont {Chen}, \citenamefont {Cheng},
  \citenamefont {Chikanian}, \citenamefont {Chou}, \citenamefont {Choumilov},
  \citenamefont {Choutko}, \citenamefont {Chung}, \citenamefont {Clark},
  \citenamefont {Clavero}, \citenamefont {Coignet}, \citenamefont {Consolandi},
  \citenamefont {Contin}, \citenamefont {Corti}, \citenamefont {Coste},
  \citenamefont {Cui}, \citenamefont {Dai}, \citenamefont {Delgado},
  \citenamefont {Della~Torre}, \citenamefont {Demirk{\"o}z}, \citenamefont
  {Derome}, \citenamefont {Di~Falco}, \citenamefont {Di~Masso}, \citenamefont
  {Dimiccoli}, \citenamefont {D{\'{\i}}az}, \citenamefont {von Doetinchem},
  \citenamefont {Du}, \citenamefont {Duranti}, \citenamefont {D'Urso},
  \citenamefont {Eline}, \citenamefont {Eppling}, \citenamefont {Eronen},
  \citenamefont {Fan}, \citenamefont {Farnesini}, \citenamefont {Feng},
  \citenamefont {Fiandrini}, \citenamefont {Fiasson}, \citenamefont {Finch},
  \citenamefont {Fisher}, \citenamefont {Galaktionov}, \citenamefont
  {Gallucci}, \citenamefont {Garc{\'{\i}}a}, \citenamefont
  {Garc{\'{\i}}a-L{\'o}pez}, \citenamefont {Gast}, \citenamefont {Gebauer},
  \citenamefont {Gervasi}, \citenamefont {Ghelfi}, \citenamefont {Gillard},
  \citenamefont {Giovacchini}, \citenamefont {Goglov}, \citenamefont {Gong},
  \citenamefont {Goy}, \citenamefont {Grabski}, \citenamefont {Grandi},
  \citenamefont {Graziani}, \citenamefont {Guandalini}, \citenamefont {Guerri},
  \citenamefont {Guo}, \citenamefont {Habiby}, \citenamefont {Haino},
  \citenamefont {Han}, \citenamefont {He}, \citenamefont {Heil}, \citenamefont
  {Hoffman}, \citenamefont {Hsieh}, \citenamefont {Huang}, \citenamefont {Huh},
  \citenamefont {Incagli}, \citenamefont {Ionica}, \citenamefont {Jang},
  \citenamefont {Jinchi}, \citenamefont {Kanishev}, \citenamefont {Kim},
  \citenamefont {Kim}, \citenamefont {Kirn}, \citenamefont {Kossakowski},
  \citenamefont {Kounina}, \citenamefont {Kounine}, \citenamefont {Koutsenko},
  \citenamefont {Krafczyk}, \citenamefont {Kunz}, \citenamefont {La~Vacca},
  \citenamefont {Laudi}, \citenamefont {Laurenti}, \citenamefont {Lazzizzera},
  \citenamefont {Lebedev}, \citenamefont {Lee}, \citenamefont {Lee},
  \citenamefont {Leluc}, \citenamefont {Li}, \citenamefont {Li}, \citenamefont
  {Li}, \citenamefont {Li}, \citenamefont {Li}, \citenamefont {Li},
  \citenamefont {Li}, \citenamefont {Li}, \citenamefont {Li}, \citenamefont
  {Lim}, \citenamefont {Lin}, \citenamefont {Lipari}, \citenamefont {Lippert},
  \citenamefont {Liu}, \citenamefont {Liu}, \citenamefont {Lomtadze},
  \citenamefont {Lu}, \citenamefont {Lu}, \citenamefont {Luebelsmeyer},
  \citenamefont {Luo}, \citenamefont {Luo}, \citenamefont {Lv}, \citenamefont
  {Majka}, \citenamefont {Malinin}, \citenamefont {Ma{\~n}{\'a}}, \citenamefont
  {Mar{\'{\i}}n}, \citenamefont {Martin}, \citenamefont {Mart{\'{\i}}nez},
  \citenamefont {Masi}, \citenamefont {Maurin}, \citenamefont {Menchaca-Rocha},
  \citenamefont {Meng}, \citenamefont {Mo}, \citenamefont {Morescalchi},
  \citenamefont {Mott}, \citenamefont {M{\"u}ller}, \citenamefont {Ni},
  \citenamefont {Nikonov}, \citenamefont {Nozzoli}, \citenamefont {Nunes},
  \citenamefont {Obermeier}, \citenamefont {Oliva}, \citenamefont {Orcinha},
  \citenamefont {Palmonari}, \citenamefont {Palomares}, \citenamefont
  {Paniccia}, \citenamefont {Papi}, \citenamefont {Pedreschi}, \citenamefont
  {Pensotti}, \citenamefont {Pereira}, \citenamefont {Pilo}, \citenamefont
  {Piluso}, \citenamefont {Pizzolotto}, \citenamefont {Plyaskin}, \citenamefont
  {Pohl}, \citenamefont {Poireau}, \citenamefont {Postaci}, \citenamefont
  {Putze}, \citenamefont {Quadrani}, \citenamefont {Qi}, \citenamefont
  {Rancoita}, \citenamefont {Rapin}, \citenamefont {Ricol}, \citenamefont
  {Rodr{\'{\i}}guez}, \citenamefont {Rosier-Lees}, \citenamefont {Rozhkov},
  \citenamefont {Rozza}, \citenamefont {Sagdeev}, \citenamefont {Sandweiss},
  \citenamefont {Saouter}, \citenamefont {Sbarra}, \citenamefont {Schael},
  \citenamefont {Schmidt}, \citenamefont {Schuckardt}, \citenamefont {von
  Dratzig}, \citenamefont {Schwering}, \citenamefont {Scolieri}, \citenamefont
  {Seo}, \citenamefont {Shan}, \citenamefont {Shan}, \citenamefont {Shi},
  \citenamefont {Shi}, \citenamefont {Shi}, \citenamefont {Siedenburg},
  \citenamefont {Son}, \citenamefont {Spada}, \citenamefont {Spinella},
  \citenamefont {Sun}, \citenamefont {Sun}, \citenamefont {Tacconi},
  \citenamefont {Tang}, \citenamefont {Tang}, \citenamefont {Tang},
  \citenamefont {Tao}, \citenamefont {Tescaro}, \citenamefont {Ting},
  \citenamefont {Ting}, \citenamefont {Tomassetti}, \citenamefont {Torsti},
  \citenamefont {T{\"u}rko{\v g}lu}, \citenamefont {Urban}, \citenamefont
  {Vagelli}, \citenamefont {Valente}, \citenamefont {Vannini}, \citenamefont
  {Valtonen}, \citenamefont {Vaurynovich}, \citenamefont {Vecchi},
  \citenamefont {Velasco}, \citenamefont {Vialle}, \citenamefont {Wang},
  \citenamefont {Wang}, \citenamefont {Wang}, \citenamefont {Wang},
  \citenamefont {Wang}, \citenamefont {Weng}, \citenamefont {Whitman},
  \citenamefont {Wienkenh{\"o}ver}, \citenamefont {Wu}, \citenamefont {Xia},
  \citenamefont {Xie}, \citenamefont {Xie}, \citenamefont {Xiong},
  \citenamefont {Xin}, \citenamefont {Xu}, \citenamefont {Xu}, \citenamefont
  {Yan}, \citenamefont {Yang}, \citenamefont {Yang}, \citenamefont {Ye},
  \citenamefont {Yi}, \citenamefont {Yu}, \citenamefont {Yu}, \citenamefont
  {Zeissler}, \citenamefont {Zhang}, \citenamefont {Zhang}, \citenamefont
  {Zhang}, \citenamefont {Zhang}, \citenamefont {Zheng}, \citenamefont
  {Zhuang}, \citenamefont {Zhukov}, \citenamefont {Zichichi}, \citenamefont
  {Zimmermann}, \citenamefont {Zuccon}, \citenamefont {Zurbach},\ and\
  \citenamefont {Collaboration}}]{AguilarEtAl2014}%
  \BibitemOpen
  \bibfield  {author} {\bibinfo {author} {\bibfnamefont {AMS-02 Collaboration}},
    \bibinfo {author} {\bibfnamefont {M.}~\bibnamefont {Aguilar}},
    \bibinfo {author} {\etal},\ }\href {\doibase
  10.1103/PhysRevLett.113.121102} {\bibfield  {journal} {\bibinfo  {journal}
  {\prl}\ }\textbf {\bibinfo {volume} {113}},\ \bibinfo {eid} {121102}
  (\bibinfo {year} {2014})}\BibitemShut {NoStop}%
\bibitem [{\citenamefont {{Adams}}\ \emph {et~al.}(1998)\citenamefont
  {{Adams}}, \citenamefont {{Sarkar}},\ and\ \citenamefont
  {{Sciama}}}]{AdamsEtAl1998}%
  \BibitemOpen
  \bibfield  {author} {\bibinfo {author} {\bibfnamefont {J.~A.}\ \bibnamefont
  {{Adams}}}, \bibinfo {author} {\bibfnamefont {S.}~\bibnamefont {{Sarkar}}}, \
  and\ \bibinfo {author} {\bibfnamefont {D.~W.}\ \bibnamefont {{Sciama}}},\
  }\href {\doibase 10.1046/j.1365-8711.1998.02017.x} {\bibfield  {journal}
  {\bibinfo  {journal} {\mnras}\ }\textbf {\bibinfo {volume} {301}},\ \bibinfo
  {pages} {210} (\bibinfo {year} {1998})},\ \Eprint
  {http://arxiv.org/abs/astro-ph/9805108} {astro-ph/9805108} \BibitemShut
  {NoStop}%
\bibitem [{\citenamefont {{Chen}}\ and\ \citenamefont
  {{Kamionkowski}}(2004)}]{ChenEtAl2004}%
  \BibitemOpen
  \bibfield  {author} {\bibinfo {author} {\bibfnamefont {X.}~\bibnamefont
  {{Chen}}}\ and\ \bibinfo {author} {\bibfnamefont {M.}~\bibnamefont
  {{Kamionkowski}}},\ }\href {\doibase 10.1103/PhysRevD.70.043502} {\bibfield
  {journal} {\bibinfo  {journal} {\prd}\ }\textbf {\bibinfo {volume} {70}},\
  \bibinfo {eid} {043502} (\bibinfo {year} {2004})},\ \Eprint
  {http://arxiv.org/abs/astro-ph/0310473} {astro-ph/0310473} \BibitemShut
  {NoStop}%
\bibitem [{\citenamefont {{Slatyer}}(2016)}]{Slatyer2016}%
  \BibitemOpen
  \bibfield  {author} {\bibinfo {author} {\bibfnamefont {T.~R.}\ \bibnamefont
  {{Slatyer}}},\ }\href {\doibase 10.1103/PhysRevD.93.023527} {\bibfield
  {journal} {\bibinfo  {journal} {\prd}\ }\textbf {\bibinfo {volume} {93}},\
  \bibinfo {eid} {023527} (\bibinfo {year} {2016})},\ \Eprint
  {http://arxiv.org/abs/1506.03811} {arXiv:1506.03811 [hep-ph]} \BibitemShut
  {NoStop}%
\bibitem [{\citenamefont {{Liu}}\ \emph {et~al.}(2016)\citenamefont {{Liu}},
  \citenamefont {{Slatyer}},\ and\ \citenamefont {{Zavala}}}]{LiuEtAl2016}%
  \BibitemOpen
  \bibfield  {author} {\bibinfo {author} {\bibfnamefont {H.}~\bibnamefont
  {{Liu}}}, \bibinfo {author} {\bibfnamefont {T.~R.}\ \bibnamefont
  {{Slatyer}}}, \ and\ \bibinfo {author} {\bibfnamefont {J.}~\bibnamefont
  {{Zavala}}},\ }\href {\doibase 10.1103/PhysRevD.94.063507} {\bibfield
  {journal} {\bibinfo  {journal} {\prd}\ }\textbf {\bibinfo {volume} {94}},\
  \bibinfo {eid} {063507} (\bibinfo {year} {2016})},\ \Eprint
  {http://arxiv.org/abs/1604.02457} {arXiv:1604.02457} \BibitemShut {NoStop}%
\bibitem [{\citenamefont {Collaboration}\ \emph {et~al.}(2018)\citenamefont
  {Collaboration}, \citenamefont {Akrami}, \citenamefont {Arroja},
  \citenamefont {Ashdown}, \citenamefont {Aumont}, \citenamefont {Baccigalupi},
  \citenamefont {Ballardini}, \citenamefont {Banday}, \citenamefont {Barreiro},
  \citenamefont {Bartolo}, \citenamefont {Basak}, \citenamefont {Battye},
  \citenamefont {Benabed}, \citenamefont {Bernard}, \citenamefont {Bersanelli},
  \citenamefont {Bielewicz}, \citenamefont {Bock}, \citenamefont {Bond},
  \citenamefont {Borrill}, \citenamefont {Bouchet}, \citenamefont {Boulanger},
  \citenamefont {Bucher}, \citenamefont {Burigana}, \citenamefont {Butler},
  \citenamefont {Calabrese}, \citenamefont {Cardoso}, \citenamefont {Carron},
  \citenamefont {Casaponsa}, \citenamefont {Challinor}, \citenamefont {Chiang},
  \citenamefont {Colombo}, \citenamefont {Combet}, \citenamefont {Contreras},
  \citenamefont {Crill}, \citenamefont {Cuttaia}, \citenamefont {de~Bernardis},
  \citenamefont {de~Zotti}, \citenamefont {Delabrouille}, \citenamefont
  {Delouis}, \citenamefont {D{\'e}sert}, \citenamefont {Di~Valentino},
  \citenamefont {Dickinson}, \citenamefont {Diego}, \citenamefont {Donzelli},
  \citenamefont {Dor{\'e}}, \citenamefont {Douspis}, \citenamefont {Ducout},
  \citenamefont {Dupac}, \citenamefont {Efstathiou}, \citenamefont {Elsner},
  \citenamefont {En{\ss}lin}, \citenamefont {Eriksen}, \citenamefont
  {Falgarone}, \citenamefont {Fantaye}, \citenamefont {Fergusson},
  \citenamefont {Fernandez-Cobos}, \citenamefont {Finelli}, \citenamefont
  {Forastieri}, \citenamefont {Frailis}, \citenamefont {Franceschi},
  \citenamefont {Frolov}, \citenamefont {Galeotta}, \citenamefont {Galli},
  \citenamefont {Ganga}, \citenamefont {G{\'e}nova-Santos}, \citenamefont
  {Gerbino}, \citenamefont {Ghosh}, \citenamefont {Gonz{\'a}lez-Nuevo},
  \citenamefont {G{\'o}rski}, \citenamefont {Gratton}, \citenamefont
  {Gruppuso}, \citenamefont {Gudmundsson}, \citenamefont {Hamann},
  \citenamefont {Handley}, \citenamefont {Hansen}, \citenamefont {Helou},
  \citenamefont {Herranz}, \citenamefont {Hivon}, \citenamefont {Huang},
  \citenamefont {Jaffe}, \citenamefont {Jones}, \citenamefont {Karakci},
  \citenamefont {Keih{\"a}nen}, \citenamefont {Keskitalo}, \citenamefont
  {Kiiveri}, \citenamefont {Kim}, \citenamefont {Kisner}, \citenamefont {Knox},
  \citenamefont {Krachmalnicoff}, \citenamefont {Kunz}, \citenamefont
  {Kurki-Suonio}, \citenamefont {Lagache}, \citenamefont {Lamarre},
  \citenamefont {Langer}, \citenamefont {Lasenby}, \citenamefont {Lattanzi},
  \citenamefont {Lawrence}, \citenamefont {Le~Jeune}, \citenamefont {Leahy},
  \citenamefont {Lesgourgues}, \citenamefont {Levrier}, \citenamefont {Lewis},
  \citenamefont {Liguori}, \citenamefont {Lilje}, \citenamefont {Lilley},
  \citenamefont {Lindholm}, \citenamefont {L{\'o}pez-Caniego}, \citenamefont
  {Lubin}, \citenamefont {Ma}, \citenamefont {Mac{\'{\i}}as-P{\'e}rez},
  \citenamefont {Maggio}, \citenamefont {Maino}, \citenamefont {Mandolesi},
  \citenamefont {Mangilli}, \citenamefont {Marcos-Caballero}, \citenamefont
  {Maris}, \citenamefont {Martin}, \citenamefont
  {Mart{\'{\i}}nez-Gonz{\'a}lez}, \citenamefont {Matarrese}, \citenamefont
  {Mauri}, \citenamefont {McEwen}, \citenamefont {Meerburg}, \citenamefont
  {Meinhold}, \citenamefont {Melchiorri}, \citenamefont {Mennella},
  \citenamefont {Migliaccio}, \citenamefont {Millea}, \citenamefont {Mitra},
  \citenamefont {Miville-Desch{\^e}nes}, \citenamefont {Molinari},
  \citenamefont {Moneti}, \citenamefont {Montier}, \citenamefont {Morgante},
  \citenamefont {Moss}, \citenamefont {Mottet}, \citenamefont {M{\"u}nchmeyer},
  \citenamefont {Natoli}, \citenamefont {N{\o}rgaard-Nielsen}, \citenamefont
  {Oxborrow}, \citenamefont {Pagano}, \citenamefont {Paoletti}, \citenamefont
  {Partridge}, \citenamefont {Patanchon}, \citenamefont {Pearson},
  \citenamefont {Peel}, \citenamefont {Peiris}, \citenamefont {Perrotta},
  \citenamefont {Pettorino}, \citenamefont {Piacentini}, \citenamefont
  {Polastri}, \citenamefont {Polenta}, \citenamefont {Puget}, \citenamefont
  {Rachen}, \citenamefont {Reinecke}, \citenamefont {Remazeilles},
  \citenamefont {Renzi}, \citenamefont {Rocha}, \citenamefont {Rosset},
  \citenamefont {Roudier}, \citenamefont {Rubi{\~n}o-Mart{\'{\i}}n},
  \citenamefont {Ruiz-Granados}, \citenamefont {Salvati}, \citenamefont
  {Sandri}, \citenamefont {Savelainen}, \citenamefont {Scott}, \citenamefont
  {Shellard}, \citenamefont {Shiraishi}, \citenamefont {Sirignano},
  \citenamefont {Sirri}, \citenamefont {Spencer}, \citenamefont {Sunyaev},
  \citenamefont {Suur-Uski}, \citenamefont {Tauber}, \citenamefont
  {Tavagnacco}, \citenamefont {Tenti}, \citenamefont {Terenzi}, \citenamefont
  {Toffolatti}, \citenamefont {Tomasi}, \citenamefont {Trombetti},
  \citenamefont {Valiviita}, \citenamefont {Van~Tent}, \citenamefont {Vibert},
  \citenamefont {Vielva}, \citenamefont {Villa}, \citenamefont {Vittorio},
  \citenamefont {Wandelt}, \citenamefont {Wehus}, \citenamefont {White},
  \citenamefont {White}, \citenamefont {Zacchei},\ and\ \citenamefont
  {Zonca}}]{PlanckCollaboration2018a}%
  \BibitemOpen
  \bibfield  {author} {\bibinfo {author} {\bibfnamefont {Planck Collaboration}},
    \bibinfo {author} {\bibfnamefont {Y.}~\bibnamefont {Akrami}},
    \bibinfo {author} {\etal},\ }\href
  {http://adsabs.harvard.edu/abs/2018arXiv180706205P} {\bibfield  {journal}
  {\bibinfo  {journal} {ArXiv e-prints}\ } (\bibinfo {year} {2018})},\ \Eprint
  {http://arxiv.org/abs/1807.06205} {arXiv:1807.06205} \BibitemShut {NoStop}%
\bibitem [{\citenamefont {{Essig}}\ \emph {et~al.}(2013)\citenamefont
  {{Essig}}, \citenamefont {{Kuflik}}, \citenamefont {{McDermott}},
  \citenamefont {{Volansky}},\ and\ \citenamefont {{Zurek}}}]{EssigEtAl2013a}%
  \BibitemOpen
  \bibfield  {author} {\bibinfo {author} {\bibfnamefont {R.}~\bibnamefont
  {{Essig}}}, \bibinfo {author} {\bibfnamefont {E.}~\bibnamefont {{Kuflik}}},
  \bibinfo {author} {\bibfnamefont {S.~D.}\ \bibnamefont {{McDermott}}},
  \bibinfo {author} {\bibfnamefont {T.}~\bibnamefont {{Volansky}}}, \ and\
  \bibinfo {author} {\bibfnamefont {K.~M.}\ \bibnamefont {{Zurek}}},\ }\href
  {\doibase 10.1007/JHEP11(2013)193} {\bibfield  {journal} {\bibinfo  {journal}
  {Journal of High Energy Physics}\ }\textbf {\bibinfo {volume} {11}},\
  \bibinfo {eid} {193} (\bibinfo {year} {2013})},\ \Eprint
  {http://arxiv.org/abs/1309.4091} {arXiv:1309.4091 [hep-ph]} \BibitemShut
  {NoStop}%
\bibitem [{\citenamefont {{Massari}}\ \emph {et~al.}(2015)\citenamefont
  {{Massari}}, \citenamefont {{Izaguirre}}, \citenamefont {{Essig}},
  \citenamefont {{Albert}}, \citenamefont {{Bloom}},\ and\ \citenamefont
  {{G{\'o}mez-Vargas}}}]{MassariEtAl2015}%
  \BibitemOpen
  \bibfield  {author} {\bibinfo {author} {\bibfnamefont {A.}~\bibnamefont
  {{Massari}}}, \bibinfo {author} {\bibfnamefont {E.}~\bibnamefont
  {{Izaguirre}}}, \bibinfo {author} {\bibfnamefont {R.}~\bibnamefont
  {{Essig}}}, \bibinfo {author} {\bibfnamefont {A.}~\bibnamefont {{Albert}}},
  \bibinfo {author} {\bibfnamefont {E.}~\bibnamefont {{Bloom}}}, \ and\
  \bibinfo {author} {\bibfnamefont {G.~A.}\ \bibnamefont
  {{G{\'o}mez-Vargas}}},\ }\href {\doibase 10.1103/PhysRevD.91.083539}
  {\bibfield  {journal} {\bibinfo  {journal} {\prd}\ }\textbf {\bibinfo
  {volume} {91}},\ \bibinfo {eid} {083539} (\bibinfo {year} {2015})},\ \Eprint
  {http://arxiv.org/abs/1503.07169} {arXiv:1503.07169 [hep-ph]} \BibitemShut
  {NoStop}%
\bibitem [{\citenamefont {B{\'e}langer}\ \emph {et~al.}(2018)\citenamefont
  {B{\'e}langer}, \citenamefont {Boudjema}, \citenamefont {Goudelis},
  \citenamefont {Pukhov},\ and\ \citenamefont
  {Zald{\'{\i}}var}}]{BelangerEtAl2018}%
  \BibitemOpen
  \bibfield  {author} {\bibinfo {author} {\bibfnamefont {G.}~\bibnamefont
  {B{\'e}langer}}, \bibinfo {author} {\bibfnamefont {F.}~\bibnamefont
  {Boudjema}}, \bibinfo {author} {\bibfnamefont {A.}~\bibnamefont {Goudelis}},
  \bibinfo {author} {\bibfnamefont {A.}~\bibnamefont {Pukhov}}, \ and\ \bibinfo
  {author} {\bibfnamefont {B.}~\bibnamefont {Zald{\'{\i}}var}},\ }\href
  {\doibase 10.1016/j.cpc.2018.04.027} {\bibfield  {journal} {\bibinfo
  {journal} {Computer Physics Communications}\ }\textbf {\bibinfo {volume}
  {231}},\ \bibinfo {pages} {173} (\bibinfo {year} {2018})},\ \Eprint
  {http://arxiv.org/abs/1801.03509} {arXiv:1801.03509 [hep-ph]} \BibitemShut
  {NoStop}%
\bibitem [{\citenamefont {Sj{\"o}strand}\ \emph {et~al.}(2015)\citenamefont
  {Sj{\"o}strand}, \citenamefont {Ask}, \citenamefont {Christiansen},
  \citenamefont {Corke}, \citenamefont {Desai}, \citenamefont {Ilten},
  \citenamefont {Mrenna}, \citenamefont {Prestel}, \citenamefont {Rasmussen},\
  and\ \citenamefont {Skands}}]{SjoestrandEtAl2015}%
  \BibitemOpen
  \bibfield  {author} {\bibinfo {author} {\bibfnamefont {T.}~\bibnamefont
  {Sj{\"o}strand}}, \bibinfo {author} {\bibfnamefont {S.}~\bibnamefont {Ask}},
  \bibinfo {author} {\bibfnamefont {J.~R.}\ \bibnamefont {Christiansen}},
  \bibinfo {author} {\bibfnamefont {R.}~\bibnamefont {Corke}}, \bibinfo
  {author} {\bibfnamefont {N.}~\bibnamefont {Desai}}, \bibinfo {author}
  {\bibfnamefont {P.}~\bibnamefont {Ilten}}, \bibinfo {author} {\bibfnamefont
  {S.}~\bibnamefont {Mrenna}}, \bibinfo {author} {\bibfnamefont
  {S.}~\bibnamefont {Prestel}}, \bibinfo {author} {\bibfnamefont {C.~O.}\
  \bibnamefont {Rasmussen}}, \ and\ \bibinfo {author} {\bibfnamefont {P.~Z.}\
  \bibnamefont {Skands}},\ }\href {\doibase 10.1016/j.cpc.2015.01.024}
  {\bibfield  {journal} {\bibinfo  {journal} {Computer Physics Communications}\
  }\textbf {\bibinfo {volume} {191}},\ \bibinfo {pages} {159} (\bibinfo {year}
  {2015})},\ \Eprint {http://arxiv.org/abs/1410.3012} {arXiv:1410.3012
  [hep-ph]} \BibitemShut {NoStop}%
\bibitem [{\citenamefont {Mo}\ \emph {et~al.}(2010)\citenamefont {Mo},
  \citenamefont {van~den Bosch},\ and\ \citenamefont {White}}]{MoEtAl2010}%
  \BibitemOpen
  \bibfield  {author} {\bibinfo {author} {\bibfnamefont {H.}~\bibnamefont
  {Mo}}, \bibinfo {author} {\bibfnamefont {F.~C.}\ \bibnamefont {van~den
  Bosch}}, \ and\ \bibinfo {author} {\bibfnamefont {S.}~\bibnamefont {White}},\
  }\href {http://adsabs.harvard.edu/abs/2010gfe..book.....M} {\emph {\bibinfo
  {title} {Galaxy Formation and Evolution}}}\ (\bibinfo  {publisher} {Cambridge
  University Press},\ \bibinfo {year} {2010})\BibitemShut {NoStop}%
\bibitem [{\citenamefont {{Binney}}\ and\ \citenamefont
  {{Tremaine}}(2008)}]{BinneyEtAl2008}%
  \BibitemOpen
  \bibfield  {author} {\bibinfo {author} {\bibfnamefont {J.}~\bibnamefont
  {{Binney}}}\ and\ \bibinfo {author} {\bibfnamefont {S.}~\bibnamefont
  {{Tremaine}}},\ }\href {http://adsabs.harvard.edu/abs/2008gady.book.....B}
  {\emph {\bibinfo {title} {Galactic Dynamics}}},\ \bibinfo {edition} {2nd}\
  ed.,\ Princeton series in astrophysics\ (\bibinfo  {publisher} {Princeton
  University Press},\ \bibinfo {address} {Princeton, NJ USA, 2008.},\ \bibinfo
  {year} {2008})\BibitemShut {NoStop}%
\bibitem [{\citenamefont {{Eddington}}(1916)}]{Eddington1916a}%
  \BibitemOpen
  \bibfield  {author} {\bibinfo {author} {\bibfnamefont {A.~S.}\ \bibnamefont
  {{Eddington}}},\ }\href@noop {} {\bibfield  {journal} {\bibinfo  {journal}
  {\mnras}\ }\textbf {\bibinfo {volume} {76}},\ \bibinfo {pages} {572}
  (\bibinfo {year} {1916})}\BibitemShut {NoStop}%
\bibitem [{\citenamefont {Lacroix}\ \emph {et~al.}(2018)\citenamefont
  {Lacroix}, \citenamefont {Stref},\ and\ \citenamefont
  {Lavalle}}]{LacroixEtAl2018}%
  \BibitemOpen
  \bibfield  {author} {\bibinfo {author} {\bibfnamefont {T.}~\bibnamefont
  {Lacroix}}, \bibinfo {author} {\bibfnamefont {M.}~\bibnamefont {Stref}}, \
  and\ \bibinfo {author} {\bibfnamefont {J.}~\bibnamefont {Lavalle}},\ }\href
  {\doibase 10.1088/1475-7516/2018/09/040} {\bibfield  {journal} {\bibinfo
  {journal} {\jcap}\ }\textbf {\bibinfo {volume} {09}},\ \bibinfo {pages} {040}
  (\bibinfo {year} {2018})},\ \Eprint {http://arxiv.org/abs/1805.02403}
  {arXiv:1805.02403} \BibitemShut {NoStop}%
\bibitem [{\citenamefont {{Ferrer}}\ and\ \citenamefont
  {{Hunter}}(2013)}]{FerrerEtAl2013}%
  \BibitemOpen
  \bibfield  {author} {\bibinfo {author} {\bibfnamefont {F.}~\bibnamefont
  {{Ferrer}}}\ and\ \bibinfo {author} {\bibfnamefont {D.~R.}\ \bibnamefont
  {{Hunter}}},\ }\href {\doibase 10.1088/1475-7516/2013/09/005} {\bibfield
  {journal} {\bibinfo  {journal} {\jcap}\ }\textbf {\bibinfo {volume} {9}},\
  \bibinfo {eid} {005} (\bibinfo {year} {2013})},\ \Eprint
  {http://arxiv.org/abs/1306.6586} {arXiv:1306.6586 [astro-ph.HE]} \BibitemShut
  {NoStop}%
\bibitem [{\citenamefont {{Hunter}}(2014)}]{Hunter2014}%
  \BibitemOpen
  \bibfield  {author} {\bibinfo {author} {\bibfnamefont {D.~R.}\ \bibnamefont
  {{Hunter}}},\ }\href {\doibase 10.1088/1475-7516/2014/02/023} {\bibfield
  {journal} {\bibinfo  {journal} {\jcap}\ }\textbf {\bibinfo {volume} {2}},\
  \bibinfo {eid} {023} (\bibinfo {year} {2014})},\ \Eprint
  {http://arxiv.org/abs/1311.0256} {arXiv:1311.0256 [astro-ph.CO]} \BibitemShut
  {NoStop}%
\bibitem [{\citenamefont {Boddy}\ \emph {et~al.}(2018)\citenamefont {Boddy},
  \citenamefont {Kumar},\ and\ \citenamefont {Strigari}}]{BoddyEtAl2018}%
  \BibitemOpen
  \bibfield  {author} {\bibinfo {author} {\bibfnamefont {K.~K.}\ \bibnamefont
  {Boddy}}, \bibinfo {author} {\bibfnamefont {J.}~\bibnamefont {Kumar}}, \ and\
  \bibinfo {author} {\bibfnamefont {L.~E.}\ \bibnamefont {Strigari}},\ }\href
  {\doibase 10.1103/PhysRevD.98.063012} {\bibfield  {journal} {\bibinfo
  {journal} {\prd}\ }\textbf {\bibinfo {volume} {98}},\ \bibinfo {pages}
  {063012} (\bibinfo {year} {2018})},\ \Eprint
  {http://arxiv.org/abs/1805.08379} {arXiv:1805.08379 [astro-ph.HE]}
  \BibitemShut {NoStop}%
\bibitem [{\citenamefont {{Osipkov}}(1979)}]{Osipkov1979}%
  \BibitemOpen
  \bibfield  {author} {\bibinfo {author} {\bibfnamefont {L.~P.}\ \bibnamefont
  {{Osipkov}}},\ }\href {http://adsabs.harvard.edu/abs/1979SvAL....5...42O}
  {\bibfield  {journal} {\bibinfo  {journal} {Soviet Astronomy Letters}\
  }\textbf {\bibinfo {volume} {5}},\ \bibinfo {pages} {42} (\bibinfo {year}
  {1979})}\BibitemShut {NoStop}%
\bibitem [{\citenamefont {{Merritt}}(1985)}]{Merritt1985a}%
  \BibitemOpen
  \bibfield  {author} {\bibinfo {author} {\bibfnamefont {D.}~\bibnamefont
  {{Merritt}}},\ }\href {\doibase 10.1086/113810} {\bibfield  {journal}
  {\bibinfo  {journal} {\aj}\ }\textbf {\bibinfo {volume} {90}},\ \bibinfo
  {pages} {1027} (\bibinfo {year} {1985})}\BibitemShut {NoStop}%
\bibitem [{\citenamefont {{Cuddeford}}(1991)}]{Cuddeford1991}%
  \BibitemOpen
  \bibfield  {author} {\bibinfo {author} {\bibfnamefont {P.}~\bibnamefont
  {{Cuddeford}}},\ }\href {\doibase 10.1093/mnras/253.3.414} {\bibfield
  {journal} {\bibinfo  {journal} {\mnras}\ }\textbf {\bibinfo {volume} {253}},\
  \bibinfo {pages} {414} (\bibinfo {year} {1991})}\BibitemShut {NoStop}%
\bibitem [{\citenamefont {{Lacroix}}\ \emph {et~al.}(2018)\citenamefont
  {{Lacroix}}, \citenamefont {{Nunez}}, \citenamefont {{Stref}}, \citenamefont
  {{Lavalle}},\ and\ \citenamefont {{Nezri}}}]{LacroixEtAl2018a}%
  \BibitemOpen
  \bibfield  {author} {\bibinfo {author} {\bibfnamefont {T.}~\bibnamefont
  {{Lacroix}}}, \bibinfo {author} {\bibfnamefont {A.}~\bibnamefont {{Nunez}}},
  \bibinfo {author} {\bibfnamefont {M.}~\bibnamefont {{Stref}}}, \bibinfo
  {author} {\bibfnamefont {J.}~\bibnamefont {{Lavalle}}}, \ and\ \bibinfo
  {author} {\bibfnamefont {E.}~\bibnamefont {{Nezri}}},\ }\href@noop {}
  {\bibfield  {journal} {\bibinfo  {journal} {In preparation}\ } (\bibinfo
  {year} {2018})}\BibitemShut {NoStop}%
\bibitem [{\citenamefont {{McMillan}}(2017)}]{McMillan2017}%
  \BibitemOpen
  \bibfield  {author} {\bibinfo {author} {\bibfnamefont {P.~J.}\ \bibnamefont
  {{McMillan}}},\ }\href {\doibase 10.1093/mnras/stw2759} {\bibfield  {journal}
  {\bibinfo  {journal} {\mnras}\ }\textbf {\bibinfo {volume} {465}},\ \bibinfo
  {pages} {76} (\bibinfo {year} {2017})},\ \Eprint
  {http://arxiv.org/abs/1608.00971} {arXiv:1608.00971} \BibitemShut {NoStop}%
\bibitem [{\citenamefont {{Navarro}}\ \emph {et~al.}(1996)\citenamefont
  {{Navarro}}, \citenamefont {{Frenk}},\ and\ \citenamefont
  {{White}}}]{NavarroEtAl1996a}%
  \BibitemOpen
  \bibfield  {author} {\bibinfo {author} {\bibfnamefont {J.~F.}\ \bibnamefont
  {{Navarro}}}, \bibinfo {author} {\bibfnamefont {C.~S.}\ \bibnamefont
  {{Frenk}}}, \ and\ \bibinfo {author} {\bibfnamefont {S.~D.~M.}\ \bibnamefont
  {{White}}},\ }\href {\doibase 10.1086/177173} {\bibfield  {journal} {\bibinfo
   {journal} {\apj}\ }\textbf {\bibinfo {volume} {462}},\ \bibinfo {pages}
  {563} (\bibinfo {year} {1996})},\ \Eprint
  {http://arxiv.org/abs/astro-ph/9508025} {astro-ph/9508025} \BibitemShut
  {NoStop}%
\bibitem [{\citenamefont {{Zhao}}(1996)}]{Zhao1996}%
  \BibitemOpen
  \bibfield  {author} {\bibinfo {author} {\bibfnamefont {H.}~\bibnamefont
  {{Zhao}}},\ }\href@noop {} {\bibfield  {journal} {\bibinfo  {journal}
  {\mnras}\ }\textbf {\bibinfo {volume} {278}},\ \bibinfo {pages} {488}
  (\bibinfo {year} {1996})},\ \Eprint {http://arxiv.org/abs/astro-ph/9509122}
  {astro-ph/9509122} \BibitemShut {NoStop}%
\bibitem [{\citenamefont {{Ginzburg}}\ and\ \citenamefont
  {{Syrovatskii}}(1964)}]{GinzburgEtAl1964}%
  \BibitemOpen
  \bibfield  {author} {\bibinfo {author} {\bibfnamefont {V.~L.}\ \bibnamefont
  {{Ginzburg}}}\ and\ \bibinfo {author} {\bibfnamefont {S.~I.}\ \bibnamefont
  {{Syrovatskii}}},\ }\href {http://adsabs.harvard.edu/abs/1964ocr..book.....G}
  {\emph {\bibinfo {title} {The Origin of Cosmic Rays}}}\ (\bibinfo
  {publisher} {New York: Macmillan},\ \bibinfo {year} {1964})\BibitemShut
  {NoStop}%
\bibitem [{\citenamefont {{Berezinskii}}\ \emph {et~al.}(1990)\citenamefont
  {{Berezinskii}}, \citenamefont {{Bulanov}}, \citenamefont {{Dogiel}},\ and\
  \citenamefont {{Ptuskin}}}]{BerezinskiiEtAl1990}%
  \BibitemOpen
  \bibfield  {author} {\bibinfo {author} {\bibfnamefont {V.~S.}\ \bibnamefont
  {{Berezinskii}}}, \bibinfo {author} {\bibfnamefont {S.~V.}\ \bibnamefont
  {{Bulanov}}}, \bibinfo {author} {\bibfnamefont {V.~A.}\ \bibnamefont
  {{Dogiel}}}, \ and\ \bibinfo {author} {\bibfnamefont {V.~S.}\ \bibnamefont
  {{Ptuskin}}},\ }\href {http://adsabs.harvard.edu/abs/1990acr..book.....B}
             {\emph {\bibinfo {title} {Astrophysics of cosmic rays}}}\
             (\bibinfo  {publisher} {North Holland},\ \bibinfo {year}
  {1990})\BibitemShut {NoStop}%
\bibitem [{\citenamefont {{Strong}}\ and\ \citenamefont
  {{Moskalenko}}(1998)}]{StrongEtAl1998}%
  \BibitemOpen
  \bibfield  {author} {\bibinfo {author} {\bibfnamefont {A.~W.}\ \bibnamefont
  {{Strong}}}\ and\ \bibinfo {author} {\bibfnamefont {I.~V.}\ \bibnamefont
  {{Moskalenko}}},\ }\href {\doibase 10.1086/306470} {\bibfield  {journal}
  {\bibinfo  {journal} {\apj}\ }\textbf {\bibinfo {volume} {509}},\ \bibinfo
  {pages} {212} (\bibinfo {year} {1998})},\ \Eprint
  {http://arxiv.org/abs/astro-ph/9807150} {astro-ph/9807150} \BibitemShut
  {NoStop}%
\bibitem [{\citenamefont {{Jones}}\ \emph {et~al.}(2001)\citenamefont
  {{Jones}}, \citenamefont {{Lukasiak}}, \citenamefont {{Ptuskin}},\ and\
  \citenamefont {{Webber}}}]{JonesEtAl2001}%
  \BibitemOpen
  \bibfield  {author} {\bibinfo {author} {\bibfnamefont {F.~C.}\ \bibnamefont
  {{Jones}}}, \bibinfo {author} {\bibfnamefont {A.}~\bibnamefont {{Lukasiak}}},
  \bibinfo {author} {\bibfnamefont {V.}~\bibnamefont {{Ptuskin}}}, \ and\
  \bibinfo {author} {\bibfnamefont {W.}~\bibnamefont {{Webber}}},\ }\href
  {\doibase 10.1086/318358} {\bibfield  {journal} {\bibinfo  {journal} {\apj}\
  }\textbf {\bibinfo {volume} {547}},\ \bibinfo {pages} {264} (\bibinfo {year}
  {2001})},\ \Eprint {http://arxiv.org/abs/astro-ph/0007293} {astro-ph/0007293}
  \BibitemShut {NoStop}%
\bibitem [{\citenamefont {{Maurin}}\ \emph {et~al.}(2001)\citenamefont
  {{Maurin}}, \citenamefont {{Donato}}, \citenamefont {{Taillet}},\ and\
  \citenamefont {{Salati}}}]{MaurinEtAl2001}%
  \BibitemOpen
  \bibfield  {author} {\bibinfo {author} {\bibfnamefont {D.}~\bibnamefont
  {{Maurin}}}, \bibinfo {author} {\bibfnamefont {F.}~\bibnamefont {{Donato}}},
  \bibinfo {author} {\bibfnamefont {R.}~\bibnamefont {{Taillet}}}, \ and\
  \bibinfo {author} {\bibfnamefont {P.}~\bibnamefont {{Salati}}},\ }\href
  {\doibase 10.1086/321496} {\bibfield  {journal} {\bibinfo  {journal} {\apj}\
  }\textbf {\bibinfo {volume} {555}},\ \bibinfo {pages} {585} (\bibinfo {year}
  {2001})},\ \Eprint {http://arxiv.org/abs/astro-ph/0101231} {astro-ph/0101231}
  \BibitemShut {NoStop}%
\bibitem [{\citenamefont {{Strong}}\ \emph {et~al.}(2007)\citenamefont
  {{Strong}}, \citenamefont {{Moskalenko}},\ and\ \citenamefont
  {{Ptuskin}}}]{StrongEtAl2007}%
  \BibitemOpen
  \bibfield  {author} {\bibinfo {author} {\bibfnamefont {A.~W.}\ \bibnamefont
  {{Strong}}}, \bibinfo {author} {\bibfnamefont {I.~V.}\ \bibnamefont
  {{Moskalenko}}}, \ and\ \bibinfo {author} {\bibfnamefont {V.~S.}\
  \bibnamefont {{Ptuskin}}},\ }\href {\doibase
  10.1146/annurev.nucl.57.090506.123011} {\bibfield  {journal} {\bibinfo
  {journal} {Annual Review of Nuclear and Particle Science}\ }\textbf {\bibinfo
  {volume} {57}},\ \bibinfo {pages} {285} (\bibinfo {year} {2007})},\ \Eprint
  {http://arxiv.org/abs/astro-ph/0701517} {astro-ph/0701517} \BibitemShut
  {NoStop}%
\bibitem [{\citenamefont {{Kissmann}}(2014)}]{Kissmann2014}%
  \BibitemOpen
  \bibfield  {author} {\bibinfo {author} {\bibfnamefont {R.}~\bibnamefont
  {{Kissmann}}},\ }\href {\doibase 10.1016/j.astropartphys.2014.02.002}
  {\bibfield  {journal} {\bibinfo  {journal} {Astroparticle Physics}\ }\textbf
  {\bibinfo {volume} {55}},\ \bibinfo {pages} {37} (\bibinfo {year} {2014})},\
  \Eprint {http://arxiv.org/abs/1401.4035} {arXiv:1401.4035 [astro-ph.HE]}
  \BibitemShut {NoStop}%
\bibitem [{\citenamefont {{Evoli}}\ \emph {et~al.}(2017)\citenamefont
  {{Evoli}}, \citenamefont {{Gaggero}}, \citenamefont {{Vittino}},
  \citenamefont {{Di Bernardo}}, \citenamefont {{Di Mauro}}, \citenamefont
  {{Ligorini}}, \citenamefont {{Ullio}},\ and\ \citenamefont
  {{Grasso}}}]{EvoliEtAl2017}%
  \BibitemOpen
  \bibfield  {author} {\bibinfo {author} {\bibfnamefont {C.}~\bibnamefont
  {{Evoli}}}, \bibinfo {author} {\bibfnamefont {D.}~\bibnamefont {{Gaggero}}},
  \bibinfo {author} {\bibfnamefont {A.}~\bibnamefont {{Vittino}}}, \bibinfo
  {author} {\bibfnamefont {G.}~\bibnamefont {{Di Bernardo}}}, \bibinfo {author}
  {\bibfnamefont {M.}~\bibnamefont {{Di Mauro}}}, \bibinfo {author}
  {\bibfnamefont {A.}~\bibnamefont {{Ligorini}}}, \bibinfo {author}
  {\bibfnamefont {P.}~\bibnamefont {{Ullio}}}, \ and\ \bibinfo {author}
  {\bibfnamefont {D.}~\bibnamefont {{Grasso}}},\ }\href {\doibase
  10.1088/1475-7516/2017/02/015} {\bibfield  {journal} {\bibinfo  {journal}
  {\jcap}\ }\textbf {\bibinfo {volume} {2}},\ \bibinfo {eid} {015} (\bibinfo
  {year} {2017})},\ \Eprint {http://arxiv.org/abs/1607.07886} {arXiv:1607.07886
  [astro-ph.HE]} \BibitemShut {NoStop}%
\bibitem [{\citenamefont {Maurin}(2018)}]{Maurin2018}%
  \BibitemOpen
  \bibfield  {author} {\bibinfo {author} {\bibfnamefont {D.}~\bibnamefont
  {Maurin}},\ }\href {http://adsabs.harvard.edu/abs/2018arXiv180702968M}
  {\bibfield  {journal} {\bibinfo  {journal} {ArXiv e-prints}\ } (\bibinfo
  {year} {2018})},\ \Eprint {http://arxiv.org/abs/1807.02968} {arXiv:1807.02968
  [astro-ph.IM]} \BibitemShut {NoStop}%
\bibitem [{\citenamefont {{Boudaud}}\ \emph
  {et~al.}(2017{\natexlab{b}})\citenamefont {{Boudaud}}, \citenamefont
  {{Bueno}}, \citenamefont {{Caroff}}, \citenamefont {{Genolini}},
  \citenamefont {{Poulin}}, \citenamefont {{Poireau}}, \citenamefont {{Putze}},
  \citenamefont {{Rosier}}, \citenamefont {{Salati}},\ and\ \citenamefont
  {{Vecchi}}}]{BoudaudEtAl2017a}%
  \BibitemOpen
  \bibfield  {author} {\bibinfo {author} {\bibfnamefont {M.}~\bibnamefont
  {{Boudaud}}}, \bibinfo {author} {\bibfnamefont {E.~F.}\ \bibnamefont
  {{Bueno}}}, \bibinfo {author} {\bibfnamefont {S.}~\bibnamefont {{Caroff}}},
  \bibinfo {author} {\bibfnamefont {Y.}~\bibnamefont {{Genolini}}}, \bibinfo
  {author} {\bibfnamefont {V.}~\bibnamefont {{Poulin}}}, \bibinfo {author}
  {\bibfnamefont {V.}~\bibnamefont {{Poireau}}}, \bibinfo {author}
  {\bibfnamefont {A.}~\bibnamefont {{Putze}}}, \bibinfo {author} {\bibfnamefont
  {S.}~\bibnamefont {{Rosier}}}, \bibinfo {author} {\bibfnamefont
  {P.}~\bibnamefont {{Salati}}}, \ and\ \bibinfo {author} {\bibfnamefont
  {M.}~\bibnamefont {{Vecchi}}},\ }\href {\doibase 10.1051/0004-6361/201630321}
  {\bibfield  {journal} {\bibinfo  {journal} {\aap}\ }\textbf {\bibinfo
  {volume} {605}},\ \bibinfo {eid} {A17} (\bibinfo {year}
  {2017}{\natexlab{b}})},\ \Eprint {http://arxiv.org/abs/1612.03924}
  {arXiv:1612.03924 [astro-ph.HE]} \BibitemShut {NoStop}%
\bibitem [{\citenamefont {{Bulanov}}\ and\ \citenamefont
  {{Dogel}}(1974)}]{BulanovEtAl1974}%
  \BibitemOpen
  \bibfield  {author} {\bibinfo {author} {\bibfnamefont {S.~V.}\ \bibnamefont
  {{Bulanov}}}\ and\ \bibinfo {author} {\bibfnamefont {V.~A.}\ \bibnamefont
  {{Dogel}}},\ }\href {\doibase 10.1007/BF02639066} {\bibfield  {journal}
  {\bibinfo  {journal} {\apss}\ }\textbf {\bibinfo {volume} {29}},\ \bibinfo
  {pages} {305} (\bibinfo {year} {1974})}\BibitemShut {NoStop}%
\bibitem [{\citenamefont {{Baltz}}\ and\ \citenamefont
  {{Edsj{\"o}}}(1998)}]{BaltzEtAl1998}%
  \BibitemOpen
  \bibfield  {author} {\bibinfo {author} {\bibfnamefont {E.~A.}\ \bibnamefont
  {{Baltz}}}\ and\ \bibinfo {author} {\bibfnamefont {J.}~\bibnamefont
  {{Edsj{\"o}}}},\ }\href {\doibase 10.1103/PhysRevD.59.023511} {\bibfield
  {journal} {\bibinfo  {journal} {\prd}\ }\textbf {\bibinfo {volume} {59}},\
  \bibinfo {eid} {023511} (\bibinfo {year} {1998})},\ \Eprint
  {http://arxiv.org/abs/astro-ph/9808243} {astro-ph/9808243} \BibitemShut
  {NoStop}%
\bibitem [{\citenamefont {{Lavalle}}\ \emph {et~al.}(2007)\citenamefont
  {{Lavalle}}, \citenamefont {{Pochon}}, \citenamefont {{Salati}},\ and\
  \citenamefont {{Taillet}}}]{LavalleEtAl2007}%
  \BibitemOpen
  \bibfield  {author} {\bibinfo {author} {\bibfnamefont {J.}~\bibnamefont
  {{Lavalle}}}, \bibinfo {author} {\bibfnamefont {J.}~\bibnamefont {{Pochon}}},
  \bibinfo {author} {\bibfnamefont {P.}~\bibnamefont {{Salati}}}, \ and\
  \bibinfo {author} {\bibfnamefont {R.}~\bibnamefont {{Taillet}}},\ }\href
  {\doibase 10.1051/0004-6361:20065312} {\bibfield  {journal} {\bibinfo
  {journal} {\aap}\ }\textbf {\bibinfo {volume} {462}},\ \bibinfo {pages} {827}
  (\bibinfo {year} {2007})},\ \Eprint
  {http://arxiv.org/abs/arXiv:astro-ph/0603796} {arXiv:astro-ph/0603796}
  \BibitemShut {NoStop}%
\bibitem [{\citenamefont {{Delahaye}}\ \emph {et~al.}(2009)\citenamefont
  {{Delahaye}}, \citenamefont {{Lineros}}, \citenamefont {{Donato}},
  \citenamefont {{Fornengo}}, \citenamefont {{Lavalle}}, \citenamefont
  {{Salati}},\ and\ \citenamefont {{Taillet}}}]{DelahayeEtAl2009}%
  \BibitemOpen
  \bibfield  {author} {\bibinfo {author} {\bibfnamefont {T.}~\bibnamefont
  {{Delahaye}}}, \bibinfo {author} {\bibfnamefont {R.}~\bibnamefont
  {{Lineros}}}, \bibinfo {author} {\bibfnamefont {F.}~\bibnamefont {{Donato}}},
  \bibinfo {author} {\bibfnamefont {N.}~\bibnamefont {{Fornengo}}}, \bibinfo
  {author} {\bibfnamefont {J.}~\bibnamefont {{Lavalle}}}, \bibinfo {author}
  {\bibfnamefont {P.}~\bibnamefont {{Salati}}}, \ and\ \bibinfo {author}
  {\bibfnamefont {R.}~\bibnamefont {{Taillet}}},\ }\href {\doibase
  10.1051/0004-6361/200811130} {\bibfield  {journal} {\bibinfo  {journal}
  {\aap}\ }\textbf {\bibinfo {volume} {501}},\ \bibinfo {pages} {821} (\bibinfo
  {year} {2009})},\ \Eprint {http://arxiv.org/abs/0809.5268} {arXiv:0809.5268}
  \BibitemShut {NoStop}%
\bibitem [{\citenamefont {{Lavalle}}\ \emph {et~al.}(2014)\citenamefont
  {{Lavalle}}, \citenamefont {{Maurin}},\ and\ \citenamefont
  {{Putze}}}]{LavalleEtAl2014}%
  \BibitemOpen
  \bibfield  {author} {\bibinfo {author} {\bibfnamefont {J.}~\bibnamefont
  {{Lavalle}}}, \bibinfo {author} {\bibfnamefont {D.}~\bibnamefont {{Maurin}}},
  \ and\ \bibinfo {author} {\bibfnamefont {A.}~\bibnamefont {{Putze}}},\ }\href
  {\doibase 10.1103/PhysRevD.90.081301} {\bibfield  {journal} {\bibinfo
  {journal} {\prd}\ }\textbf {\bibinfo {volume} {90}},\ \bibinfo {eid} {081301}
  (\bibinfo {year} {2014})},\ \Eprint {http://arxiv.org/abs/1407.2540}
  {arXiv:1407.2540 [astro-ph.HE]} \BibitemShut {NoStop}%
\bibitem [{\citenamefont {Reinert}\ and\ \citenamefont
  {Winkler}(2018)}]{ReinertEtAl2018}%
  \BibitemOpen
  \bibfield  {author} {\bibinfo {author} {\bibfnamefont {A.}~\bibnamefont
  {Reinert}}\ and\ \bibinfo {author} {\bibfnamefont {M.~W.}\ \bibnamefont
  {Winkler}},\ }\href {\doibase 10.1088/1475-7516/2018/01/055} {\bibfield
  {journal} {\bibinfo  {journal} {\jcap}\ }\textbf {\bibinfo {volume} {1}},\
  \bibinfo {eid} {055} (\bibinfo {year} {2018})},\ \Eprint
  {http://arxiv.org/abs/1712.00002} {arXiv:1712.00002 [astro-ph.HE]}
  \BibitemShut {NoStop}%
\bibitem [{\citenamefont {Boudaud}\ and\ \citenamefont
  {Cirelli}(2018)}]{BoudaudEtAl2018}%
  \BibitemOpen
  \bibfield  {author} {\bibinfo {author} {\bibfnamefont {M.}~\bibnamefont
  {Boudaud}}\ and\ \bibinfo {author} {\bibfnamefont {M.}~\bibnamefont
  {Cirelli}},\ }\href {http://adsabs.harvard.edu/abs/2018arXiv180703075B}
  {\bibfield  {journal} {\bibinfo  {journal} {ArXiv e-prints}\ } (\bibinfo
  {year} {2018})},\ \Eprint {http://arxiv.org/abs/1807.03075} {arXiv:1807.03075
  [astro-ph.HE]} \BibitemShut {NoStop}%
\bibitem [{\citenamefont {{Delahaye}}\ \emph {et~al.}(2010)\citenamefont
  {{Delahaye}}, \citenamefont {{Lavalle}}, \citenamefont {{Lineros}},
  \citenamefont {{Donato}},\ and\ \citenamefont
  {{Fornengo}}}]{DelahayeEtAl2010}%
  \BibitemOpen
  \bibfield  {author} {\bibinfo {author} {\bibfnamefont {T.}~\bibnamefont
  {{Delahaye}}}, \bibinfo {author} {\bibfnamefont {J.}~\bibnamefont
  {{Lavalle}}}, \bibinfo {author} {\bibfnamefont {R.}~\bibnamefont
  {{Lineros}}}, \bibinfo {author} {\bibfnamefont {F.}~\bibnamefont {{Donato}}},
  \ and\ \bibinfo {author} {\bibfnamefont {N.}~\bibnamefont {{Fornengo}}},\
  }\href {\doibase 10.1051/0004-6361/201014225} {\bibfield  {journal} {\bibinfo
   {journal} {\aap}\ }\textbf {\bibinfo {volume} {524}},\ \bibinfo {eid} {A51}
  (\bibinfo {year} {2010})},\ \Eprint {http://arxiv.org/abs/1002.1910}
  {arXiv:1002.1910 [astro-ph.HE]} \BibitemShut {NoStop}%
\bibitem [{\citenamefont {Lavalle}(2011)}]{Lavalle2011b}%
  \BibitemOpen
  \bibfield  {author} {\bibinfo {author} {\bibfnamefont {J.}~\bibnamefont
  {Lavalle}},\ }\href {\doibase 10.1111/j.1365-2966.2011.18294.x} {\bibfield
  {journal} {\bibinfo  {journal} {\mnras}\ }\textbf {\bibinfo {volume} {414}},\
  \bibinfo {pages} {985L} (\bibinfo {year} {2011})},\ \Eprint
  {http://arxiv.org/abs/1011.3063} {arXiv:1011.3063 [astro-ph.HE]} \BibitemShut
  {NoStop}%
\bibitem [{\citenamefont {{Shen}}(1970)}]{Shen1970}%
  \BibitemOpen
  \bibfield  {author} {\bibinfo {author} {\bibfnamefont {C.~S.}\ \bibnamefont
  {{Shen}}},\ }\href {\doibase 10.1086/180650} {\bibfield  {journal} {\bibinfo
  {journal} {\apjl}\ }\textbf {\bibinfo {volume} {162}},\ \bibinfo {pages}
  {L181+} (\bibinfo {year} {1970})}\BibitemShut {NoStop}%
\bibitem [{\citenamefont {{Aharonian}}\ \emph {et~al.}(1995)\citenamefont
  {{Aharonian}}, \citenamefont {{Atoyan}},\ and\ \citenamefont
  {{Voelk}}}]{AharonianEtAl1995}%
  \BibitemOpen
  \bibfield  {author} {\bibinfo {author} {\bibfnamefont {F.~A.}\ \bibnamefont
  {{Aharonian}}}, \bibinfo {author} {\bibfnamefont {A.~M.}\ \bibnamefont
  {{Atoyan}}}, \ and\ \bibinfo {author} {\bibfnamefont {H.~J.}\ \bibnamefont
  {{Voelk}}},\ }\href {http://adsabs.harvard.edu/abs/1995A%26A...294L..41A}
  {\bibfield  {journal} {\bibinfo  {journal} {\aap}\ }\textbf {\bibinfo
  {volume} {294}},\ \bibinfo {pages} {L41} (\bibinfo {year}
  {1995})}\BibitemShut {NoStop}%
\bibitem [{\citenamefont {{Hooper}}\ \emph {et~al.}(2009)\citenamefont
  {{Hooper}}, \citenamefont {{Blasi}},\ and\ \citenamefont {{Dario
  Serpico}}}]{HooperEtAl2009}%
  \BibitemOpen
  \bibfield  {author} {\bibinfo {author} {\bibfnamefont {D.}~\bibnamefont
  {{Hooper}}}, \bibinfo {author} {\bibfnamefont {P.}~\bibnamefont {{Blasi}}}, \
  and\ \bibinfo {author} {\bibfnamefont {P.}~\bibnamefont {{Dario Serpico}}},\
  }\href {\doibase 10.1088/1475-7516/2009/01/025} {\bibfield  {journal}
  {\bibinfo  {journal} {\jcap}\ }\textbf {\bibinfo {volume} {1}},\ \bibinfo
  {eid} {025} (\bibinfo {year} {2009})},\ \Eprint
  {http://arxiv.org/abs/0810.1527} {arXiv:0810.1527} \BibitemShut {NoStop}%
\bibitem [{\citenamefont {Profumo}(2011)}]{Profumo2011}%
  \BibitemOpen
  \bibfield  {author} {\bibinfo {author} {\bibfnamefont {S.}~\bibnamefont
  {Profumo}},\ }\href {\doibase 10.2478/s11534-011-0099-z} {\bibfield
  {journal} {\bibinfo  {journal} {Central Eur.J.Phys.}\ }\textbf {\bibinfo
  {volume} {10}},\ \bibinfo {pages} {1} (\bibinfo {year} {2012})},\ \Eprint
  {http://arxiv.org/abs/0812.4457} {arXiv:0812.4457} \BibitemShut {NoStop}%
\bibitem [{\citenamefont {{Boudaud}}\ \emph
  {et~al.}(2015{\natexlab{a}})\citenamefont {{Boudaud}}, \citenamefont
  {{Aupetit}}, \citenamefont {{Caroff}}, \citenamefont {{Putze}}, \citenamefont
  {{Belanger}}, \citenamefont {{Genolini}}, \citenamefont {{Goy}},
  \citenamefont {{Poireau}}, \citenamefont {{Poulin}}, \citenamefont
  {{Rosier}}, \citenamefont {{Salati}}, \citenamefont {{Tao}},\ and\
  \citenamefont {{Vecchi}}}]{BoudaudEtAl2015}%
  \BibitemOpen
  \bibfield  {author} {\bibinfo {author} {\bibfnamefont {M.}~\bibnamefont
  {{Boudaud}}}, \bibinfo {author} {\bibfnamefont {S.}~\bibnamefont
  {{Aupetit}}}, \bibinfo {author} {\bibfnamefont {S.}~\bibnamefont {{Caroff}}},
  \bibinfo {author} {\bibfnamefont {A.}~\bibnamefont {{Putze}}}, \bibinfo
  {author} {\bibfnamefont {G.}~\bibnamefont {{Belanger}}}, \bibinfo {author}
  {\bibfnamefont {Y.}~\bibnamefont {{Genolini}}}, \bibinfo {author}
  {\bibfnamefont {C.}~\bibnamefont {{Goy}}}, \bibinfo {author} {\bibfnamefont
  {V.}~\bibnamefont {{Poireau}}}, \bibinfo {author} {\bibfnamefont
  {V.}~\bibnamefont {{Poulin}}}, \bibinfo {author} {\bibfnamefont
  {S.}~\bibnamefont {{Rosier}}}, \bibinfo {author} {\bibfnamefont
  {P.}~\bibnamefont {{Salati}}}, \bibinfo {author} {\bibfnamefont
  {L.}~\bibnamefont {{Tao}}}, \ and\ \bibinfo {author} {\bibfnamefont
  {M.}~\bibnamefont {{Vecchi}}},\ }\href {\doibase 10.1051/0004-6361/201425197}
  {\bibfield  {journal} {\bibinfo  {journal} {\aap}\ }\textbf {\bibinfo
  {volume} {575}},\ \bibinfo {eid} {A67} (\bibinfo {year}
  {2015}{\natexlab{a}})},\ \Eprint {http://arxiv.org/abs/1410.3799}
  {arXiv:1410.3799 [astro-ph.HE]} \BibitemShut {NoStop}%
\bibitem [{\citenamefont {{DuVernois}}\ \emph {et~al.}(2001)\citenamefont
  {{DuVernois}}, \citenamefont {{Barwick}}, \citenamefont {{Beatty}},
  \citenamefont {{Bhattacharyya}}, \citenamefont {{Bower}}, \citenamefont
  {{Chaput}}, \citenamefont {{Coutu}}, \citenamefont {{de Nolfo}},
  \citenamefont {{Lowder}}, \citenamefont {{McKee}}, \citenamefont
  {{M{\"u}ller}}, \citenamefont {{Musser}}, \citenamefont {{Nutter}},
  \citenamefont {{Schneider}}, \citenamefont {{Swordy}}, \citenamefont
  {{Tarl{\'e}}}, \citenamefont {{Tomasch}},\ and\ \citenamefont
  {{Torbet}}}]{DuVernoisEtAl2001}%
  \BibitemOpen
  \bibfield  {author} {\bibinfo {author} {\bibfnamefont {HEAT Collaboration}},
    \bibinfo {author} {\bibfnamefont {M.~A.}\ \bibnamefont {{DuVernois}}},
    \bibinfo {author} {\etal},\ }\href {\doibase
  10.1086/322324} {\bibfield  {journal} {\bibinfo  {journal} {\apj}\ }\textbf
  {\bibinfo {volume} {559}},\ \bibinfo {pages} {296} (\bibinfo {year}
  {2001})}\BibitemShut {NoStop}%
\bibitem [{\citenamefont {{Adriani}}\ \emph {et~al.}(2009)\citenamefont
  {{Adriani}}, \citenamefont {{Barbarino}}, \citenamefont {{Bazilevskaya}},
  \citenamefont {{Bellotti}}, \citenamefont {{Boezio}}, \citenamefont
  {{Bogomolov}}, \citenamefont {{Bonechi}}, \citenamefont {{Bongi}},
  \citenamefont {{Bonvicini}}, \citenamefont {{Bottai}}, \citenamefont
  {{Bruno}}, \citenamefont {{Cafagna}}, \citenamefont {{Campana}},
  \citenamefont {{Carlson}}, \citenamefont {{Casolino}}, \citenamefont
  {{Castellini}}, \citenamefont {{de Pascale}}, \citenamefont {{de Rosa}},
  \citenamefont {{de Simone}}, \citenamefont {{di Felice}}, \citenamefont
  {{Galper}}, \citenamefont {{Grishantseva}}, \citenamefont {{Hofverberg}},
  \citenamefont {{Koldashov}}, \citenamefont {{Krutkov}}, \citenamefont
  {{Kvashnin}}, \citenamefont {{Leonov}}, \citenamefont {{Malvezzi}},
  \citenamefont {{Marcelli}}, \citenamefont {{Menn}}, \citenamefont
  {{Mikhailov}}, \citenamefont {{Mocchiutti}}, \citenamefont {{Orsi}},
  \citenamefont {{Osteria}}, \citenamefont {{Papini}}, \citenamefont
  {{Pearce}}, \citenamefont {{Picozza}}, \citenamefont {{Ricci}}, \citenamefont
  {{Ricciarini}}, \citenamefont {{Simon}}, \citenamefont {{Sparvoli}},
  \citenamefont {{Spillantini}}, \citenamefont {{Stozhkov}}, \citenamefont
  {{Vacchi}}, \citenamefont {{Vannuccini}}, \citenamefont {{Vasilyev}},
  \citenamefont {{Voronov}}, \citenamefont {{Yurkin}}, \citenamefont {{Zampa}},
  \citenamefont {{Zampa}},\ and\ \citenamefont {{Zverev}}}]{AdrianiEtAl2009}%
  \BibitemOpen
  \bibfield  {author} {\bibinfo {author} {\bibfnamefont {PAMELA Collaboration}},
    \bibinfo {author} {\bibfnamefont {O.}~\bibnamefont{{Adriani}}},\bibinfo {author} {\etal},\
  }\href {\doibase 10.1038/nature07942} {\bibfield  {journal} {\bibinfo
  {journal} {Nature}\ }\textbf {\bibinfo {volume} {458}},\ \bibinfo {pages}
  {607} (\bibinfo {year} {2009})},\ \Eprint {http://arxiv.org/abs/0810.4995}
  {arXiv:0810.4995} \BibitemShut {NoStop}%
\bibitem [{\citenamefont {{Aguilar}}\ \emph {et~al.}(2013)\citenamefont
  {{Aguilar}}, \citenamefont {{Alberti}}, \citenamefont {{Alpat}},
  \citenamefont {{Alvino}}, \citenamefont {{Ambrosi}}, \citenamefont
  {{Andeen}}, \citenamefont {{Anderhub}}, \citenamefont {{Arruda}},
  \citenamefont {{Azzarello}}, \citenamefont {{Bachlechner}}, \citenamefont
  {{Barao}}, \citenamefont {{Baret}}, \citenamefont {{Barrau}}, \citenamefont
  {{Barrin}}, \citenamefont {{Bartoloni}}, \citenamefont {{Basara}},
  \citenamefont {{Basili}}, \citenamefont {{Batalha}}, \citenamefont {{Bates}},
  \citenamefont {{Battiston}}, \citenamefont {{Bazo}}, \citenamefont
  {{Becker}}, \citenamefont {{Becker}}, \citenamefont {{Behlmann}},
  \citenamefont {{Beischer}}, \citenamefont {{Berdugo}}, \citenamefont
  {{Berges}}, \citenamefont {{Bertucci}}, \citenamefont {{Bigongiari}},
  \citenamefont {{Biland}}, \citenamefont {{Bindi}}, \citenamefont
  {{Bizzaglia}}, \citenamefont {{Boella}}, \citenamefont {{de Boer}},
  \citenamefont {{Bollweg}}, \citenamefont {{Bolmont}}, \citenamefont
  {{Borgia}}, \citenamefont {{Borsini}}, \citenamefont {{Boschini}},
  \citenamefont {{Boudoul}}, \citenamefont {{Bourquin}}, \citenamefont
  {{Brun}}, \citenamefont {{Bu{\'e}nerd}}, \citenamefont {{Burger}},
  \citenamefont {{Burger}}, \citenamefont {{Cadoux}}, \citenamefont {{Cai}},
  \citenamefont {{Capell}}, \citenamefont {{Casadei}}, \citenamefont
  {{Casaus}}, \citenamefont {{Cascioli}}, \citenamefont {{Castellini}},
  \citenamefont {{Cernuda}}, \citenamefont {{Cervelli}}, \citenamefont
  {{Chae}}, \citenamefont {{Chang}}, \citenamefont {{Chen}}, \citenamefont
  {{Chen}}, \citenamefont {{Chen}}, \citenamefont {{Cheng}}, \citenamefont
  {{Chen}}, \citenamefont {{Cheng}}, \citenamefont {{Chernoplyiokov}},
  \citenamefont {{Chikanian}}, \citenamefont {{Choumilov}}, \citenamefont
  {{Choutko}}, \citenamefont {{Chung}}, \citenamefont {{Clark}}, \citenamefont
  {{Clavero}}, \citenamefont {{Coignet}}, \citenamefont {{Commichau}},
  \citenamefont {{Consolandi}}, \citenamefont {{Contin}}, \citenamefont
  {{Corti}}, \citenamefont {{Costado Dios}}, \citenamefont {{Coste}},
  \citenamefont {{Crespo}}, \citenamefont {{Cui}}, \citenamefont {{Dai}},
  \citenamefont {{Delgado}}, \citenamefont {{Della Torre}}, \citenamefont
  {{Demirkoz}}, \citenamefont {{Dennett}}, \citenamefont {{Derome}},
  \citenamefont {{Di Falco}}, \citenamefont {{Diao}}, \citenamefont {{Diago}},
  \citenamefont {{Djambazov}}, \citenamefont {{D{\'{\i}}az}}, \citenamefont
  {{von Doetinchem}}, \citenamefont {{Du}}, \citenamefont {{Dubois}},
  \citenamefont {{Duperay}}, \citenamefont {{Duranti}}, \citenamefont
  {{D'Urso}}, \citenamefont {{Egorov}}, \citenamefont {{Eline}}, \citenamefont
  {{Eppling}}, \citenamefont {{Eronen}}, \citenamefont {{van Es}},
  \citenamefont {{Esser}}, \citenamefont {{Falvard}}, \citenamefont
  {{Fiandrini}}, \citenamefont {{Fiasson}}, \citenamefont {{Finch}},
  \citenamefont {{Fisher}}, \citenamefont {{Flood}}, \citenamefont {{Foglio}},
  \citenamefont {{Fohey}}, \citenamefont {{Fopp}}, \citenamefont {{Fouque}},
  \citenamefont {{Galaktionov}}, \citenamefont {{Gallilee}}, \citenamefont
  {{Gallin-Martel}}, \citenamefont {{Gallucci}}, \citenamefont
  {{Garc{\'{\i}}a}}, \citenamefont {{Garc{\'{\i}}a}}, \citenamefont
  {{Garc{\'{\i}}a-L{\'o}pez}}, \citenamefont {{Garc{\'{\i}}a-Tabares}},
  \citenamefont {{Gargiulo}}, \citenamefont {{Gast}}, \citenamefont
  {{Gebauer}}, \citenamefont {{Gentile}}, \citenamefont {{Gervasi}},
  \citenamefont {{Gillard}}, \citenamefont {{Giovacchini}}, \citenamefont
  {{Girard}}, \citenamefont {{Goglov}}, \citenamefont {{Gong}}, \citenamefont
  {{Goy-Henningsen}}, \citenamefont {{Grandi}}, \citenamefont {{Graziani}},
  \citenamefont {{Grechko}}, \citenamefont {{Gross}}, \citenamefont {{Guerri}},
  \citenamefont {{de la Gu{\'{\i}}a}}, \citenamefont {{Guo}}, \citenamefont
  {{Habiby}}, \citenamefont {{Haino}}, \citenamefont {{Hauler}}, \citenamefont
  {{He}}, \citenamefont {{Heil}}, \citenamefont {{Heilig}}, \citenamefont
  {{Hermel}}, \citenamefont {{Hofer}}, \citenamefont {{Huang}}, \citenamefont
  {{Hungerford}}, \citenamefont {{Incagli}}, \citenamefont {{Ionica}},
  \citenamefont {{Jacholkowska}}, \citenamefont {{Jang}}, \citenamefont
  {{Jinchi}}, \citenamefont {{Jongmanns}}, \citenamefont {{Journet}},
  \citenamefont {{Jungermann}}, \citenamefont {{Karpinski}}, \citenamefont
  {{Kim}}, \citenamefont {{Kim}}, \citenamefont {{Kirn}}, \citenamefont
  {{Kossakowski}}, \citenamefont {{Koulemzine}}, \citenamefont {{Kounina}},
  \citenamefont {{Kounine}}, \citenamefont {{Koutsenko}}, \citenamefont
  {{Krafczyk}}, \citenamefont {{Laudi}}, \citenamefont {{Laurenti}},
  \citenamefont {{Lauritzen}}, \citenamefont {{Lebedev}}, \citenamefont
  {{Lee}}, \citenamefont {{Lee}}, \citenamefont {{Leluc}}, \citenamefont
  {{Le{\'o}n Vargas}}, \citenamefont {{Lepareur}}, \citenamefont {{Li}},
  \citenamefont {{Li}}, \citenamefont {{Li}}, \citenamefont {{Li}},
  \citenamefont {{Li}}, \citenamefont {{Lipari}}, \citenamefont {{Lin}},
  \citenamefont {{Liu}}, \citenamefont {{Liu}}, \citenamefont {{Lomtadze}},
  \citenamefont {{Lu}}, \citenamefont {{Lucidi}}, \citenamefont
  {{L{\"u}belsmeyer}}, \citenamefont {{Luo}}, \citenamefont {{Lustermann}},
  \citenamefont {{Lv}}, \citenamefont {{Madsen}}, \citenamefont {{Majka}},
  \citenamefont {{Malinin}}, \citenamefont {{Ma{\~n}{\'a}}}, \citenamefont
  {{Mar{\'{\i}}n}}, \citenamefont {{Martin}}, \citenamefont
  {{Mart{\'{\i}}nez}}, \citenamefont {{Masciocchi}}, \citenamefont {{Masi}},
  \citenamefont {{Maurin}}, \citenamefont {{McInturff}}, \citenamefont
  {{McIntyre}}, \citenamefont {{Menchaca-Rocha}}, \citenamefont {{Meng}},
  \citenamefont {{Menichelli}}, \citenamefont {{Mereu}}, \citenamefont
  {{Millinger}}, \citenamefont {{Mo}}, \citenamefont {{Molina}}, \citenamefont
  {{Mott}}, \citenamefont {{Mujunen}}, \citenamefont {{Natale}}, \citenamefont
  {{Nemeth}}, \citenamefont {{Ni}}, \citenamefont {{Nikonov}}, \citenamefont
  {{Nozzoli}}, \citenamefont {{Nunes}}, \citenamefont {{Obermeier}},
  \citenamefont {{Oh}}, \citenamefont {{Oliva}}, \citenamefont {{Palmonari}},
  \citenamefont {{Palomares}}, \citenamefont {{Paniccia}}, \citenamefont
  {{Papi}}, \citenamefont {{Park}}, \citenamefont {{Pauluzzi}}, \citenamefont
  {{Pauss}}, \citenamefont {{Pauw}}, \citenamefont {{Pedreschi}}, \citenamefont
  {{Pensotti}}, \citenamefont {{Pereira}}, \citenamefont {{Perrin}},
  \citenamefont {{Pessina}}, \citenamefont {{Pierschel}}, \citenamefont
  {{Pilo}}, \citenamefont {{Piluso}}, \citenamefont {{Pizzolotto}},
  \citenamefont {{Plyaskin}}, \citenamefont {{Pochon}}, \citenamefont {{Pohl}},
  \citenamefont {{Poireau}}, \citenamefont {{Porter}}, \citenamefont {{Pouxe}},
  \citenamefont {{Putze}}, \citenamefont {{Quadrani}}, \citenamefont {{Qi}},
  \citenamefont {{Rancoita}}, \citenamefont {{Rapin}}, \citenamefont {{Ren}},
  \citenamefont {{Ricol}}, \citenamefont {{Riihonen}}, \citenamefont
  {{Rodr{\'{\i}}guez}}, \citenamefont {{Roeser}}, \citenamefont
  {{Rosier-Lees}}, \citenamefont {{Rossi}}, \citenamefont {{Rozhkov}},
  \citenamefont {{Rozza}}, \citenamefont {{Sabellek}}, \citenamefont
  {{Sagdeev}}, \citenamefont {{Sandweiss}}, \citenamefont {{Santos}},
  \citenamefont {{Saouter}}, \citenamefont {{Sarchioni}}, \citenamefont
  {{Schael}}, \citenamefont {{Schinzel}}, \citenamefont {{Schmanau}},
  \citenamefont {{Schwering}}, \citenamefont {{Schulz von Dratzig}},
  \citenamefont {{Scolieri}}, \citenamefont {{Seo}}, \citenamefont {{Shan}},
  \citenamefont {{Shi}}, \citenamefont {{Shi}}, \citenamefont {{Siedenburg}},
  \citenamefont {{Siedling}}, \citenamefont {{Son}}, \citenamefont {{Spada}},
  \citenamefont {{Spinella}}, \citenamefont {{Steuer}}, \citenamefont
  {{Stiff}}, \citenamefont {{Sun}}, \citenamefont {{Sun}}, \citenamefont
  {{Sun}}, \citenamefont {{Tacconi}}, \citenamefont {{Tang}}, \citenamefont
  {{Tang}}, \citenamefont {{Tang}}, \citenamefont {{Tao}}, \citenamefont
  {{Tassan-Viol}}, \citenamefont {{Ting}}, \citenamefont {{Ting}},
  \citenamefont {{Titus}}, \citenamefont {{Tomassetti}}, \citenamefont
  {{Toral}}, \citenamefont {{Torsti}}, \citenamefont {{Tsai}}, \citenamefont
  {{Tutt}}, \citenamefont {{Ulbricht}}, \citenamefont {{Urban}}, \citenamefont
  {{Vagelli}}, \citenamefont {{Valente}}, \citenamefont {{Vannini}},
  \citenamefont {{Valtonen}}, \citenamefont {{Vargas Trevino}}, \citenamefont
  {{Vaurynovich}}, \citenamefont {{Vecchi}}, \citenamefont {{Vergain}},
  \citenamefont {{Verlaat}}, \citenamefont {{Vescovi}}, \citenamefont
  {{Vialle}}, \citenamefont {{Viertel}}, \citenamefont {{Volpini}},
  \citenamefont {{Wang}}, \citenamefont {{Wang}}, \citenamefont {{Wang}},
  \citenamefont {{Wang}}, \citenamefont {{Wang}}, \citenamefont {{Wang}},
  \citenamefont {{Wallraff}}, \citenamefont {{Weng}}, \citenamefont
  {{Willenbrock}}, \citenamefont {{Wlochal}}, \citenamefont {{Wu}},
  \citenamefont {{Wu}}, \citenamefont {{Wu}}, \citenamefont {{Xiao}},
  \citenamefont {{Xie}}, \citenamefont {{Xiong}}, \citenamefont {{Xin}},
  \citenamefont {{Xu}}, \citenamefont {{Xu}}, \citenamefont {{Yan}},
  \citenamefont {{Yang}}, \citenamefont {{Yang}}, \citenamefont {{Ye}},
  \citenamefont {{Yi}}, \citenamefont {{Yu}}, \citenamefont {{Yu}},
  \citenamefont {{Zeissler}}, \citenamefont {{Zhang}}, \citenamefont {{Zhang}},
  \citenamefont {{Zhang}}, \citenamefont {{Zheng}}, \citenamefont {{Zhuang}},
  \citenamefont {{Zhukov}}, \citenamefont {{Zichichi}}, \citenamefont
  {{Zuccon}},\ and\ \citenamefont {{Zurbach}}}]{AguilarEtAl2013}%
  \BibitemOpen
  \bibfield  {author} {
    \bibinfo {author} {\bibinfo {author} {\bibfnamefont{AMS-02 Collaboration}},
      \bibfnamefont {M.}~\bibnamefont{{Aguilar}}},
    \bibinfo {author} {\etal},\ }\href {\doibase 10.1103/PhysRevLett.110.141102} {\bibfield
  {journal} {\bibinfo  {journal} {\prl}\ }\textbf {\bibinfo {volume} {110}},\
  \bibinfo {eid} {141102} (\bibinfo {year} {2013})}\BibitemShut {NoStop}%
\bibitem [{\citenamefont {Zhao}\ \emph {et~al.}(2016)\citenamefont {Zhao},
  \citenamefont {Bi}, \citenamefont {Jia}, \citenamefont {Yin},\ and\
  \citenamefont {Zhu}}]{ZhaoEtAl2016}%
  \BibitemOpen
  \bibfield  {author} {\bibinfo {author} {\bibfnamefont {Y.}~\bibnamefont
  {Zhao}}, \bibinfo {author} {\bibfnamefont {X.-J.}\ \bibnamefont {Bi}},
  \bibinfo {author} {\bibfnamefont {H.-Y.}\ \bibnamefont {Jia}}, \bibinfo
  {author} {\bibfnamefont {P.-F.}\ \bibnamefont {Yin}}, \ and\ \bibinfo
  {author} {\bibfnamefont {F.-R.}\ \bibnamefont {Zhu}},\ }\href {\doibase
  10.1103/PhysRevD.93.083513} {\bibfield  {journal} {\bibinfo  {journal}
  {\prd}\ }\textbf {\bibinfo {volume} {93}},\ \bibinfo {eid} {083513} (\bibinfo
  {year} {2016})},\ \Eprint {http://arxiv.org/abs/1601.02181} {arXiv:1601.02181
  [astro-ph.HE]} \BibitemShut {NoStop}%
\bibitem [{\citenamefont {{Donato}}\ \emph {et~al.}(2004)\citenamefont
  {{Donato}}, \citenamefont {{Fornengo}}, \citenamefont {{Maurin}},
  \citenamefont {{Salati}},\ and\ \citenamefont {{Taillet}}}]{DonatoEtAl2004}%
  \BibitemOpen
  \bibfield  {author} {\bibinfo {author} {\bibfnamefont {F.}~\bibnamefont
  {{Donato}}}, \bibinfo {author} {\bibfnamefont {N.}~\bibnamefont
  {{Fornengo}}}, \bibinfo {author} {\bibfnamefont {D.}~\bibnamefont
  {{Maurin}}}, \bibinfo {author} {\bibfnamefont {P.}~\bibnamefont {{Salati}}},
  \ and\ \bibinfo {author} {\bibfnamefont {R.}~\bibnamefont {{Taillet}}},\
  }\href {\doibase 10.1103/PhysRevD.69.063501} {\bibfield  {journal} {\bibinfo
  {journal} {\prd}\ }\textbf {\bibinfo {volume} {69}},\ \bibinfo {eid} {063501}
  (\bibinfo {year} {2004})},\ \Eprint {http://arxiv.org/abs/astro-ph/0306207}
  {astro-ph/0306207} \BibitemShut {NoStop}%
\bibitem [{\citenamefont {Boudaud}\ \emph {et~al.}(2018)\citenamefont
  {Boudaud}, \citenamefont {Caroff}, \citenamefont {G{\'e}nolini},
  \citenamefont {Poulin}, \citenamefont {Silva~Batista}, \citenamefont
  {Derome}, \citenamefont {Lavalle}, \citenamefont {Maurin}, \citenamefont
  {Poireau}, \citenamefont {Rosier}, \citenamefont {Salati}, \citenamefont
  {Serpico},\ and\ \citenamefont {Vecchi}}]{BoudaudEtAl2018a}%
  \BibitemOpen
  \bibfield  {author} {\bibinfo {author} {\bibfnamefont {M.}~\bibnamefont
  {Boudaud}}, \bibinfo {author} {\bibfnamefont {S.}~\bibnamefont {Caroff}},
  \bibinfo {author} {\bibfnamefont {Y.}~\bibnamefont {G{\'e}nolini}}, \bibinfo
  {author} {\bibfnamefont {V.}~\bibnamefont {Poulin}}, \bibinfo {author}
  {\bibfnamefont {P.-I.}\ \bibnamefont {Silva~Batista}}, \bibinfo {author}
  {\bibfnamefont {L.}~\bibnamefont {Derome}}, \bibinfo {author} {\bibfnamefont
  {J.}~\bibnamefont {Lavalle}}, \bibinfo {author} {\bibfnamefont
  {D.}~\bibnamefont {Maurin}}, \bibinfo {author} {\bibfnamefont
  {V.}~\bibnamefont {Poireau}}, \bibinfo {author} {\bibfnamefont
  {S.}~\bibnamefont {Rosier}}, \bibinfo {author} {\bibfnamefont
  {P.}~\bibnamefont {Salati}}, \bibinfo {author} {\bibfnamefont {P.~D.}\
  \bibnamefont {Serpico}}, \ and\ \bibinfo {author} {\bibfnamefont
  {M.}~\bibnamefont {Vecchi}},\ }\href@noop {} {\bibfield  {journal} {\bibinfo
  {journal} {In preparation}\ } (\bibinfo {year} {2018})}\BibitemShut {NoStop}%
\bibitem [{\citenamefont {Aguilar}\ \emph {et~al.}(2016)\citenamefont
  {Aguilar}, \citenamefont {Ali~Cavasonza}, \citenamefont {Ambrosi},
  \citenamefont {Arruda}, \citenamefont {Attig}, \citenamefont {Aupetit},
  \citenamefont {Azzarello}, \citenamefont {Bachlechner}, \citenamefont
  {Barao}, \citenamefont {Barrau}, \citenamefont {Barrin}, \citenamefont
  {Bartoloni}, \citenamefont {Basara}, \citenamefont {Ba{\c s}e{\v g}mez-du
  Pree}, \citenamefont {Battarbee}, \citenamefont {Battiston}, \citenamefont
  {Becker}, \citenamefont {Behlmann}, \citenamefont {Beischer}, \citenamefont
  {Berdugo}, \citenamefont {Bertucci}, \citenamefont {Bindel}, \citenamefont
  {Bindi}, \citenamefont {Boella}, \citenamefont {de~Boer}, \citenamefont
  {Bollweg}, \citenamefont {Bonnivard}, \citenamefont {Borgia}, \citenamefont
  {Boschini}, \citenamefont {Bourquin}, \citenamefont {Bueno}, \citenamefont
  {Burger}, \citenamefont {Cadoux}, \citenamefont {Cai}, \citenamefont
  {Capell}, \citenamefont {Caroff}, \citenamefont {Casaus}, \citenamefont
  {Castellini}, \citenamefont {Cervelli}, \citenamefont {Chae}, \citenamefont
  {Chang}, \citenamefont {Chen}, \citenamefont {Chen}, \citenamefont {Chen},
  \citenamefont {Cheng}, \citenamefont {Chou}, \citenamefont {Choumilov},
  \citenamefont {Choutko}, \citenamefont {Chung}, \citenamefont {Clark},
  \citenamefont {Clavero}, \citenamefont {Coignet}, \citenamefont {Consolandi},
  \citenamefont {Contin}, \citenamefont {Corti}, \citenamefont {Creus},
  \citenamefont {Crispoltoni}, \citenamefont {Cui}, \citenamefont {Dai},
  \citenamefont {Delgado}, \citenamefont {Della~Torre}, \citenamefont
  {Demakov}, \citenamefont {Demirk{\"o}z}, \citenamefont {Derome},
  \citenamefont {Di~Falco}, \citenamefont {Dimiccoli}, \citenamefont
  {D{\'{\i}}az}, \citenamefont {von Doetinchem}, \citenamefont {Dong},
  \citenamefont {Donnini}, \citenamefont {Duranti}, \citenamefont {D'Urso},
  \citenamefont {Egorov}, \citenamefont {Eline}, \citenamefont {Eronen},
  \citenamefont {Feng}, \citenamefont {Fiandrini}, \citenamefont {Finch},
  \citenamefont {Fisher}, \citenamefont {Formato}, \citenamefont {Galaktionov},
  \citenamefont {Gallucci}, \citenamefont {Garc{\'{\i}}a}, \citenamefont
  {Garc{\'{\i}}a-L{\'o}pez}, \citenamefont {Gargiulo}, \citenamefont {Gast},
  \citenamefont {Gebauer}, \citenamefont {Gervasi}, \citenamefont {Ghelfi},
  \citenamefont {Giovacchini}, \citenamefont {Goglov}, \citenamefont
  {G{\'o}mez-Coral}, \citenamefont {Gong}, \citenamefont {Goy}, \citenamefont
  {Grabski}, \citenamefont {Grandi}, \citenamefont {Graziani}, \citenamefont
  {Guo}, \citenamefont {Haino}, \citenamefont {Han}, \citenamefont {He},
  \citenamefont {Heil}, \citenamefont {Hoffman}, \citenamefont {Hsieh},
  \citenamefont {Huang}, \citenamefont {Huang}, \citenamefont {Huh},
  \citenamefont {Incagli}, \citenamefont {Ionica}, \citenamefont {Jang},
  \citenamefont {Jinchi}, \citenamefont {Kang}, \citenamefont {Kanishev},
  \citenamefont {Kim}, \citenamefont {Kim}, \citenamefont {Kirn}, \citenamefont
  {Konak}, \citenamefont {Kounina}, \citenamefont {Kounine}, \citenamefont
  {Koutsenko}, \citenamefont {Krafczyk}, \citenamefont {La~Vacca},
  \citenamefont {Laudi}, \citenamefont {Laurenti}, \citenamefont {Lazzizzera},
  \citenamefont {Lebedev}, \citenamefont {Lee}, \citenamefont {Lee},
  \citenamefont {Leluc}, \citenamefont {Li}, \citenamefont {Li}, \citenamefont
  {Li}, \citenamefont {Li}, \citenamefont {Li}, \citenamefont {Li},
  \citenamefont {Li}, \citenamefont {Li}, \citenamefont {Li}, \citenamefont
  {Lim}, \citenamefont {Lin}, \citenamefont {Lipari}, \citenamefont {Lippert},
  \citenamefont {Liu}, \citenamefont {Liu}, \citenamefont {Lordello},
  \citenamefont {Lu}, \citenamefont {Lu}, \citenamefont {Luebelsmeyer},
  \citenamefont {Luo}, \citenamefont {Luo}, \citenamefont {Lv}, \citenamefont
  {Machate}, \citenamefont {Majka}, \citenamefont {Ma{\~n}{\'a}}, \citenamefont
  {Mar{\'{\i}}n}, \citenamefont {Martin}, \citenamefont {Mart{\'{\i}}nez},
  \citenamefont {Masi}, \citenamefont {Maurin}, \citenamefont {Menchaca-Rocha},
  \citenamefont {Meng}, \citenamefont {Mikuni}, \citenamefont {Mo},
  \citenamefont {Morescalchi}, \citenamefont {Mott}, \citenamefont {Nelson},
  \citenamefont {Ni}, \citenamefont {Nikonov}, \citenamefont {Nozzoli},
  \citenamefont {Oliva}, \citenamefont {Orcinha}, \citenamefont {Palmonari},
  \citenamefont {Palomares}, \citenamefont {Paniccia}, \citenamefont
  {Pauluzzi}, \citenamefont {Pensotti}, \citenamefont {Pereira}, \citenamefont
  {Picot-Clemente}, \citenamefont {Pilo}, \citenamefont {Pizzolotto},
  \citenamefont {Plyaskin}, \citenamefont {Pohl}, \citenamefont {Poireau},
  \citenamefont {Putze}, \citenamefont {Quadrani}, \citenamefont {Qi},
  \citenamefont {Qin}, \citenamefont {Qu}, \citenamefont {R{\"a}ih{\"a}},
  \citenamefont {Rancoita}, \citenamefont {Rapin}, \citenamefont {Ricol},
  \citenamefont {Rosier-Lees}, \citenamefont {Rozhkov}, \citenamefont {Rozza},
  \citenamefont {Sagdeev}, \citenamefont {Sandweiss}, \citenamefont {Saouter},
  \citenamefont {Schael}, \citenamefont {Schmidt}, \citenamefont {Schulz~von
  Dratzig}, \citenamefont {Schwering}, \citenamefont {Seo}, \citenamefont
  {Shan}, \citenamefont {Shi}, \citenamefont {Siedenburg}, \citenamefont {Son},
  \citenamefont {Song}, \citenamefont {Sun}, \citenamefont {Tacconi},
  \citenamefont {Tang}, \citenamefont {Tang}, \citenamefont {Tao},
  \citenamefont {Tescaro}, \citenamefont {Ting}, \citenamefont {Ting},
  \citenamefont {Tomassetti}, \citenamefont {Torsti}, \citenamefont
  {T{\"u}rko{\v g}lu}, \citenamefont {Urban}, \citenamefont {Vagelli},
  \citenamefont {Valente}, \citenamefont {Vannini}, \citenamefont {Valtonen},
  \citenamefont {V{\'a}zquez~Acosta}, \citenamefont {Vecchi}, \citenamefont
  {Velasco}, \citenamefont {Vialle}, \citenamefont {Vitale}, \citenamefont
  {Vitillo}, \citenamefont {Wang}, \citenamefont {Wang}, \citenamefont {Wang},
  \citenamefont {Wang}, \citenamefont {Wang}, \citenamefont {Wang},
  \citenamefont {Wei}, \citenamefont {Weng}, \citenamefont {Whitman},
  \citenamefont {Wienkenh{\"o}ver}, \citenamefont {Wu}, \citenamefont {Wu},
  \citenamefont {Xia}, \citenamefont {Xiong}, \citenamefont {Xu}, \citenamefont
  {Yan}, \citenamefont {Yang}, \citenamefont {Yang}, \citenamefont {Yang},
  \citenamefont {Yi}, \citenamefont {Yu}, \citenamefont {Yu}, \citenamefont
  {Zeissler}, \citenamefont {Zhang}, \citenamefont {Zhang}, \citenamefont
  {Zhang}, \citenamefont {Zhang}, \citenamefont {Zhang}, \citenamefont {Zhang},
  \citenamefont {Zheng}, \citenamefont {Zhu}, \citenamefont {Zhuang},
  \citenamefont {Zhukov}, \citenamefont {Zichichi}, \citenamefont {Zimmermann},
  \citenamefont {Zuccon},\ and\ \citenamefont
  {Collaboration}}]{AguilarEtAl2016a}%
  \BibitemOpen
  \bibfield  {author} {\bibinfo {author} {\bibfnamefont {AMS-02 Collaboration}},
    \bibinfo {author} {\bibfnamefont {M.}~\bibnamefont
  {Aguilar}}, \bibinfo {author} {\etal},\ }\href {\doibase
  10.1103/PhysRevLett.117.231102} {\bibfield  {journal} {\bibinfo  {journal}
  {\prl}\ }\textbf {\bibinfo {volume} {117}},\ \bibinfo {eid} {231102}
  (\bibinfo {year} {2016})}\BibitemShut {NoStop}%
\bibitem [{\citenamefont {Ahn}\ \emph {et~al.}(2010)\citenamefont {Ahn},
  \citenamefont {Allison}, \citenamefont {Bagliesi}, \citenamefont {Beatty},
  \citenamefont {Bigongiari}, \citenamefont {Childers}, \citenamefont
  {Conklin}, \citenamefont {Coutu}, \citenamefont {DuVernois}, \citenamefont
  {Ganel}, \citenamefont {Han}, \citenamefont {Jeon}, \citenamefont {Kim},
  \citenamefont {Lee}, \citenamefont {Lutz}, \citenamefont {Maestro},
  \citenamefont {Malinin}, \citenamefont {Marrocchesi}, \citenamefont
  {Minnick}, \citenamefont {Mognet}, \citenamefont {Nam}, \citenamefont {Nam},
  \citenamefont {Nutter}, \citenamefont {Park}, \citenamefont {Park},
  \citenamefont {Seo}, \citenamefont {Sina}, \citenamefont {Wu}, \citenamefont
  {Yang}, \citenamefont {Yoon}, \citenamefont {Zei},\ and\ \citenamefont
  {Zinn}}]{AhnEtAl2010}%
  \BibitemOpen
  \bibfield  {author} {\bibinfo {author} {\bibfnamefont {CREAM Collaboration}},
  \bibinfo {author} {\bibfnamefont {H.~S.}\ \bibnamefont
  {Ahn}}, \bibinfo {author} {\etal},\ }\href {\doibase
  10.1088/2041-8205/714/1/L89} {\bibfield  {journal} {\bibinfo  {journal}
  {\apjl}\ }\textbf {\bibinfo {volume} {714}},\ \bibinfo {pages} {L89}
  (\bibinfo {year} {2010})},\ \Eprint {http://arxiv.org/abs/1004.1123}
  {arXiv:1004.1123 [astro-ph.HE]} \BibitemShut {NoStop}%
\bibitem [{\citenamefont {{Adriani}}\ \emph {et~al.}(2011)\citenamefont
  {{Adriani}}, \citenamefont {{Barbarino}}, \citenamefont {{Bazilevskaya}},
  \citenamefont {{Bellotti}}, \citenamefont {{Boezio}}, \citenamefont
  {{Bogomolov}}, \citenamefont {{Bonechi}}, \citenamefont {{Bongi}},
  \citenamefont {{Bonvicini}}, \citenamefont {{Borisov}}, \citenamefont
  {{Bottai}}, \citenamefont {{Bruno}}, \citenamefont {{Cafagna}}, \citenamefont
  {{Campana}}, \citenamefont {{Carbone}}, \citenamefont {{Carlson}},
  \citenamefont {{Casolino}}, \citenamefont {{Castellini}}, \citenamefont
  {{Consiglio}}, \citenamefont {{De Pascale}}, \citenamefont {{De Santis}},
  \citenamefont {{De Simone}}, \citenamefont {{Di Felice}}, \citenamefont
  {{Galper}}, \citenamefont {{Gillard}}, \citenamefont {{Grishantseva}},
  \citenamefont {{Jerse}}, \citenamefont {{Karelin}}, \citenamefont
  {{Koldashov}}, \citenamefont {{Krutkov}}, \citenamefont {{Kvashnin}},
  \citenamefont {{Leonov}}, \citenamefont {{Malakhov}}, \citenamefont
  {{Malvezzi}}, \citenamefont {{Marcelli}}, \citenamefont {{Mayorov}},
  \citenamefont {{Menn}}, \citenamefont {{Mikhailov}}, \citenamefont
  {{Mocchiutti}}, \citenamefont {{Monaco}}, \citenamefont {{Mori}},
  \citenamefont {{Nikonov}}, \citenamefont {{Osteria}}, \citenamefont
  {{Palma}}, \citenamefont {{Papini}}, \citenamefont {{Pearce}}, \citenamefont
  {{Picozza}}, \citenamefont {{Pizzolotto}}, \citenamefont {{Ricci}},
  \citenamefont {{Ricciarini}}, \citenamefont {{Rossetto}}, \citenamefont
  {{Sarkar}}, \citenamefont {{Simon}}, \citenamefont {{Sparvoli}},
  \citenamefont {{Spillantini}}, \citenamefont {{Stozhkov}}, \citenamefont
  {{Vacchi}}, \citenamefont {{Vannuccini}}, \citenamefont {{Vasilyev}},
  \citenamefont {{Voronov}}, \citenamefont {{Yurkin}}, \citenamefont {{Wu}},
  \citenamefont {{Zampa}}, \citenamefont {{Zampa}},\ and\ \citenamefont
  {{Zverev}}}]{AdrianiEtAl2011a}%
  \BibitemOpen
  \bibfield  {author} {\bibinfo {author} {\bibfnamefont {PAMELA Collaboration}},
    \bibinfo {author} {\bibfnamefont {O.}~\bibnamefont
      {{Adriani}}}, \bibinfo {author} {\etal},\ }\href
  {\doibase 10.1126/science.1199172} {\bibfield  {journal} {\bibinfo  {journal}
  {Science}\ }\textbf {\bibinfo {volume} {332}},\ \bibinfo {pages} {69}
  (\bibinfo {year} {2011})},\ \Eprint {http://arxiv.org/abs/1103.4055}
  {arXiv:1103.4055 [astro-ph.HE]} \BibitemShut {NoStop}%
\bibitem [{\citenamefont {{Aguilar}}\ \emph {et~al.}(2015)\citenamefont
  {{Aguilar}}, \citenamefont {{Aisa}}, \citenamefont {{Alpat}}, \citenamefont
  {{Alvino}}, \citenamefont {{Ambrosi}}, \citenamefont {{Andeen}},
  \citenamefont {{Arruda}}, \citenamefont {{Attig}}, \citenamefont
  {{Azzarello}}, \citenamefont {{Bachlechner}}, \citenamefont {{Barao}},
  \citenamefont {{Barrau}}, \citenamefont {{Barrin}}, \citenamefont
  {{Bartoloni}}, \citenamefont {{Basara}}, \citenamefont {{Battarbee}},
  \citenamefont {{Battiston}}, \citenamefont {{Bazo}}, \citenamefont
  {{Becker}}, \citenamefont {{Behlmann}}, \citenamefont {{Beischer}},
  \citenamefont {{Berdugo}}, \citenamefont {{Bertucci}}, \citenamefont
  {{Bigongiari}}, \citenamefont {{Bindi}}, \citenamefont {{Bizzaglia}},
  \citenamefont {{Bizzarri}}, \citenamefont {{Boella}}, \citenamefont {{de
  Boer}}, \citenamefont {{Bollweg}}, \citenamefont {{Bonnivard}}, \citenamefont
  {{Borgia}}, \citenamefont {{Borsini}}, \citenamefont {{Boschini}},
  \citenamefont {{Bourquin}}, \citenamefont {{Burger}}, \citenamefont
  {{Cadoux}}, \citenamefont {{Cai}}, \citenamefont {{Capell}}, \citenamefont
  {{Caroff}}, \citenamefont {{Casaus}}, \citenamefont {{Cascioli}},
  \citenamefont {{Castellini}}, \citenamefont {{Cernuda}}, \citenamefont
  {{Cerreta}}, \citenamefont {{Cervelli}}, \citenamefont {{Chae}},
  \citenamefont {{Chang}}, \citenamefont {{Chen}}, \citenamefont {{Chen}},
  \citenamefont {{Cheng}}, \citenamefont {{Chen}}, \citenamefont {{Cheng}},
  \citenamefont {{Chou}}, \citenamefont {{Choumilov}}, \citenamefont
  {{Choutko}}, \citenamefont {{Chung}}, \citenamefont {{Clark}}, \citenamefont
  {{Clavero}}, \citenamefont {{Coignet}}, \citenamefont {{Consolandi}},
  \citenamefont {{Contin}}, \citenamefont {{Corti}}, \citenamefont {{Gil}},
  \citenamefont {{Coste}}, \citenamefont {{Creus}}, \citenamefont
  {{Crispoltoni}}, \citenamefont {{Cui}}, \citenamefont {{Dai}}, \citenamefont
  {{Delgado}}, \citenamefont {{Della Torre}}, \citenamefont {{Demirk{\"o}z}},
  \citenamefont {{Derome}}, \citenamefont {{Di Falco}}, \citenamefont {{Di
  Masso}}, \citenamefont {{Dimiccoli}}, \citenamefont {{D{\'{\i}}az}},
  \citenamefont {{von Doetinchem}}, \citenamefont {{Donnini}}, \citenamefont
  {{Du}}, \citenamefont {{Duranti}}, \citenamefont {{D'Urso}}, \citenamefont
  {{Eline}}, \citenamefont {{Eppling}}, \citenamefont {{Eronen}}, \citenamefont
  {{Fan}}, \citenamefont {{Farnesini}}, \citenamefont {{Feng}}, \citenamefont
  {{Fiandrini}}, \citenamefont {{Fiasson}}, \citenamefont {{Finch}},
  \citenamefont {{Fisher}}, \citenamefont {{Galaktionov}}, \citenamefont
  {{Gallucci}}, \citenamefont {{Garc{\'{\i}}a}}, \citenamefont
  {{Garc{\'{\i}}a-L{\'o}pez}}, \citenamefont {{Gargiulo}}, \citenamefont
  {{Gast}}, \citenamefont {{Gebauer}}, \citenamefont {{Gervasi}}, \citenamefont
  {{Ghelfi}}, \citenamefont {{Gillard}}, \citenamefont {{Giovacchini}},
  \citenamefont {{Goglov}}, \citenamefont {{Gong}}, \citenamefont {{Goy}},
  \citenamefont {{Grabski}}, \citenamefont {{Grandi}}, \citenamefont
  {{Graziani}}, \citenamefont {{Guandalini}}, \citenamefont {{Guerri}},
  \citenamefont {{Guo}}, \citenamefont {{Haas}}, \citenamefont {{Habiby}},
  \citenamefont {{Haino}}, \citenamefont {{Han}}, \citenamefont {{He}},
  \citenamefont {{Heil}}, \citenamefont {{Hoffman}}, \citenamefont {{Hsieh}},
  \citenamefont {{Huang}}, \citenamefont {{Huh}}, \citenamefont {{Incagli}},
  \citenamefont {{Ionica}}, \citenamefont {{Jang}}, \citenamefont {{Jinchi}},
  \citenamefont {{Kanishev}}, \citenamefont {{Kim}}, \citenamefont {{Kim}},
  \citenamefont {{Kirn}}, \citenamefont {{Kossakowski}}, \citenamefont
  {{Kounina}}, \citenamefont {{Kounine}}, \citenamefont {{Koutsenko}},
  \citenamefont {{Krafczyk}}, \citenamefont {{La Vacca}}, \citenamefont
  {{Laudi}}, \citenamefont {{Laurenti}}, \citenamefont {{Lazzizzera}},
  \citenamefont {{Lebedev}}, \citenamefont {{Lee}}, \citenamefont {{Lee}},
  \citenamefont {{Leluc}}, \citenamefont {{Levi}}, \citenamefont {{Li}},
  \citenamefont {{Li}}, \citenamefont {{Li}}, \citenamefont {{Li}},
  \citenamefont {{Li}}, \citenamefont {{Li}}, \citenamefont {{Li}},
  \citenamefont {{Li}}, \citenamefont {{Li}}, \citenamefont {{Lim}},
  \citenamefont {{Lin}}, \citenamefont {{Lipari}}, \citenamefont {{Lippert}},
  \citenamefont {{Liu}}, \citenamefont {{Liu}}, \citenamefont {{Lolli}},
  \citenamefont {{Lomtadze}}, \citenamefont {{Lu}}, \citenamefont {{Lu}},
  \citenamefont {{Lu}}, \citenamefont {{Luebelsmeyer}}, \citenamefont {{Luo}},
  \citenamefont {{Lv}}, \citenamefont {{Majka}}, \citenamefont
  {{Ma{\~n}{\'a}}}, \citenamefont {{Mar{\'{\i}}n}}, \citenamefont {{Martin}},
  \citenamefont {{Mart{\'{\i}}nez}}, \citenamefont {{Masi}}, \citenamefont
  {{Maurin}}, \citenamefont {{Menchaca-Rocha}}, \citenamefont {{Meng}},
  \citenamefont {{Mo}}, \citenamefont {{Morescalchi}}, \citenamefont {{Mott}},
  \citenamefont {{M{\"u}ller}}, \citenamefont {{Ni}}, \citenamefont
  {{Nikonov}}, \citenamefont {{Nozzoli}}, \citenamefont {{Nunes}},
  \citenamefont {{Obermeier}}, \citenamefont {{Oliva}}, \citenamefont
  {{Orcinha}}, \citenamefont {{Palmonari}}, \citenamefont {{Palomares}},
  \citenamefont {{Paniccia}}, \citenamefont {{Papi}}, \citenamefont
  {{Pauluzzi}}, \citenamefont {{Pedreschi}}, \citenamefont {{Pensotti}},
  \citenamefont {{Pereira}}, \citenamefont {{Picot-Clemente}}, \citenamefont
  {{Pilo}}, \citenamefont {{Piluso}}, \citenamefont {{Pizzolotto}},
  \citenamefont {{Plyaskin}}, \citenamefont {{Pohl}}, \citenamefont
  {{Poireau}}, \citenamefont {{Postaci}}, \citenamefont {{Putze}},
  \citenamefont {{Quadrani}}, \citenamefont {{Qi}}, \citenamefont {{Qin}},
  \citenamefont {{Qu}}, \citenamefont {{R{\"a}ih{\"a}}}, \citenamefont
  {{Rancoita}}, \citenamefont {{Rapin}}, \citenamefont {{Ricol}}, \citenamefont
  {{Rodr{\'{\i}}guez}}, \citenamefont {{Rosier-Lees}}, \citenamefont
  {{Rozhkov}}, \citenamefont {{Rozza}}, \citenamefont {{Sagdeev}},
  \citenamefont {{Sandweiss}}, \citenamefont {{Saouter}}, \citenamefont
  {{Sbarra}}, \citenamefont {{Schael}}, \citenamefont {{Schmidt}},
  \citenamefont {{von Dratzig}}, \citenamefont {{Schwering}}, \citenamefont
  {{Scolieri}}, \citenamefont {{Seo}}, \citenamefont {{Shan}}, \citenamefont
  {{Shan}}, \citenamefont {{Shi}}, \citenamefont {{Shi}}, \citenamefont
  {{Shi}}, \citenamefont {{Siedenburg}}, \citenamefont {{Son}}, \citenamefont
  {{Spada}}, \citenamefont {{Spinella}}, \citenamefont {{Sun}}, \citenamefont
  {{Sun}}, \citenamefont {{Tacconi}}, \citenamefont {{Tang}}, \citenamefont
  {{Tang}}, \citenamefont {{Tang}}, \citenamefont {{Tao}}, \citenamefont
  {{Tescaro}}, \citenamefont {{Ting}}, \citenamefont {{Ting}}, \citenamefont
  {{Tomassetti}}, \citenamefont {{Torsti}}, \citenamefont {{T{\"u}rko{\v
  g}lu}}, \citenamefont {{Urban}}, \citenamefont {{Vagelli}}, \citenamefont
  {{Valente}}, \citenamefont {{Vannini}}, \citenamefont {{Valtonen}},
  \citenamefont {{Vaurynovich}}, \citenamefont {{Vecchi}}, \citenamefont
  {{Velasco}}, \citenamefont {{Vialle}}, \citenamefont {{Vitale}},
  \citenamefont {{Vitillo}}, \citenamefont {{Wang}}, \citenamefont {{Wang}},
  \citenamefont {{Wang}}, \citenamefont {{Wang}}, \citenamefont {{Wang}},
  \citenamefont {{Wang}}, \citenamefont {{Weng}}, \citenamefont {{Whitman}},
  \citenamefont {{Wienkenh{\"o}ver}}, \citenamefont {{Wu}}, \citenamefont
  {{Wu}}, \citenamefont {{Xia}}, \citenamefont {{Xie}}, \citenamefont {{Xie}},
  \citenamefont {{Xiong}}, \citenamefont {{Xin}}, \citenamefont {{Xu}},
  \citenamefont {{Xu}}, \citenamefont {{Yan}}, \citenamefont {{Yang}},
  \citenamefont {{Yang}}, \citenamefont {{Ye}}, \citenamefont {{Yi}},
  \citenamefont {{Yu}}, \citenamefont {{Yu}}, \citenamefont {{Zeissler}},
  \citenamefont {{Zhang}}, \citenamefont {{Zhang}}, \citenamefont {{Zhang}},
  \citenamefont {{Zhang}}, \citenamefont {{Zheng}}, \citenamefont {{Zhuang}},
  \citenamefont {{Zhukov}}, \citenamefont {{Zichichi}}, \citenamefont
  {{Zimmermann}}, \citenamefont {{Zuccon}}, \citenamefont {{Zurbach}},\ and\
  \citenamefont {{AMS Collaboration}}}]{AguilarEtAl2015}%
  \BibitemOpen
  \bibfield  {author} {\bibinfo {author} {\bibfnamefont {AMS-02 Collaboration}},
  \bibinfo {author} {\bibfnamefont {M.}~\bibnamefont
  {{Aguilar}}}, \bibinfo {author} {\etal},\ }\href
  {\doibase 10.1103/PhysRevLett.114.171103} {\bibfield  {journal} {\bibinfo
  {journal} {\prl}\ }\textbf {\bibinfo {volume} {114}},\ \bibinfo {eid}
  {171103} (\bibinfo {year} {2015})}\BibitemShut {NoStop}%
\bibitem [{\citenamefont {Aguilar}\ \emph {et~al.}(2015)\citenamefont
  {Aguilar}, \citenamefont {Aisa}, \citenamefont {Alpat}, \citenamefont
  {Alvino}, \citenamefont {Ambrosi}, \citenamefont {Andeen}, \citenamefont
  {Arruda}, \citenamefont {Attig}, \citenamefont {Azzarello}, \citenamefont
  {Bachlechner}, \citenamefont {Barao}, \citenamefont {Barrau}, \citenamefont
  {Barrin}, \citenamefont {Bartoloni}, \citenamefont {Basara}, \citenamefont
  {Battarbee}, \citenamefont {Battiston}, \citenamefont {Bazo}, \citenamefont
  {Becker}, \citenamefont {Behlmann}, \citenamefont {Beischer}, \citenamefont
  {Berdugo}, \citenamefont {Bertucci}, \citenamefont {Bindi}, \citenamefont
  {Bizzaglia}, \citenamefont {Bizzarri}, \citenamefont {Boella}, \citenamefont
  {de~Boer}, \citenamefont {Bollweg}, \citenamefont {Bonnivard}, \citenamefont
  {Borgia}, \citenamefont {Borsini}, \citenamefont {Boschini}, \citenamefont
  {Bourquin}, \citenamefont {Burger}, \citenamefont {Cadoux}, \citenamefont
  {Cai}, \citenamefont {Capell}, \citenamefont {Caroff}, \citenamefont
  {Casaus}, \citenamefont {Castellini}, \citenamefont {Cernuda}, \citenamefont
  {Cerreta}, \citenamefont {Cervelli}, \citenamefont {Chae}, \citenamefont
  {Chang}, \citenamefont {Chen}, \citenamefont {Chen}, \citenamefont {Chen},
  \citenamefont {Chen}, \citenamefont {Cheng}, \citenamefont {Chou},
  \citenamefont {Choumilov}, \citenamefont {Choutko}, \citenamefont {Chung},
  \citenamefont {Clark}, \citenamefont {Clavero}, \citenamefont {Coignet},
  \citenamefont {Consolandi}, \citenamefont {Contin}, \citenamefont {Corti},
  \citenamefont {Gil}, \citenamefont {Coste}, \citenamefont {Creus},
  \citenamefont {Crispoltoni}, \citenamefont {Cui}, \citenamefont {Dai},
  \citenamefont {Delgado}, \citenamefont {Della~Torre}, \citenamefont
  {Demirk{\"o}z}, \citenamefont {Derome}, \citenamefont {Di~Falco},
  \citenamefont {Di~Masso}, \citenamefont {Dimiccoli}, \citenamefont
  {D{\'{\i}}az}, \citenamefont {von Doetinchem}, \citenamefont {Donnini},
  \citenamefont {Duranti}, \citenamefont {D'Urso}, \citenamefont {Egorov},
  \citenamefont {Eline}, \citenamefont {Eppling}, \citenamefont {Eronen},
  \citenamefont {Fan}, \citenamefont {Farnesini}, \citenamefont {Feng},
  \citenamefont {Fiandrini}, \citenamefont {Fiasson}, \citenamefont {Finch},
  \citenamefont {Fisher}, \citenamefont {Formato}, \citenamefont {Galaktionov},
  \citenamefont {Gallucci}, \citenamefont {Garc{\'{\i}}a}, \citenamefont
  {Garc{\'{\i}}a-L{\'o}pez}, \citenamefont {Gargiulo}, \citenamefont {Gast},
  \citenamefont {Gebauer}, \citenamefont {Gervasi}, \citenamefont {Ghelfi},
  \citenamefont {Giovacchini}, \citenamefont {Goglov}, \citenamefont {Gong},
  \citenamefont {Goy}, \citenamefont {Grabski}, \citenamefont {Grandi},
  \citenamefont {Graziani}, \citenamefont {Guandalini}, \citenamefont {Guerri},
  \citenamefont {Guo}, \citenamefont {Haas}, \citenamefont {Habiby},
  \citenamefont {Haino}, \citenamefont {Han}, \citenamefont {He}, \citenamefont
  {Heil}, \citenamefont {Hoffman}, \citenamefont {Hsieh}, \citenamefont
  {Huang}, \citenamefont {Huh}, \citenamefont {Incagli}, \citenamefont
  {Ionica}, \citenamefont {Jang}, \citenamefont {Jinchi}, \citenamefont
  {Kanishev}, \citenamefont {Kim}, \citenamefont {Kim}, \citenamefont {Kirn},
  \citenamefont {Korkmaz}, \citenamefont {Kossakowski}, \citenamefont
  {Kounina}, \citenamefont {Kounine}, \citenamefont {Koutsenko}, \citenamefont
  {Krafczyk}, \citenamefont {La~Vacca}, \citenamefont {Laudi}, \citenamefont
  {Laurenti}, \citenamefont {Lazzizzera}, \citenamefont {Lebedev},
  \citenamefont {Lee}, \citenamefont {Lee}, \citenamefont {Leluc},
  \citenamefont {Li}, \citenamefont {Li}, \citenamefont {Li}, \citenamefont
  {Li}, \citenamefont {Li}, \citenamefont {Li}, \citenamefont {Li},
  \citenamefont {Li}, \citenamefont {Li}, \citenamefont {Li}, \citenamefont
  {Lim}, \citenamefont {Lin}, \citenamefont {Lipari}, \citenamefont {Lippert},
  \citenamefont {Liu}, \citenamefont {Liu}, \citenamefont {Liu}, \citenamefont
  {Lolli}, \citenamefont {Lomtadze}, \citenamefont {Lu}, \citenamefont {Lu},
  \citenamefont {Lu}, \citenamefont {Luebelsmeyer}, \citenamefont {Luo},
  \citenamefont {Luo}, \citenamefont {Lv}, \citenamefont {Majka}, \citenamefont
  {Ma{\~n}{\'a}}, \citenamefont {Mar{\'{\i}}n}, \citenamefont {Martin},
  \citenamefont {Mart{\'{\i}}nez}, \citenamefont {Masi}, \citenamefont
  {Maurin}, \citenamefont {Menchaca-Rocha}, \citenamefont {Meng}, \citenamefont
  {Mo}, \citenamefont {Morescalchi}, \citenamefont {Mott}, \citenamefont
  {M{\"u}ller}, \citenamefont {Nelson}, \citenamefont {Ni}, \citenamefont
  {Nikonov}, \citenamefont {Nozzoli}, \citenamefont {Nunes}, \citenamefont
  {Obermeier}, \citenamefont {Oliva}, \citenamefont {Orcinha}, \citenamefont
  {Palmonari}, \citenamefont {Palomares}, \citenamefont {Paniccia},
  \citenamefont {Papi}, \citenamefont {Pauluzzi}, \citenamefont {Pedreschi},
  \citenamefont {Pensotti}, \citenamefont {Pereira}, \citenamefont
  {Picot-Clemente}, \citenamefont {Pilo}, \citenamefont {Piluso}, \citenamefont
  {Pizzolotto}, \citenamefont {Plyaskin}, \citenamefont {Pohl}, \citenamefont
  {Poireau}, \citenamefont {Putze}, \citenamefont {Quadrani}, \citenamefont
  {Qi}, \citenamefont {Qin}, \citenamefont {Qu}, \citenamefont {R{\"a}ih{\"a}},
  \citenamefont {Rancoita}, \citenamefont {Rapin}, \citenamefont {Ricol},
  \citenamefont {Rodr{\'{\i}}guez}, \citenamefont {Rosier-Lees}, \citenamefont
  {Rozhkov}, \citenamefont {Rozza}, \citenamefont {Sagdeev}, \citenamefont
  {Sandweiss}, \citenamefont {Saouter}, \citenamefont {Schael}, \citenamefont
  {Schmidt}, \citenamefont {von Dratzig}, \citenamefont {Schwering},
  \citenamefont {Scolieri}, \citenamefont {Seo}, \citenamefont {Shan},
  \citenamefont {Shan}, \citenamefont {Shi}, \citenamefont {Shi}, \citenamefont
  {Shi}, \citenamefont {Siedenburg}, \citenamefont {Son}, \citenamefont {Song},
  \citenamefont {Spada}, \citenamefont {Spinella}, \citenamefont {Sun},
  \citenamefont {Sun}, \citenamefont {Tacconi}, \citenamefont {Tang},
  \citenamefont {Tang}, \citenamefont {Tang}, \citenamefont {Tao},
  \citenamefont {Tescaro}, \citenamefont {Ting}, \citenamefont {Ting},
  \citenamefont {Tomassetti}, \citenamefont {Torsti}, \citenamefont
  {T{\"u}rko{\v g}lu}, \citenamefont {Urban}, \citenamefont {Vagelli},
  \citenamefont {Valente}, \citenamefont {Vannini}, \citenamefont {Valtonen},
  \citenamefont {Vaurynovich}, \citenamefont {Vecchi}, \citenamefont {Velasco},
  \citenamefont {Vialle}, \citenamefont {Vitale}, \citenamefont {Vitillo},
  \citenamefont {Wang}, \citenamefont {Wang}, \citenamefont {Wang},
  \citenamefont {Wang}, \citenamefont {Wang}, \citenamefont {Wang},
  \citenamefont {Weng}, \citenamefont {Whitman}, \citenamefont
  {Wienkenh{\"o}ver}, \citenamefont {Willenbrock}, \citenamefont {Wu},
  \citenamefont {Wu}, \citenamefont {Xia}, \citenamefont {Xie}, \citenamefont
  {Xie}, \citenamefont {Xiong}, \citenamefont {Xu}, \citenamefont {Xu},
  \citenamefont {Yan}, \citenamefont {Yang}, \citenamefont {Yang},
  \citenamefont {Yang}, \citenamefont {Ye}, \citenamefont {Yi}, \citenamefont
  {Yu}, \citenamefont {Yu}, \citenamefont {Zeissler}, \citenamefont {Zhang},
  \citenamefont {Zhang}, \citenamefont {Zhang}, \citenamefont {Zhang},
  \citenamefont {Zhang}, \citenamefont {Zhang}, \citenamefont {Zhang},
  \citenamefont {Zheng}, \citenamefont {Zhuang}, \citenamefont {Zhukov},
  \citenamefont {Zichichi}, \citenamefont {Zimmermann}, \citenamefont
  {Zuccon},\ and\ \citenamefont {Collaboration}}]{AguilarEtAl2015a}%
  \BibitemOpen
  \bibfield  {author} {\bibinfo {author} {\bibfnamefont {AMS-02 Collaboration}},
  \bibinfo {author} {\bibfnamefont {M.}~\bibnamefont
  {Aguilar}}, \bibinfo {author} {\etal},\ }\href {\doibase
  10.1103/PhysRevLett.115.211101} {\bibfield  {journal} {\bibinfo  {journal}
  {\prl}\ }\textbf {\bibinfo {volume} {115}},\ \bibinfo {eid} {211101}
  (\bibinfo {year} {2015})}\BibitemShut {NoStop}%
\bibitem [{\citenamefont {{Blasi}}\ \emph {et~al.}(2012)\citenamefont
  {{Blasi}}, \citenamefont {{Amato}},\ and\ \citenamefont
  {{Serpico}}}]{BlasiEtal2012a}%
  \BibitemOpen
  \bibfield  {author} {\bibinfo {author} {\bibfnamefont {P.}~\bibnamefont
  {{Blasi}}}, \bibinfo {author} {\bibfnamefont {E.}~\bibnamefont {{Amato}}}, \
  and\ \bibinfo {author} {\bibfnamefont {P.~D.}\ \bibnamefont {{Serpico}}},\
  }\href {\doibase 10.1103/PhysRevLett.109.061101} {\bibfield  {journal}
  {\bibinfo  {journal} {\prl}\ }\textbf {\bibinfo {volume}
  {109}},\ \bibinfo {eid} {061101} (\bibinfo {year} {2012})},\ \Eprint
  {http://arxiv.org/abs/1207.3706} {arXiv:1207.3706 [astro-ph.HE]} \BibitemShut
  {NoStop}%
\bibitem [{\citenamefont {Blasi}(2017)}]{Blasi2017}%
  \BibitemOpen
  \bibfield  {author} {\bibinfo {author} {\bibfnamefont {P.}~\bibnamefont
  {Blasi}},\ }\href {\doibase 10.1093/mnras/stx1696} {\bibfield  {journal}
  {\bibinfo  {journal} {\mnras}\ }\textbf {\bibinfo {volume} {471}},\ \bibinfo
  {pages} {1662} (\bibinfo {year} {2017})},\ \Eprint
  {http://arxiv.org/abs/1707.00525} {arXiv:1707.00525 [astro-ph.HE]}
  \BibitemShut {NoStop}%
\bibitem [{\citenamefont {G{\'e}nolini}\ \emph {et~al.}(2017)\citenamefont
  {G{\'e}nolini}, \citenamefont {Serpico}, \citenamefont {Boudaud},
  \citenamefont {Caroff}, \citenamefont {Poulin}, \citenamefont {Derome},
  \citenamefont {Lavalle}, \citenamefont {Maurin}, \citenamefont {Poireau},
  \citenamefont {Rosier}, \citenamefont {Salati},\ and\ \citenamefont
  {Vecchi}}]{GenoliniEtAl2017}%
  \BibitemOpen
  \bibfield  {author} {\bibinfo {author} {\bibfnamefont {Y.}~\bibnamefont
  {G{\'e}nolini}}, \bibinfo {author} {\bibfnamefont {P.~D.}\ \bibnamefont
  {Serpico}}, \bibinfo {author} {\bibfnamefont {M.}~\bibnamefont {Boudaud}},
  \bibinfo {author} {\bibfnamefont {S.}~\bibnamefont {Caroff}}, \bibinfo
  {author} {\bibfnamefont {V.}~\bibnamefont {Poulin}}, \bibinfo {author}
  {\bibfnamefont {L.}~\bibnamefont {Derome}}, \bibinfo {author} {\bibfnamefont
  {J.}~\bibnamefont {Lavalle}}, \bibinfo {author} {\bibfnamefont
  {D.}~\bibnamefont {Maurin}}, \bibinfo {author} {\bibfnamefont
  {V.}~\bibnamefont {Poireau}}, \bibinfo {author} {\bibfnamefont
  {S.}~\bibnamefont {Rosier}}, \bibinfo {author} {\bibfnamefont
  {P.}~\bibnamefont {Salati}}, \ and\ \bibinfo {author} {\bibfnamefont
  {M.}~\bibnamefont {Vecchi}},\ }\href {\doibase
  10.1103/PhysRevLett.119.241101} {\bibfield  {journal} {\bibinfo  {journal}
  {\prl}\ }\textbf {\bibinfo {volume} {119}},\ \bibinfo {eid} {241101}
  (\bibinfo {year} {2017})},\ \Eprint {http://arxiv.org/abs/1706.09812}
  {arXiv:1706.09812 [astro-ph.HE]} \BibitemShut {NoStop}%
\bibitem [{\citenamefont {Evoli}\ \emph {et~al.}(2018)\citenamefont {Evoli},
  \citenamefont {Blasi}, \citenamefont {Morlino},\ and\ \citenamefont
  {Aloisio}}]{EvoliEtAl2018}%
  \BibitemOpen
  \bibfield  {author} {\bibinfo {author} {\bibfnamefont {C.}~\bibnamefont
  {Evoli}}, \bibinfo {author} {\bibfnamefont {P.}~\bibnamefont {Blasi}},
  \bibinfo {author} {\bibfnamefont {G.}~\bibnamefont {Morlino}}, \ and\
  \bibinfo {author} {\bibfnamefont {R.}~\bibnamefont {Aloisio}},\ }\href
  {\doibase 10.1103/PhysRevLett.121.021102} {\bibfield  {journal} {\bibinfo
  {journal} {\prl}\ }\textbf {\bibinfo {volume} {121}},\ \bibinfo {eid}
  {021102} (\bibinfo {year} {2018})},\ \Eprint
  {http://arxiv.org/abs/1806.04153} {arXiv:1806.04153 [astro-ph.HE]}
  \BibitemShut {NoStop}%
\bibitem [{\citenamefont {{Boudaud}}\ \emph
  {et~al.}(2015{\natexlab{b}})\citenamefont {{Boudaud}}, \citenamefont
  {{Cirelli}}, \citenamefont {{Giesen}},\ and\ \citenamefont
  {{Salati}}}]{BoudaudEtAl2015a}%
  \BibitemOpen
  \bibfield  {author} {\bibinfo {author} {\bibfnamefont {M.}~\bibnamefont
  {{Boudaud}}}, \bibinfo {author} {\bibfnamefont {M.}~\bibnamefont
  {{Cirelli}}}, \bibinfo {author} {\bibfnamefont {G.}~\bibnamefont {{Giesen}}},
  \ and\ \bibinfo {author} {\bibfnamefont {P.}~\bibnamefont {{Salati}}},\
  }\href {\doibase 10.1088/1475-7516/2015/05/013} {\bibfield  {journal}
  {\bibinfo  {journal} {\jcap}\ }\textbf {\bibinfo {volume} {5}},\ \bibinfo
  {eid} {013} (\bibinfo {year} {2015}{\natexlab{b}})},\ \Eprint
  {http://arxiv.org/abs/1412.5696} {arXiv:1412.5696 [astro-ph.HE]} \BibitemShut
  {NoStop}%
\bibitem [{\citenamefont {{Ferri{\`e}re}}(2001)}]{Ferriere2001}%
  \BibitemOpen
  \bibfield  {author} {\bibinfo {author} {\bibfnamefont {K.~M.}\ \bibnamefont
  {{Ferri{\`e}re}}},\ }\href {\doibase 10.1103/RevModPhys.73.1031} {\bibfield
  {journal} {\bibinfo  {journal} {Reviews of Modern Physics}\ }\textbf
  {\bibinfo {volume} {73}},\ \bibinfo {pages} {1031} (\bibinfo {year}
  {2001})},\ \Eprint {http://arxiv.org/abs/astro-ph/0106359} {astro-ph/0106359}
  \BibitemShut {NoStop}%
\bibitem [{\citenamefont {J{\'o}hannesson}\ \emph {et~al.}(2018)\citenamefont
  {J{\'o}hannesson}, \citenamefont {Porter},\ and\ \citenamefont
  {Moskalenko}}]{JohannessonEtAl2018}%
  \BibitemOpen
  \bibfield  {author} {\bibinfo {author} {\bibfnamefont {G.}~\bibnamefont
  {J{\'o}hannesson}}, \bibinfo {author} {\bibfnamefont {T.~A.}\ \bibnamefont
  {Porter}}, \ and\ \bibinfo {author} {\bibfnamefont {I.~V.}\ \bibnamefont
  {Moskalenko}},\ }\href {\doibase 10.3847/1538-4357/aab26e} {\bibfield
  {journal} {\bibinfo  {journal} {\apj}\ }\textbf {\bibinfo {volume} {856}},\
  \bibinfo {eid} {45} (\bibinfo {year} {2018})},\ \Eprint
  {http://arxiv.org/abs/1802.08646} {arXiv:1802.08646 [astro-ph.HE]}
  \BibitemShut {NoStop}%
\bibitem [{\citenamefont {{Potgieter}}(2013)}]{Potgieter2013}%
  \BibitemOpen
  \bibfield  {author} {\bibinfo {author} {\bibfnamefont {M.}~\bibnamefont
  {{Potgieter}}},\ }\href {\doibase 10.12942/lrsp-2013-3} {\bibfield  {journal}
  {\bibinfo  {journal} {Living Reviews in Solar Physics}\ }\textbf {\bibinfo
  {volume} {10}},\ \bibinfo {pages} {3} (\bibinfo {year} {2013})},\ \Eprint
  {http://arxiv.org/abs/1306.4421} {arXiv:1306.4421 [physics.space-ph]}
  \BibitemShut {NoStop}%
\bibitem [{\citenamefont {{Gleeson}}\ and\ \citenamefont
  {{Axford}}(1968)}]{GleesonEtAl1968a}%
  \BibitemOpen
  \bibfield  {author} {\bibinfo {author} {\bibfnamefont {L.~J.}\ \bibnamefont
  {{Gleeson}}}\ and\ \bibinfo {author} {\bibfnamefont {W.~I.}\ \bibnamefont
  {{Axford}}},\ }\href {\doibase 10.1086/149822} {\bibfield  {journal}
  {\bibinfo  {journal} {\apj}\ }\textbf {\bibinfo {volume} {154}},\ \bibinfo
  {pages} {1011} (\bibinfo {year} {1968})}\BibitemShut {NoStop}%
\bibitem [{\citenamefont {{Fisk}}(1971)}]{Fisk1971}%
  \BibitemOpen
  \bibfield  {author} {\bibinfo {author} {\bibfnamefont {L.~A.}\ \bibnamefont
  {{Fisk}}},\ }\href {\doibase 10.1029/JA076i001p00221} {\bibfield  {journal}
  {\bibinfo  {journal} {\jgr}\ }\textbf {\bibinfo {volume} {76}},\ \bibinfo
  {pages} {221} (\bibinfo {year} {1971})}\BibitemShut {NoStop}%
\bibitem [{\citenamefont {{Ghelfi}}\ \emph {et~al.}(2016)\citenamefont
  {{Ghelfi}}, \citenamefont {{Barao}}, \citenamefont {{Derome}},\ and\
  \citenamefont {{Maurin}}}]{GhelfiEtAl2016}%
  \BibitemOpen
  \bibfield  {author} {\bibinfo {author} {\bibfnamefont {A.}~\bibnamefont
  {{Ghelfi}}}, \bibinfo {author} {\bibfnamefont {F.}~\bibnamefont {{Barao}}},
  \bibinfo {author} {\bibfnamefont {L.}~\bibnamefont {{Derome}}}, \ and\
  \bibinfo {author} {\bibfnamefont {D.}~\bibnamefont {{Maurin}}},\ }\href
  {\doibase 10.1051/0004-6361/201527852} {\bibfield  {journal} {\bibinfo
  {journal} {\aap}\ }\textbf {\bibinfo {volume} {591}},\ \bibinfo {eid} {A94}
  (\bibinfo {year} {2016})},\ \Eprint {http://arxiv.org/abs/1511.08650}
  {arXiv:1511.08650 [astro-ph.HE]} \BibitemShut {NoStop}%
\bibitem [{\citenamefont {{Sun}}\ and\ \citenamefont
  {{Reich}}(2010)}]{SunEtAl2010}%
  \BibitemOpen
  \bibfield  {author} {\bibinfo {author} {\bibfnamefont {X.-H.}\ \bibnamefont
  {{Sun}}}\ and\ \bibinfo {author} {\bibfnamefont {W.}~\bibnamefont
  {{Reich}}},\ }\href {\doibase 10.1088/1674-4527/10/12/009} {\bibfield
  {journal} {\bibinfo  {journal} {Research in Astronomy and Astrophysics}\
  }\textbf {\bibinfo {volume} {10}},\ \bibinfo {pages} {1287} (\bibinfo {year}
  {2010})},\ \Eprint {http://arxiv.org/abs/1010.4394} {arXiv:1010.4394}
  \BibitemShut {NoStop}%
\bibitem [{\citenamefont {Jansson}\ and\ \citenamefont
  {Farrar}(2012)}]{JanssonEtAl2012}%
  \BibitemOpen
  \bibfield  {author} {\bibinfo {author} {\bibfnamefont {R.}~\bibnamefont
  {Jansson}}\ and\ \bibinfo {author} {\bibfnamefont {G.~R.}\ \bibnamefont
  {Farrar}},\ }\href {\doibase 10.1088/2041-8205/761/1/L11} {\bibfield
  {journal} {\bibinfo  {journal} {\apjl}\ }\textbf {\bibinfo {volume} {761}},\
  \bibinfo {eid} {L11} (\bibinfo {year} {2012})},\ \Eprint
  {http://arxiv.org/abs/1210.7820} {arXiv:1210.7820 [astro-ph.GA]} \BibitemShut
  {NoStop}%
\bibitem [{\citenamefont {Porter}\ \emph {et~al.}(2017)\citenamefont {Porter},
  \citenamefont {J{\'o}hannesson},\ and\ \citenamefont
  {Moskalenko}}]{PorterEtAl2017}%
  \BibitemOpen
  \bibfield  {author} {\bibinfo {author} {\bibfnamefont {T.~A.}\ \bibnamefont
  {Porter}}, \bibinfo {author} {\bibfnamefont {G.}~\bibnamefont
  {J{\'o}hannesson}}, \ and\ \bibinfo {author} {\bibfnamefont {I.~V.}\
  \bibnamefont {Moskalenko}},\ }\href {\doibase 10.3847/1538-4357/aa844d}
  {\bibfield  {journal} {\bibinfo  {journal} {\apj}\ }\textbf {\bibinfo
  {volume} {846}},\ \bibinfo {eid} {67} (\bibinfo {year} {2017})},\ \Eprint
  {http://arxiv.org/abs/1708.00816} {arXiv:1708.00816 [astro-ph.HE]}
  \BibitemShut {NoStop}%
\bibitem [{\citenamefont {Choquette}\ \emph {et~al.}(2016)\citenamefont
  {Choquette}, \citenamefont {Cline},\ and\ \citenamefont
  {Cornell}}]{ChoquetteEtAl2016}%
  \BibitemOpen
  \bibfield  {author} {\bibinfo {author} {\bibfnamefont {J.}~\bibnamefont
  {Choquette}}, \bibinfo {author} {\bibfnamefont {J.~M.}\ \bibnamefont
  {Cline}}, \ and\ \bibinfo {author} {\bibfnamefont {J.~M.}\ \bibnamefont
  {Cornell}},\ }\href {\doibase 10.1103/PhysRevD.94.015018} {\bibfield
  {journal} {\bibinfo  {journal} {\prd}\ }\textbf {\bibinfo {volume} {94}},\
  \bibinfo {eid} {015018} (\bibinfo {year} {2016})},\ \Eprint
  {http://arxiv.org/abs/1604.01039} {arXiv:1604.01039 [hep-ph]} \BibitemShut
  {NoStop}%
\bibitem [{\citenamefont {Boschini}\ \emph {et~al.}(2018)\citenamefont
  {Boschini}, \citenamefont {Della~Torre}, \citenamefont {Gervasi},
  \citenamefont {Grandi}, \citenamefont {J{\'o}hannesson}, \citenamefont
  {La~Vacca}, \citenamefont {Masi}, \citenamefont {Moskalenko}, \citenamefont
  {Pensotti}, \citenamefont {Porter}, \citenamefont {Quadrani}, \citenamefont
  {Rancoita}, \citenamefont {Rozza},\ and\ \citenamefont
  {Tacconi}}]{BoschiniEtAl2018}%
  \BibitemOpen
  \bibfield  {author} {\bibinfo {author} {\bibfnamefont {M.~J.}\ \bibnamefont
  {Boschini}}, \bibinfo {author} {\bibfnamefont {S.}~\bibnamefont
  {Della~Torre}}, \bibinfo {author} {\bibfnamefont {M.}~\bibnamefont
      {Gervasi}}, \bibinfo {author} {\bibfnamefont {D.}~\bibnamefont {Grandi}},
  \bibinfo {author} {\bibfnamefont {G.}~\bibnamefont {J{\'o}hannesson}},
  \bibinfo {author} {\bibfnamefont {G.}~\bibnamefont {La~Vacca}}, \bibinfo
  {author} {\bibfnamefont {N.}~\bibnamefont {Masi}}, \bibinfo {author}
  {\bibfnamefont {I.~V.}\ \bibnamefont {Moskalenko}}, \bibinfo {author}
  {\bibfnamefont {S.}~\bibnamefont {Pensotti}}, \bibinfo {author}
  {\bibfnamefont {T.~A.}\ \bibnamefont {Porter}}, \bibinfo {author}
  {\bibfnamefont {L.}~\bibnamefont {Quadrani}}, \bibinfo {author}
  {\bibfnamefont {P.~G.}\ \bibnamefont {Rancoita}}, \bibinfo {author}
  {\bibfnamefont {D.}~\bibnamefont {Rozza}}, \ and\ \bibinfo {author}
  {\bibfnamefont {M.}~\bibnamefont {Tacconi}},\ }\href {\doibase
  10.3847/1538-4357/aaa75e} {\bibfield  {journal} {\bibinfo  {journal} {\apj}\
  }\textbf {\bibinfo {volume} {854}},\ \bibinfo {eid} {94} (\bibinfo {year}
  {2018})},\ \Eprint {http://arxiv.org/abs/1801.04059} {arXiv:1801.04059
  [astro-ph.HE]} \BibitemShut {NoStop}%
\end{thebibliography}%

\end{document}